\documentclass[a4paper,11pt,onecolumn]{quantumarticle}
\pdfoutput=1
\usepackage[utf8]{inputenc}
\usepackage[english]{babel}
\usepackage[T1]{fontenc}

\usepackage{hyperref}
\usepackage{lipsum}
\usepackage{array}
\usepackage{amsmath}  
\usepackage{amsfonts} 
\usepackage{amssymb}  
\usepackage{amsthm}   
\usepackage{bm}       
\usepackage{stmaryrd} 
\usepackage{booktabs} 
\usepackage{multirow} 
\usepackage{here}     
\usepackage{ifthen}   
\usepackage{tikz}     
\usepackage{pgfplots} 
\usepackage{appendix}
\usepackage{cite}

\usetikzlibrary{patterns}
\usetikzlibrary{calc}
\usetikzlibrary{math}
\usetikzlibrary{positioning}
\usetikzlibrary{arrows.meta}
\usetikzlibrary{shapes.geometric}
\usetikzlibrary{shapes.misc}
\usetikzlibrary{decorations.pathreplacing}

\theoremstyle{definition}
\newtheoremstyle{tight}
    {3pt}                          
    {3pt}                           
    {\addtolength{\leftskip}{2em}} 
    {}                             
    {\bfseries}                    
    {.}                            
    {.5em}                         
    {}                             
\theoremstyle{tight}

\newtheorem*{myDefinition*}{Definition}

\newtheorem{myAssumptionA}{}
\newtheorem{myAssumptionB}{}
\newtheorem{myAssumptionC}{}

\newcommand{\bettershortstack}[4][c]{%
  \renewcommand{\arraystretch}{#2}
  \begin{tabular}[b]{@{}#1@{}}
  #4
  \end{tabular}
  \renewcommand{\arraystretch}{#3}
}

\newcommand{\mvariable}[3]{%
  \ifthenelse{\equal{#3}{}}{{#1}(\mbox{#2})}{}
  \ifthenelse{\equal{#3}{leftright}}{{#1}\left(\mbox{#2}\right)}{}
  \ifthenelse{\equal{#3}{big}}{{#1}\big(\mbox{#2}\big)}{}
  \ifthenelse{\equal{#3}{Big}}{{#1}\Big(\mbox{#2}\Big)}{} }

\newcommand{\xvar}[2]{\mvariable{x}{#1}{#2}}
\newcommand{\xjvar}[2]{\mvariable{x^{(j)}}{#1}{#2}}
\newcommand{\xop}[2]{\mvariable{\hat{x}}{#1}{#2}}
\newcommand{\zop}[2]{\mvariable{\hat{z}}{#1}{#2}}
\newcommand{\zopd}[2]{\mvariable{\hat{z}^{\dagger}}{#1}{#2}}

\newcommand{\stateset}[1]{\{\!\!\{{#1}\}\!\!\}}

\newcommand{\state}[2]{%
  \ifthenelse{\equal{#2}{}}{{\{{#1}\}}}{{\{{#1}\}^{{#2}}}} }

\newcommand{\occupation}[2]{%
  \ifthenelse{\equal{#2}{}}{{\mathrm{o}}({#1})}{}
  \ifthenelse{\equal{#2}{leftright}}{{\mathrm{o}}\left({#1}\right)}{}
  \ifthenelse{\equal{#2}{big}}{{\mathrm{o}}\big({#1}\big)}{}
  \ifthenelse{\equal{#2}{Big}}{{\mathrm{o}}\Big({#1}\Big)}{} }

\begin{document}
\title{Hidden variable theory for non-relativistic QED: the critical role of selection rules}
\author{Hiroshi Ishikawa}
\email{HiroshiIshikawaHVT@gmail.com, hishihvt@outlook.jp}
\affiliation{Member, Japanese Society of Applied Physics, Japan}
\maketitle

\begin{abstract}
Quantum theory, despite its remarkable success, struggles to represent certain experimental data, particularly those involving integer functions and deterministic relations between quantum jumps.
We address this limitation by proposing a hidden variable theory compatible with non-relativistic quantum electrodynamics (QED). 
Our approach introduces logical variables to describe propositions about the occupation of stationary states, based on three key assumptions: (i) the probabilities of propositions are predictable by quantum theory, (ii) the truth values of propositions conform to Boolean logic, and (iii) propositions satisfy a novel selection rule: ${\rm{Tr}}[\hat{K}(t) \hat{\rho} \hat{K}^{\dagger}(t)]\,{=}\,{\rm{Tr}}[({\mathcal{P}}\hat{K}(t)) \hat{\rho} ({\mathcal{P}}\hat{K}$ $(t))^{\dagger}]$.
Here, $\hat{K}(t)$ is the product of projection operators for the given propositions, and ${\mathcal{P}}\hat{K}(t)$ denotes $\hat{K}(t)$ rearranged arbitrarily.
This selection rule extends the consistency condition in the Consistent Histories approach, relaxing constraints imposed by no-go theorems on logical variable representation, joint probability existence, and measurement contextuality.
Consequently, our theory successfully describes the essential properties of individual trials, including the discreteness of quantum jumps, the continuity of classical trajectories, and deterministic relations between entangled subsystems. It achieves this without conflicting with quantum equations of motion, canonical commutation relations, or the operator ordering in quantum coherence functions.
The direct description of individual trials made possible by this approach sheds new light on the discrete nature of quantum phenomena and the essence of reality.
\end{abstract}

\setcounter{tocdepth}{2}
\tableofcontents

\section{Introduction}
\label{cIntro}
The difficulties quantum theory encounters in describing integer functions are widely recognized. 
This issue, however, has often been understood in the context of paradoxes, making a clear initial problem statement critical for avoiding logical pitfalls.
To better understand the paradoxical nature of this problem, it is instructive to revisit Einstein's seminal question: ``Can quantum-mechanical description of physical reality be considered complete?''~\cite{EPR,Einstein1949}. 
Today, this inquiry requires reframing, as subsequent research~\cite{Bell,Bell1966,CHSH,Aspect} has refuted Einstein's assumption of local point particles as the basis for modeling physical reality.
Revisiting the experimental observations that motivated Einstein's question~\cite{Einstein1905} reveals his primary concern: the discontinuous change in energy inferred from its dependence on the interger function, the number of photoemitted electrons during light-matter interactions.
This realization leads us to a more pertinent question for our times: ``Can the quantum-mechanical description of \emph{integer functions} be considered complete?''
\par
Contrary to what one might expect, the answer to this revised question is an unequivocal no. 
This definitive conclusion is rooted in the fundamental restrictions inherent to the standard formulation of quantum theory, as evidenced by various no-go theorems concerning the representation of integer functions and individual trials~\cite{Peres,Breuer}. 
In this respect, Einstein's critique about the incompleteness of quantum theory was essentially correct, albeit based on inappropriate assumptions about physical reality. 
We now recognize that the core issue lies in the critical gap between experimental observations and theoretical descriptions of integer functions. 
This limitation underscores the urgent need for a novel approach that transcends the constraints imposed by existing no-go theorems.
\par
To address this fundamental problem, we clearly restate the experimental data to be explaned and modify the selection rules for logical propositions in quantum theory, which have been either implicitly assumed or treated as auxiliary conditions.
Specifically, we introduce an alternative to the orthogonality condition in von Neumann-L\"{u}ders' theory~\cite{Lueders1951} or the consistency condition in the Consistent Histories approach~\cite{Griffiths2001}.
This allows us to construct a novel hidden variable theory that can represent integer functions without contradicting the laws of quantum theory.
This seemingly minor adjustment yields profound consequences: it harmonizes quantum theory with Boolean logic, enables the construction of a stationary-state-based model of reality, and allows for the definition of instantaneous values of physical quantities in individual trials. 
This paper thus elucidates the physics brought about by this new selection rule.
\par
The paper is organized as follows. 
Section \ref{cPreliminary} establishes the physical background for extending the theory beyond the standard formulation of quantum theory. 
We identify experimental data unexplained by the standard formulation, outline limitations of existing theories, and briefly summarize key features of our approach. 
Section \ref{cLogical} derives a new selection rule to harmonize Boolean logic with quantum theory, enabling the definition of logical variables whose expectation values are predictable by quantum theory.
Section \ref{cReality} presents our model of reality and delineates the assumptions necessary for constructing the hidden variable theory. 
Section \ref{cMeasurement} summarizes the key features of the resulting theory, including: (i) derivation of deterministic relations in entangled subsystems, (ii) definition of instantaneous values of physical quantities, (iii) explanation of measurement contextuality through the new selection rule, (iv) formulation of a classical approximation for individual trials, (v) revised explanation of the Bell and CHSH inequalities, and (vi) advantage over the light quantum hypothesis. 
Section \ref{cDiscussion} addresses unresolved aspects of our theory, while Section \ref{cConclusion} provides overall conclusions.
Appendices offer in-depth explanations of specific topics, and supplementary materials present calculation examples and various deterministic relations in well-known quantum paradoxes.
\par
This paper focuses on the phenomenological representation of integer functions observed in quantum optics experiments. 
Therefore, we minimize references to literature not directly related to the representation of integer functions.
In particular, we believe that connections to quantum information theory~\cite{Jammer} or the interpretations of quantum theory~\cite{Nielsen2000} require separate discussion.
We also omit the discussion on relativistic theory, instead highlighting unanswered questions about the definition of stationary states in non-relativistic theories.
As an initial attempt to construct a physical theory, we prioritize conceptual development over mathematical rigor. 
This approach should be justified at this stage, given that Hilbert space construction is common to conventional theories~\cite{Breuer,Griffiths2001,Messiah}. 
On the other hand, we have confirmed that our theory~\cite{Theory2} reproduces standard formulas~\cite{Mandel,Kelly1964} for light detection probability, quantum coherence functions, and counting statistics. 
While most calculations are straightforward, some care is needed to describe double-slit experiments since it requires a realistic model of measuring instruments (an array of photodetectors, each with multiple absorption centers).

\section{Preliminary considerations and methodology summary}
\label{cPreliminary}
\subsection{Experimental data unexplained by standard quantum theory}
\label{sMotivation}
\begin{figure}[t]
\centering
\includegraphics{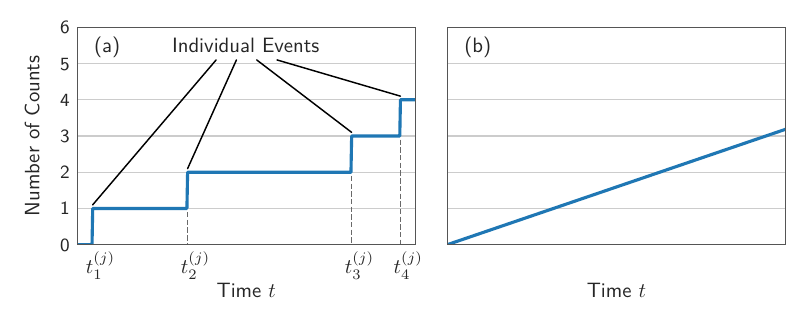}
\caption{Schematic representation of two types of experimental data in a photon counting experiment.
(a) An integer-valued function obtained from an individual trial, as described by Eq.\eqref{eeIntro1}. The trial number is denoted by $j$, and individual events are characterized by discontinuities in these integer functions. 
(b) A real-valued function defined as the expectation value of the integer function, as given by Eq.\eqref{eeIntro2}. For ensembles where individual events occur at random times, the time evolution of the expectation value appears continuous. This continuity arises from the averaging of discontinuous changes in the individual events through the limit operation $\lim_{N\to\infty} \sum_{j=1}^{N} ({1}/{N})$.}
\label{figRealityExpectation}
\end{figure}
The standard formulation of quantum theory encounters a significant challenge when attempting to represent the number of quantum jumps observed in photon counting experiments~\cite{Hamamatsu3,Hamamatsu1} and quantum jump experiments~\cite{QuantumJump1,QuantumJump2,QuantumJump3,Haroche1,Haroche2}. 
A distinctive feature of these experimental data lies in their unique mathematical structure. 
Specifically, these experiments typically yield data on the occurrence times of quantum jumps, denoted by $t^{(j)}_{i}$.
The number of quantum jumps in the $j$-th trial is defined by substituting $t^{(j)}_{i}$ into the following equation:
\begin{align}
n^{(j)}(t) \equiv \sum_{i} \Theta(t-t^{(j)}_{i})
,\quad 
\Theta(t) \equiv 
\begin{cases} 
  0, & t < 0 \\
  1, & t \geqq 0
\end{cases}
\label{eeIntro1}
\end{align}
(Fig.\ref{figRealityExpectation}(a)). 
Here, $j\,{\in}\,{\mathbb{N}}$ is the trial number, $i\,{\in}\,{\mathbb{N}}$ is the sequential number of quantum jumps within a trial ($t^{(j)}_{i}\,{<}\,t^{(j)}_{i+1}$), and $\Theta(t)$ is an integer-valued step function. 
The expectation value of this integer function is defined as
\begin{align}
\langle n(t) \rangle \equiv \lim_{N\to\infty} \frac{1}{N} \sum_{j=1}^{N} n^{(j)}(t).
\label{eeIntro2}
\end{align}
This expectation value typically manifests as a continuous real-valued function due to the stochastic variation of $t^{(j)}_{i}$ (Fig.\ref{figRealityExpectation}(b)). 
For this continuous experimental data, the equation of motion, combined with the Born rule, establishes the following relationship:
\begin{align}
\langle n(t) \rangle = {\mathrm{Tr}}\Big[ \hat{n} \hat{\rho}(t) \Big]
,\quad
\frac{d}{dt}\hat{\rho}(t) = {\mathcal{L}} \hat{\rho}(t),
\label{eeIntro3}
\end{align}
where $\hat{\rho}(t)$ is the density operator for the given ensemble, ${\mathcal{L}}$ is the Liouville super-operator specifying the von Neumann equation~\cite{Breuer}, and $\hat{n}$ is an experiment-specific operator~\cite{Theory2}.
\par
The problem becomes evident when we introduce the assumption that the integer function $n^{(j)}(t)$ can be represented by a density operator $\check{\rho}^{(j)}(t)$ as
\begin{align}
n^{(j)}(t) \overset{?}{=} {\rm{Tr}}\Big[ \hat{n} \check{\rho}^{(j)}(t) \Big]
,\quad \forall j.
\label{eeIntro4}
\end{align}
Firstly, individual integer functions are non-differentiable at quantum jump occurrences, $t\,{=}\,t^{(j)}_{i}$.
Therefore, it is unreasonable to assume that  $\check{\rho}^{(j)}(t)$ satisfies an equation of motion of the same form as Eq.\eqref{eeIntro3}:
\begin{align}
\frac{d}{dt}\check{\rho}^{(j)}(t) \overset{?}{=} {\mathcal{L}} \check{\rho}^{(j)}(t).
\label{eeIntro5}
\end{align}
Secondly, even without assuming the differentiability of $\check{\rho}^{(j)}(t)$, a problem arises when $n^{(j)}(t)$ is a logical variable describing the occupation of orthonormalized quantum states $\{|\phi_{1}\rangle,\ldots,|\phi_{M}\rangle\}$:
\begin{align}
n^{(j)}_{\mu}(t)
\equiv \begin{cases}
1, & \mbox{when $|\phi^{(j)}(t)\rangle$ is $|\phi_{\mu}\rangle$}\\
0, & \mbox{when $|\phi^{(j)}(t)\rangle$ is $|\phi_{\mu}\rangle$}
\end{cases}
,\quad 1\,{\leqq}\,\mu\,{\leqq}\,M.
\label{eeIntro6}
\end{align}
Appendix \ref{sDensity} demonstrates that the combination of Eq.\eqref{eeIntro4} and Eq.\eqref{eeIntro6} leads to a conclusion contradicting the superposition principle.
The standard interpretation~\cite[{\S}7.2]{Peres} of Gleason's theorem~\cite{Gleason1957} also supports the impossibility of assuming the combination of Eq.\eqref{eeIntro4} and Eq.\eqref{eeIntro6}.
This paradox reveals a fundamental limitation of the standard formulation of quantum theory: while it can predict the expectation values of integer functions, it fails to represent individual integer functions themselves.
In other words, quantum theory---operating under the premise of describing all experimental data via density operators---cannot represent \emph{integer functions whose expectation values are predictable by quantum theory}.

\subsection{Problems with conventional theories}
\label{sProblem}
To address these fundamental issues with sufficient specificity, we focus our analysis on experimental observations in quantum optics~\cite{Agarwal}. 
In particular, we consider how to describe the aforementioned integer functions within the framework of non-relativistic QED.
We find that the discontinuous behavior of these integer functions has traditionally been explained using the concept of photons~\cite{Einstein1905,Einstein1906,Einstein1917,Bose1924,Einstein1924,Lewis1926}. 
However, modern quantum optics theory~\cite{Mandel,Glauber2006,Cohen,Loudon,Milonni1993,Milonni2004} has largely rendered the photon concept unnecessary for most applications.
As Lamb described in his ``anti-photon'' paper~\cite{Antiphoton}, this shift results from two key developments: (i) the quantization of both atomic systems and radiation fields by non-relativistic QED~\cite{Dirac1927,DiracQM,Fermi1932} and (ii) the introduction of modern wave representations through the quantum theory of coherence~\cite{Glauber1963} and the semi-classical theory of radiation~\cite{Jaynes1963,Lamb1964,Lamb1968}.
Given this shift away from the photon concept, it seems reasonable to explain the discontinuous behavior of integer functions based solely on quantum jumps between stationary states~\cite{Dirac1927,Bohr1913}.
Contrary to expectations, however, this approach does not work in the standard formulation.
This is because if all experimental facts must be represented within Hilbert space, then quantum jumps cannot be a discontinuous phenomenon~\cite{Zeh1993} ({\S}\ref{cIntro}). 
From this perspective, the conceptual confusion surrounding photons and quantum jumps~\cite{Zeh1993,Knight1983,Muthukrishnan} is intrinsically linked to the lack of a formalism capable of representing integer functions.
\par
To address these conceptual challenges and identify the core issues in conventional theory, it is beneficial to introduce phenomenological concepts. 
First, if we define \emph{individual events} as discontinuous changes in integer functions (Fig.\ref{figRealityExpectation}(a)), the discontinuous changes can be described without assuming the existence of photons and quantum jumps.
We can also see that individual events are non-differentiable by definition, meaning that they cannot be derived from the quantum equations of motion, especially under ordinary mathematics where no derivative exists for integer functions ({\S}\ref{sMotivation}).
Second, to highlight the existence of two types of experimental data for a given time $t$ (Fig.\ref{figRealityExpectation}), it is useful to refer to the integer functions observed in individual trials (Fig.\ref{figRealityExpectation}(a)) as the \emph{world of reality}, while the real-valued function defined in a statistical limit (Fig.\ref{figRealityExpectation}(b)) as the \emph{world of expectation}.
This distinction reveals a fundamental dichotomy: the quantum equation of motion holds in the world of expectation but fails in the world of reality.
Note that individual events occur only in the world of reality.
We then find that quantum theory is a theory of the world of expectation (i.e., real-valued function), requiring \emph{a hidden variable theory to describe the world of reality (i.e., individual integer functions)}.
Crucially, to break away from the contradictions surrounding photons and local reality, a hidden variable theory in this sense must focus on stationary states rather than local point particles such as photons.
\par
According to the above discussion, special attention must be paid to the no-go theorem for theories that describe the stationary state occupation in terms of logical variables.
From our point of view, this type of no-go theorem makes sense when one assumes a combination of Bohr's interpretation of quantum theory (rejecting hidden variables)~\cite{Bohr1913,Bohr1924,EPR-Bohr,Bohr1950,Stapp1972} with Einstein's model of physical reality (local point particles)~\cite{EPR,Einstein1949}. 
Under this combination, any theory that represents logical variables using Hilbert space concepts (density operators or projection operators) is prohibited by Gleason's theorem ({\S}\ref{sMotivation}). 
Additionally, hidden variable theories based on the locality assumption are rejected by Bell's theorem and related experiments~\cite{Bell,Bell1966,CHSH,Aspect}, while theories assigning integer values to individual trials face contradictions from the CHSH theorem~\cite{CHSH} and the Kochen-Specker theorem~\cite{Peres,Budroni2022}.
In contrast, a hidden variable theory aligned with the stationary state model would likely combine Einstein's interpretation of quantum theory (allowing hidden variables)~\cite{Einstein1949,Ballentine} with Bohr's model of physical reality (assuming stationary states)~\cite{Dirac1927,Bohr1913}. 
This inverse combination is independent of Bell's theorem's locality assumption, resolving Gleason's theorem's constraints by assuming only Eq.\eqref{eeIntro6} (See Appendix \ref{sDensity}).
In this respect, the essential challenge for hidden variable theories lies in avoiding contradictions related to measurement contextuality, as exemplified by the CHSH and Kochen-Specker theorems~\cite{Peres}.
\par
Meanwhile, quantum jump theories not assuming hidden variables were extensively studied after quantum jumps were experimentally observed in the 1980s. 
Among these theories, Quantum Jump Approach in Quantum Optics~\cite{Carmichael,Plenio1998} focused on single quantum systems described by master equations, developing numerical methods that explicitly incorporate the state vector collapse due to decoherence~\cite{Zurek2003}.
Furthermore, Consistent Histories Approach to Quantum Theory~\cite{Griffiths2001,Omnes1992} focused on the chains of quantum jumps (a.k.a.\,``Histories''), demonstrating how to calculate the probability that an individual History is realized.
Among these developments, the Consistent Histories approach made a significant contribution by showing that it is possible to construct a theory compatible with both quantum theory and Boolean logic, by limiting the range of Histories using selection rules (consistency conditions)~\cite{Griffiths2001}.
Nevertheless, these theories did not acknowledge the existence of hidden variables, assigning continuous values (0 to 1) to stationary state populations~\cite{Griffiths2001,Plenio1998}.
This is why conventional quantum jump theories cannot represent integer functions---the fundamental limitation of theories describing all experimental data using Hilbert space concepts.
Although significant progress has been made in the theory of open quantum systems~\cite{Breuer,Paris} and quantum information theory~\cite{Nielsen2000,Uola2023}, this limitation still seems to persist in more recent theories.

\subsection{Key features of our approach}
\label{sMethod}
\begin{figure}[t]
\centering
\begin{tikzpicture}
  \tikzset{Box1/.style={rectangle, draw, text centered, text width=3.9cm, minimum height=1.2cm, inner xsep=0.3}};
  \tikzset{Box2/.style={rectangle, draw, text centered, text width=2.3cm, minimum height=1.2cm, inner xsep=0.3}};
  \tikzset{Box3/.style={rectangle, draw, text centered, text width=3.4cm, minimum height=1.2cm, inner xsep=0.3}};
  \node[Box1] (Logical) at (-6,2.7) {\begin{tabular}{c}\small Instantaneous value\\\small of logical variables \end{tabular}};
  \node[Box2] (LawBool) at (-2.7,2.7) {\begin{tabular}{c}\small Stationary\\\small states \end{tabular}};
  \node[Box3] (Occupation) at (0.4,2.7) {\begin{tabular}{c}\small Boolean logic\\\small Exclusivity \end{tabular}};
  \draw[-,very thick] (Logical.east) to[out=0,in=180] (LawBool.west);
  \draw[-,very thick] (LawBool.east) to[out=0,in=180] (Occupation.west);
  \draw[thick] (-8.2,1.9) rectangle (2.3,3.5);
  \node (Reality) at (-10.4,2.7) {\begin{tabular}{c}\bf\small World of reality\\\small (Integer function) \end{tabular}};
  \node[Box1] (Probability) at (-6,0) {\begin{tabular}{c}\small Expectation value\\\small of logical variables \end{tabular}};
  \node[Box2] (LawQuantum) at (-2.7,0) {\begin{tabular}{c}\small Density\\\small operator \end{tabular}};
  \node[Box3] (Density) at (0.4,0) {\begin{tabular}{c}\small Born rule\\\small Equation of motion\end{tabular}};
  \draw[-,very thick] (Probability.east) to[out=0,in=180] (LawQuantum.west);
  \draw[-,very thick] (LawQuantum.east) to[out=0,in=180] (Density.west);
  \draw[thick] (-8.2,-0.8) rectangle (2.3,0.8);
  \node (Reality) at (-10.4,0) {\begin{tabular}{c}\bf\small World of expectation\\\small (Real-valued function) \end{tabular}};
  \draw[-,very thick] (Probability.north) to[out=90,in=270] (Logical.south);
  \draw[-,very thick] (Density.north) to[out=90,in=270] (Occupation.south);
  \node (Definition) at (-4.4,1.35) {\begin{tabular}{c}\small Definition of\\\small expectation values\end{tabular}};
  
  \node (Auxiliary) at (-0.9,1.35) {\begin{tabular}{c}\small Compatibility\\\small condition \end{tabular}};
\end{tikzpicture}
\caption{{\bf The overall structure of the theory}. A two-layer theory is considered to capture the mathematical structure of propositions whose probabilities are predictable by quantum theory. Upper frame: The truth values of propositions are represented by the instantaneous values of logical variables, which describe the stationary state occupations. The truth values of propositions are assumed to follow the laws of Boolean logic and exclusivity. Lower frame: The probabilities of propositions are represented by the expectation values of logical variables, which are determined by the density operator. The probabilities of propositions are assumed to obey the laws of quantum theory (i.e., the Born rule and the equations of motion). Left line: The truth values and probabilities of propositions are linked through the definition of expectation values of logical variables. Right line: The contradiction between the laws of Boolean logic and quantum theory is avoided by a new selection rule (i.e., compatibility condition).}
\label{figRealityModel}
\end{figure}
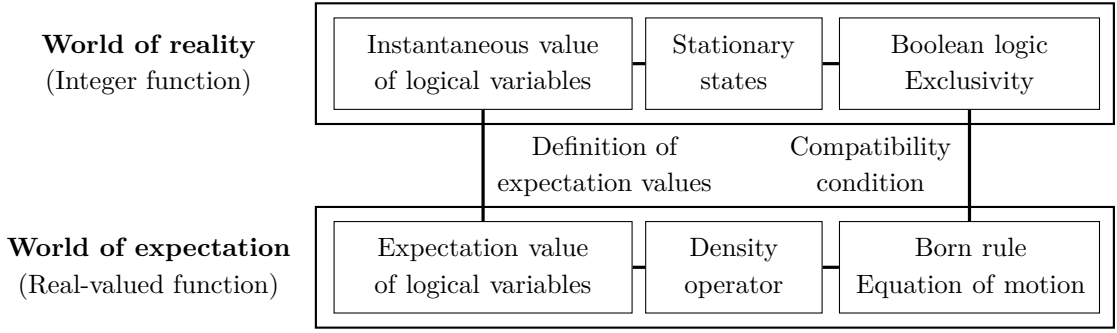

To overcome this limitation of the standard formulation, we propose a novel theoretical framework that integrates (i) Einstein's interpretation of quantum theory, (ii) Bohr's model of physical reality, and (iii) the methodology of the Consistent Histories approach.
In other words, we consider a hidden variable theory that employs Boolean logical variables to describe stationary state occupation, introducing a novel selection rule to reconcile Boolean logic and quantum theory.
More specifically, our theory centers on \emph{propositions whose probabilities are predictable by quantum theory}, constituting a dual framework such that their truth values are represented by Boolean logical variables and their probabilities are calculated using the density operator (Fig.\ref{figRealityModel}).
A key feature of this theoretical framework is the selection rule for determining allowable propositions, the form of which is determined based on two fundamental requirements: (i) the uniqueness of the probability of the result of a Boolean operation and (ii) the consistency of the time ordering of the operators involved.
This selection rule, which we refer to below as the compatibility condition, serves as a necessary condition for consistently defining the conjunction and disjunction of propositions.
Remarkably, it allows for a wider range of propositions than conventional selection rules, such as the orthogonality condition of von Neumann-L\"{u}ders' theory and the consistency condition of the Consistent Histories approach.
\par
The modification of the selection rule has the following effects:
\begin{enumerate}
\item The existence of joint probability is guaranteed for propositions that are true only for a particular ensemble and at a given instant of time.
This means that even for propositions with non-commutative operators, joint probability can be defined by reducing the precision of proposition representation.
\item If logical variables are used to define instantaneous values of physical quantities, arithmetic operations are prohibited between physical quantities that involve propositions not satisfying the compatibility condition.
This overcomes the inconsistencies regarding measurement contextuality (e.g., the CHSH and Kochen-Specker theorems).
\item Canonical variables and canonical momenta meet the compatibility condition when the precision of representation is adjusted under classical approximation. 
This eliminates the constraint on joint probabilities by Nelson's theorem~\cite{Breuer}, allowing us to derive the complementarity principle and classical equations of motion.
\item The resolution of the known contradictions in major no-go theorems justifies the use of logical variables (hidden variables) to represent propositions whose probabilities are predictable by quantum theory.
This also makes it possible to describe integer functions whose expectation values are predictable by quantum theory.
\item Deterministic relations between individual events can be explicitly expressed as equations between logical variables. 
In fundamental problems such as the photoionization of hydrogen atoms and spin measurements in a spin-singlet system, deterministic relations in entangled subsystems can be derived from the first principles.
\item For propositions whose joint probabilities can be calculated by quantum theory, the same joint probability formulas as in quantum theory are obtained.
The definition of conjunction ensures the time and normal ordering of operators, leading to the standard formulas for quantum coherence functions in quantum optics~\cite{Mandel,Kelly1964}.
\item For propositions whose joint probabilities cannot be calculated by quantum theory, this theory can represent deterministic relations that cannot be represented by the standard formulation.
Relatedly, non-deterministic hidden variables may exhibit deterministic behavior when the precision of proposition representation is reduced.
\end{enumerate}
\par
Among these key features, the ability to express deterministic relations in individual trials (\#7) is not found in previous theories, so its basic properties will be briefly described below.
First of all, the most fundamental hidden variable in our theory is the logical variable that describes the occupation of the stationary state $|\phi\rangle$ of the entire system:
\begin{align}
\xjvar{`$\occupation{\state{\phi}{}}{},t$'}{big}
= \begin{cases}
1, & \mbox{Proposition `$\occupation{\state{\phi}{}}{},t$' is True}, \\
0, & \mbox{Proposition `$\occupation{\state{\phi}{}}{},t$' is False}.
\end{cases}
\label{eeIntro11}
\end{align}
Here, the symbol `$\occupation{\state{\phi}{}}{},t$' represents a proposition `A stationary state $|\phi\rangle$ is occupied at time $t$'.
Denoting the stationary state occupied at time $t$ in the $j$-th trial as $|\Omega^{(j)}(t)\rangle$, stationary state $|\Omega^{(j)}(t)\rangle$ will vary stochastically according to the laws of quantum theory.
Consequently, there is no special advantage in considering this type of hidden variable.
On the other hand, the stationary states of many-body systems are usually too complex to calculate, suggesting that a proposition with reduced precision is necessary to express experimental results.
Now, let $\stateset{{\mathrm{C}}_{\Omega}}$ denote the subspace of Hilbert space always containing $|\Omega^{(j)}(t)\rangle$ and let `$\occupation{\stateset{{\mathrm{C}}}}{},t$' denotes a proposition `The stationary states occupied at time $t$ is contained in subspace $\stateset{{\mathrm{C}}}$'.
Then, proposition `$\occupation{\stateset{{\mathrm{C}}_{\Omega}}}{},t$' is always true by definition, meaning that a logical variable
\begin{align}
\xjvar{`$\occupation{\stateset{{\mathrm{C}}_{\Omega}}}{},t$'}{big} 
= \begin{cases}
1, & \mbox{Proposition `$\occupation{\stateset{{\mathrm{C}}_{\Omega}}}{},t$' is True} \\
0, & \mbox{Proposition `$\occupation{\stateset{{\mathrm{C}}_{\Omega}}}{},t$' is False}
\end{cases}
\label{eeIntro12}
\end{align}
will always have the constant value 1.
It is this type of propositions that become describable by introducing the compatibility condition.
Significantly, the new hidden variable in Eq.\eqref{eeIntro12} is deterministic while the hidden variable in Eq.\eqref{eeIntro11} is non-deterministic.
Furthermore, when the subspace $\stateset{{\mathrm{C}}_{\Omega}}$ of Hilbert space can be decomposed into a direct product $\stateset{{\mathrm{C}}_{\Omega,1}}\,{\otimes}\,\stateset{{\mathrm{C}}_{\Omega,2}}$, the deterministic relation
\begin{align}
\xjvar{`$\occupation{\stateset{{\mathrm{C}}_{\Omega,1}}}{},t$'}{big} 
= \xjvar{`$\occupation{\stateset{{\mathrm{C}}_{\Omega,2}}}{},t$'}{big} 
= 1.
\label{eeIntro13}
\end{align}
is established between the propositions `$\occupation{\stateset{{\mathrm{C}}_{\Omega,1}}}{},t$' and `$\occupation{\stateset{{\mathrm{C}}_{\Omega,2}}}{},t$'.
In this sense, the compatibility condition is a selection rule to describe deterministic relations in the world of reality, which sometimes allows us to derive deterministic laws for individual trials from the probabilistic laws of quantum theory.
Importantly, what matters in the laws of reality is no longer the time variation of logical variables, but conditions ${\mathrm{C}}_{\Omega,1}$ and ${\mathrm{C}}_{\Omega,2}$ that specify propositions `$\occupation{\stateset{{\mathrm{C}}_{\Omega,1}}}{},t$' and `$\occupation{\stateset{{\mathrm{C}}_{\Omega,2}}}{},t$' that always become true.
\par
It should also be pointed out that the standard formulation sometimes fails to represent this type of deterministic relation.
To see this, suppose that the logical variables for propositions `$\mathrm{A}_{1},\,t_{1}$' and `$\mathrm{A}_{2},\,t_{2}$', each describing subsystems 1 and 2 and not necessarily equal 1, can be consistently defined as
\begin{align}
\xjvar{`${\mathrm{A}}_{l},t_{l}$'}{big}
= \begin{cases}
1 & \mbox{`${\mathrm{A}}_{l}$' is True at time $t_{l}$} \\
0 & \mbox{`${\mathrm{A}}_{l}$' is False at time $t_{l}$}
\end{cases}
,\quad
l \in \{1,2\}.
\label{eeIntro14}
\end{align}
We will later show that this is possible if the interaction between the subsystems is negligible at the time of interest.
When the subsystems are entangled, we can then derive a deterministic relation between propositions `${\mathrm{A}}_{1},t_{1}$' and `${\mathrm{A}}_{2},t_{2}$', namely
\begin{align}
\xjvar{`$\mathrm{A}_{1},\,t_{1}$'}{} = \xjvar{`$\mathrm{A}_{2},\,t_{2}$'}{}
,\quad \forall{j}.
\label{eeIntro15}
\end{align}
Denoting the conjunction (AND) of `$\mathrm{A}_{1},\,t_{1}$' and `$\mathrm{A}_{2},\,t_{2}$' as `$\mathrm{A}_{1},\,t_{1}$'$\wedge$`$\mathrm{A}_{2},\,t_{2}$', this equation is equivalent to
\begin{align}
\langle \xvar{`$\mathrm{A}_{1},\,t_{1}$'$\wedge$`$\mathrm{A}_{2},\,t_{2}$'}{} \rangle 
= \langle \xvar{`$\mathrm{A}_{1},\,t_{1}$'}{} \rangle 
= \langle \xvar{`$\mathrm{A}_{2},\,t_{2}$'}{} \rangle,
\label{eeIntro16}
\end{align}
which implies that the conditional probabilities are 1.
In previous theories, it was not possible to define the left-hand side of Eq.\eqref{eeIntro16} when operators $\xop{`$\mathrm{A}_{1},\,t_{1}$'}{}$ and $\xop{`$\mathrm{A}_{2},\,t_{2}$'}{}$ do not commute (as per Nelson's theorem).
Therefore, the only relation that can be safely expressed was
\begin{align}
\big\langle \xvar{`$\mathrm{A},\,t_{\mathrm{A}}$'}{} \big\rangle 
= \big\langle \xvar{`$\mathrm{B},\,t_{\mathrm{B}}$'}{} \big\rangle.
\label{eeIntro17}
\end{align}
We can now see that Eq.\eqref{eeIntro17} can be derived from Eq.\eqref{eeIntro15}, but not vice versa.
For this reason, the equation linking two or more logical variables may express a law of reality that the standard formulation has failed to represent.
Specific examples of the laws of reality are Eq.\eqref{eeIntro11} and Eq.\eqref{eeIntro12}, and more specific examples are the classical equations of motion and the light quantum hypothesis.
Interestingly, however, when the light quantum hypothesis is expressed in the form of Eq.\eqref{eeIntro15}, the corresponding operators do not have a consistent expression for different initial conditions.
This suggests that the stationary-state type model of reality, as postulated in our theory, is more reasonable than the light quantum hypothesis as a substructure of non-relativistic QED.
\begin{table}[b]
\caption{Representation of individual trials and ensembles.}
\label{tMethods}
\centering
\begin{tabular}{p{1mm}p{54mm}p{41mm}l}
\toprule
{\small $\#$} &
  \raisebox{0em}{\bettershortstack[l]{0.7}{2.5}{~~~~~~~~~~~Approach}} &
  \raisebox{0em}{\bettershortstack[l]{0.7}{2.5}{~~~~~Individual trial}} &
  \raisebox{0em}{\bettershortstack[l]{0.7}{2.5}{~~~~~~Ensemble}} \\
\midrule
1 &
  \raisebox{0em}{\bettershortstack[l]{0.7}{2.5}{Light-quantum hypothesis~\cite{Einstein1917}}} &
  \raisebox{0em}{\bettershortstack[l]{0.7}{2.5}{Energy and momentum}} &
  \raisebox{0em}{\bettershortstack[l]{0.7}{2.5}{Classical EM field}} \\
2 &
  \raisebox{0em}{\bettershortstack[l]{0.7}{2.5}{Einstein-Podolsky-Rosen~\cite{EPR}}} &
  \raisebox{0em}{\bettershortstack[l]{0.7}{2.5}{Local canonical variables}} &
  \raisebox{0em}{\bettershortstack[l]{0.7}{2.5}{Wave function}} \\
3 &
  \raisebox{0em}{\bettershortstack[l]{0.7}{2.5}{Copenhagen (Bohr)\cite{Bohr1950,Stapp1972}}} &
  \raisebox{0em}{\bettershortstack[l]{0.7}{2.5}{---}} &
  \raisebox{0em}{\bettershortstack[l]{0.7}{2.5}{Wave function}} \\
4 &
  \raisebox{0em}{\bettershortstack[l]{0.7}{2.5}{Copenhagen (modern)\cite{Peres}}} &
  \raisebox{0em}{\bettershortstack[l]{0.7}{2.5}{---}} &
  \raisebox{0em}{\bettershortstack[l]{0.7}{2.5}{Density operator}} \\
5 &
  \raisebox{0em}{\bettershortstack[l]{0.7}{2.5}{Quantum Jump~\cite{Carmichael,Plenio1998}}} &
  \raisebox{0em}{\bettershortstack[l]{0.7}{2.5}{(State vectors)}} &
  \raisebox{0em}{\bettershortstack[l]{0.7}{2.5}{Density operator}} \\
6 &
  \raisebox{0em}{\bettershortstack[l]{0.7}{2.5}{Consistent Histories~\cite{Griffiths2001,Omnes1992}}} &
  \raisebox{0em}{\bettershortstack[l]{0.7}{2.5}{(Histories)}} &
  \raisebox{0em}{\bettershortstack[l]{0.7}{2.5}{Density operator}} \\
7 &
  \raisebox{0em}{\bettershortstack[l]{0.7}{2.5}{This work}} &
  \raisebox{0em}{\bettershortstack[l]{0.7}{2.5}{Boolean logical variables}} &
  \raisebox{0em}{\bettershortstack[l]{0.7}{2.5}{Density operator}} \\
\bottomrule
\end{tabular}
\end{table}

\subsection{Brief summary of methodology}
\label{sMethod2}
\begin{table}[t]
\renewcommand{\arraystretch}{1.25}
\caption{List of assumptions. See Section \ref{cReality} for exact formula.}
\label{tAssumptions}
\centering
\begin{tabular}{cll}
\toprule
\raisebox{0em}{\bettershortstack[c]{0.7}{2.5}{\small \!\!\!\!\!\#}} & 
  \raisebox{0em}{\bettershortstack[l]{0.7}{2.5}{\!\!\!\!\!\!\!\!\!\!\small ~~~~Assumption}} & 
  \raisebox{0em}{\bettershortstack[l]{0.7}{2.5}{\!\!\!\!\!\!\!\!\!\!\small ~~~~~~~~~~~~~~~~~~~~~~~~~~~~~~~~~Description}} \\
\midrule
\raisebox{0em}{\bettershortstack[c]{0.7}{2.5}{\small \!\!\!\!\!1}} & 
  \raisebox{0em}{\bettershortstack[l]{0.7}{2.5}{\!\!\!\!\!\!\!\!\!\!\small Target system}} & 
  \raisebox{0em}{\bettershortstack[l]{0.7}{2.5}{\!\!\!\!\!\!\!\!\!\!\small $\hat{H}^{\hat{\mathsf{W}}}\,{=}\,{\hat{\mathsf{W}}}^{\dagger}\,(\hat{H}_{\mathrm{A}}\,{+}\,\hat{H}_{\mathrm{F}}\,{+}\,\hat{H}_{\mathrm{I}})\,{\hat{\mathsf{W}}},~\hat{1}\,{=}\,{\hat{\mathsf{W}}}^{\dagger}\,{\hat{\mathsf{W}}}\,{=}\,{\hat{\mathsf{W}}}\,{\hat{\mathsf{W}}}^{\dagger}$}}\!\!\!\!\!\! \\
\raisebox{0em}{\bettershortstack[c]{0.7}{2.5}{\small \!\!\!\!\!2}} & 
  \raisebox{0em}{\bettershortstack[l]{0.7}{2.5}{\!\!\!\!\!\!\!\!\!\!\small Stationary states}} & 
  \raisebox{0em}{\bettershortstack[l]{0.7}{2.5}{\!\!\!\!\!\!\!\!\!\!\small $\hat{H}^{\hat{\mathsf{U}}}_{0}\,{=}\,{\hat{\mathsf{U}}}^{\dagger}\,\big(\hat{H}_{\mathrm{A}}\,{+}\,\hat{H}_{\mathrm{F}}\big){\hat{\mathsf{U}}},~\hat{H}^{\hat{\mathsf{U}}}_{0}|\phi\rangle\,{=}\,E|\phi\rangle,~\langle\phi|\phi'\rangle\,{=}\,\delta_{\phi \phi'},~\hat{1}\,{=}\,{\hat{\mathsf{U}}}^{\dagger}\,{\hat{\mathsf{U}}}\,{=}\,{\hat{\mathsf{U}}}\,{\hat{\mathsf{U}}}^{\dagger}$}}\!\!\!\!\!\! \\
\raisebox{0em}{\bettershortstack[c]{0.7}{2.5}{\small \!\!\!\!\!3}} & 
  \raisebox{0em}{\bettershortstack[l]{0.7}{2.5}{\!\!\!\!\!\!\!\!\!\!\small Stationary state sets}} & 
  \raisebox{0em}{\bettershortstack[l]{0.7}{2.5}{\!\!\!\!\!\!\!\!\!\!\small $\stateset{\mathrm{C}}_{\hat{\mathsf{U}}}\,{\equiv}\,\{\,\forall |\phi\rangle\,|~\hat{H}^{\hat{\mathsf{U}}}_{0}|\phi\rangle\,{=}\,E(\phi)|\phi\rangle,~\langle\phi|\phi'\rangle\,{=}\,\delta_{\phi,\phi'},~\mbox{$|\phi\rangle$ satisfy C\,}\}$}}\!\!\!\!\!\! \\
\raisebox{0em}{\bettershortstack[c]{0.7}{2.5}{\small \!\!\!\!\!4}} & 
  \raisebox{0em}{\bettershortstack[l]{0.7}{2.5}{\!\!\!\!\!\!\!\!\!\!\small Element propositions}} & 
  \raisebox{0em}{\bettershortstack[l]{0.7}{2.5}{\!\!\!\!\!\!\!\!\!\!\small `$\occupation{\stateset{{\mathrm{C}}}}{}$'$\equiv$`One of the stationary states in $\stateset{\mathrm{C}}$ is occupied at time $t$'}}\!\!\!\!\!\! \\
\raisebox{0em}{\bettershortstack[c]{0.7}{2.5}{\small \!\!\!\!\!5}} & 
  \raisebox{0em}{\bettershortstack[l]{0.7}{2.5}{\!\!\!\!\!\!\!\!\!\!\small Logical variables}} & 
  \raisebox{0em}{\bettershortstack[l]{0.7}{2.5}{\!\!\!\!\!\!\!\!\!\!\small $\xjvar{`o($\stateset{\mathrm{C}}$)'}{big}\,{\in}\,\{0,1\}$}}\!\!\!\!\!\! \\
\raisebox{0em}{\bettershortstack[c]{0.7}{2.5}{\small \!\!\!\!\!6}} & 
  \raisebox{0em}{\bettershortstack[l]{0.7}{2.5}{\!\!\!\!\!\!\!\!\!\!\small Laws of Boolean logic}} & 
  \raisebox{0em}{\bettershortstack[l]{0.7}{2.5}{\!\!\!\!\!\!\!\!\!\!\scriptsize $\xjvar{`$\occupation{\stateset{{\mathrm{C}}_{1}}}{},t_{1}$'$\wedge$`$\occupation{\stateset{{\mathrm{C}}_{2}}}{},t_{2}$'}{}\,{=}\,\xjvar{`$\occupation{\stateset{{\mathrm{C}}_{1}}}{},t_{1}$'}{}\,\xjvar{`$\occupation{\stateset{{\mathrm{C}}_{2}}}{},t_{2}$'}{}$ etc.}}\!\!\!\!\!\! \\
\raisebox{0em}{\bettershortstack[c]{0.7}{2.5}{\small \!\!\!\!\!7}} & 
  \raisebox{0em}{\bettershortstack[l]{0.7}{2.5}{\!\!\!\!\!\!\!\!\!\!\small Exclusivity}} & 
  \raisebox{0em}{\bettershortstack[l]{0.7}{2.5}{\!\!\!\!\!\!\!\!\!\!\small $\xjvar{`$\occupation{\state{\phi_{1}}{}}{},t$'$\wedge$`$\occupation{\state{\phi_{2}}{}}{},t$'}{}\,{=}\,0$~~for~~$\phi_{1}\,{\neq}\,\phi_{2}$}}\!\!\!\!\!\! \\
\raisebox{0em}{\bettershortstack[c]{0.7}{2.5}{\small \!\!\!\!\!8}} & 
  \raisebox{0em}{\bettershortstack[l]{0.7}{2.5}{\!\!\!\!\!\!\!\!\!\!\small Complex propositions}} & 
  \raisebox{0em}{\bettershortstack[l]{0.7}{2.5}{\!\!\!\!\!\!\!\!\!\!\small `A,$\{t\}$'$\,{\equiv}\,\bigwedge^{I}_{i=1} \mbox{`${\mathrm{A}}_{i},t_{i}$'}$, `${\mathrm{A}}_{i},t_{i}$': $i$-th elementary proposition}}\!\!\!\!\!\! \\
\raisebox{0em}{\bettershortstack[c]{0.7}{2.5}{\small \!\!\!\!\!9}} & 
  \raisebox{0em}{\bettershortstack[l]{0.7}{2.5}{\!\!\!\!\!\!\!\!\!\!\small Expectation values}} & 
  \raisebox{0em}{\bettershortstack[l]{0.7}{2.5}{\!\!\!\!\!\!\!\!\!\!\small $\langle \xvar{`A,$\{t\}$'}{} \rangle\,{\equiv}\,{\sum_{j}^{\mbox{\scriptsize `${\mathrm{A}}_{0},t_{0}$'}} \xjvar{`A,$\{t\}$'}{}}/{\sum_{j}^{\mbox{\scriptsize `${\mathrm{A}}_{0},t_{0}$'}} \,1}$}}\!\!\!\!\!\! \\
\raisebox{0em}{\bettershortstack[c]{0.7}{2.5}{\small \!\!\!\!\!10}} & 
  \raisebox{0em}{\bettershortstack[l]{0.7}{2.5}{\!\!\!\!\!\!\!\!\!\!\small Born rule (1)}} & 
  \raisebox{0em}{\bettershortstack[l]{0.7}{2.5}{\!\!\!\!\!\!\!\!\!\!\small $\langle \xvar{`A,$\{t\}$'}{} \rangle\,{=}\,{\mathrm{Tr}}[\,\xop{`A,$\{t\}$'}{}\,\hat{\rho}(\mbox{`${\mathrm{A}}_{0}$'})\,]$, $\xop{`A,$\{t\}$'}{}\,{=}\,\zopd{`A,$\{t\}$'}{} \zop{`A,$\{t\}$'}{}$}}\!\!\!\!\!\! \\
\raisebox{0em}{\bettershortstack[c]{0.7}{2.5}{\small \!\!\!\!\!11}} & 
  \raisebox{0em}{\bettershortstack[l]{0.7}{2.5}{\!\!\!\!\!\!\!\!\!\!\small Born rule (2)}} & 
  \raisebox{0em}{\bettershortstack[l]{0.7}{2.5}{\!\!\!\!\!\!\!\!\!\!\small $\zop{`$\occupation{\stateset{{\mathrm{C}}}}{}$'}{}\,{=}\,\sum_{|\phi\rangle}^{\stateset{\mathrm{C}}} |\phi\rangle\langle\phi|$}}\!\!\!\!\!\! \\
\raisebox{0em}{\bettershortstack[c]{0.7}{2.5}{\small \!\!\!\!\!12}} & 
  \raisebox{0em}{\bettershortstack[l]{0.7}{2.5}{\!\!\!\!\!\!\!\!\!\!\small Conversion rules}} & 
  \raisebox{0em}{\bettershortstack[l]{0.7}{2.5}{\!\!\!\!\!\!\!\!\!\!\scriptsize $\xop{`$\occupation{\stateset{{\mathrm{C}}_{1}}}{},t_{1}$'$\wedge$`$\occupation{\stateset{{\mathrm{C}}_{2}}}{},t_{2}$'}{}\,{=}\,\mathcal{T}::\xop{`$\occupation{\stateset{{\mathrm{C}}_{1}}}{},t_{1}$'}{}\,\xop{`$\occupation{\stateset{{\mathrm{C}}_{2}}}{},t_{2}$'}{}::$, etc.}}\!\!\!\!\!\! \\
\raisebox{0em}{\bettershortstack[c]{0.7}{2.5}{\small \!\!\!\!\!13}} & 
  \raisebox{0em}{\bettershortstack[l]{0.7}{2.5}{\!\!\!\!\!\!\!\!\!\!\small Equation of motion}} & 
  \raisebox{0em}{\bettershortstack[l]{0.7}{2.5}{\!\!\!\!\!\!\!\!\!\!\small $\zop{`$\occupation{\stateset{{\mathrm{C}}}}{},t$'}{}\,{=}\,\hat{U}^{\dagger}(t)\,\zop{`$\occupation{\stateset{{\mathrm{C}}}}{}$'}{}\,\hat{U}(t)$, $\hat{U}(t)\,{\equiv}\,\exp({t}\hat{H}/{{\rm{i}}\hbar})$}}\!\!\!\!\!\! \\
\raisebox{0em}{\bettershortstack[c]{0.7}{2.5}{\small \!\!\!\!\!14}} & 
  \raisebox{0em}{\bettershortstack[l]{0.7}{2.5}{\!\!\!\!\!\!\!\!\!\!\small Comatibility condition}} & 
  \raisebox{0em}{\bettershortstack[l]{0.7}{2.5}{\!\!\!\!\!\!\!\!\!\!\scriptsize ${\rm{Tr}}[\hat{K} \hat{\rho}(\mbox{`${\mathrm{A}}_{0},t_{0}$'}) \hat{K}^{\dagger}]\,{=}\,{\rm{Tr}}[({\mathcal{P}}\hat{K}) \hat{\rho}(\mbox{`${\mathrm{A}}_{0},t_{0}$'}) ({\mathcal{P}}\hat{K})^{\dagger}]$, $\hat{K}(t)\,{\equiv}\,\zop{`${\mathrm{A}}_{i_{1}},t$'}{}\,{\cdots}\,\zop{`${\mathrm{A}}_{i_{\mathcal{I}}},t$'}{}$}} \!\!\!\!\!\!\!\! \\
\raisebox{0em}{\bettershortstack[c]{0.7}{2.5}{\small \!\!\!\!\!0}} & 
  \raisebox{0em}{\bettershortstack[l]{0.7}{2.5}{\!\!\!\!\!\!\!\!\!\!\small Model of reality}} & 
  \raisebox{0em}{\bettershortstack[l]{0.7}{2.5}{\!\!\!\!\!\!\!\!\!\!\small $\xjvar{`$\occupation{\stateset{{\mathrm{C}}_{\Omega}}}{},t$'}{}\,{=}\,1,~\Omega^{(j)}(t)\,{\in}\,\stateset{{\mathrm{C}}_{\Omega}}$}} \!\!\!\!\!\!\!\! \\
\bottomrule
\end{tabular}
\renewcommand{\arraystretch}{1}
\end{table}
To summarize the discussion up to this point, we present in Table \ref{tMethods} a comparison of how various approaches to quantum theory represent the world of reality (i.e., individual integer functions) and the world of expectation (i.e., real-valued functions).
The standard formulation of quantum theory (\#4) does not recognize the existence of hidden variables, excluding integer functions from theoretical descriptions~\cite{Peres} (Appendix \ref{sDensity}).
Similarly, recent quantum jump theories (\#5 and \#6) do not admit hidden variables, describing quantum jumps based on continuous changes in Hilbert space.
In contrast, our theory (\#7) describes the world of expectation using density operators and the world of reality using Boolean logical variables.
Such a theory is only possible under a combination of (i) Einstein's interpretation of quantum theory, (ii) Bohr's model of physical reality, and (iii) a selection rule to harmonize Boolean logic and quantum theory.
A comprehensive list of necessary assumptions is shown in Table \ref{tAssumptions}.

\subsection{Notation}
In the following, a special notation is used to express propositions as concisely as possible.
First, propositions whose probabilities are predictable by quantum theory are distinguished by single quotation marks (`$\cdots$').
A time variable is added at the end if necessary.
For instance, the proposition `A is true at time $t$' is denoted as `A,\,$t$'.
Second, when expressing propositions, a single stationary state $|\phi\rangle$ is denoted as $\state{\phi}{}$ by enclosing the quantum number in single braces.
Similarly, a set of stationary states satisfying a condition C is denoted as $\stateset{\mathrm{C}}$ by enclosing the condition label in double curly braces.
To save symbols, the subspace of the Hilbert space spanned by the stationary states in set $\stateset{\mathrm{C}}$ is also represented using the same symbol $\stateset{\mathrm{C}}$.
Third, the proposition describing stationary state occupation is denoted as `$\occupation{~}{}$'.
Specifically, `$\occupation{\state{\phi}{}}{},t$' represents a proposition `A stationary state $|\phi\rangle$ is occupied at time $t$', and `$\occupation{\stateset{\mathrm{C}}}{},t$' represents a proposition `The stationary state occupied at time $t$ is contained in subspace $\stateset{\mathrm{C}}$'.
Other symbols are defined according to quantum optics conventions whenever possible~\cite{Mandel,Cohen}.
The meanings of all symbols are summarized in Appendix \ref{sAppNotation}, but will be explained as necessary.

\section{Definition of logical variables}
\label{cLogical}
\subsection{Basic assumptions}
\label{sBasic}
In this section, we identify the combination of assumptions necessary to define Boolean logical variables whose expectation values are predictable by quantum theory.
To this end, we first make the following assumption:
\begin{myAssumptionA}
The truth value of a given proposition `A,\,$t$' in individual trials can be expressed using the logical variable
\begin{align}
\xjvar{`A,\,$t$'}{}\equiv
\begin{cases}
1, & \mbox{`A' is True at time $t$}\\
0, & \mbox{`A' is False at time $t$}
\end{cases}
,\quad j\,{\in}\,\mathbb{N},
\label{eeBasic1}
\end{align}
where $j$ is the trial number and $\mathbb{N}$ denotes natural numbers (i.e., integers greater than 0).
\end{myAssumptionA}
\begin{myAssumptionA}
The expectation value of the logical variable can be uniquely determined as a real-valued function by 
\begin{align}
\big\langle \xvar{`A,\,$t$'}{} \big\rangle
\equiv \lim_{N\to\infty} \frac{1}{N} \! \sum_{j}^{\mbox{\scriptsize `$\mathrm{A}_{0},t_{0}$'}} \xjvar{`A,\,$t$'}{}
\in {\mathbb{R}},
\label{eeBasic2}
\end{align}
where `$\mathrm{A}_{0},t_{0}$' is a proposition that becomes true for a given ensemble, and $\sum_{j}^{\mbox{\scriptsize `$\mathrm{A}_{0},t_{0}$'}}$ stands for the sum over the given ensemble satisfying `$\mathrm{A}_{0},t_{0}$'.
The sample number is defined by $N\,{\equiv}\,\sum_{j}^{\mbox{\scriptsize `$\mathrm{A}_{0},t_{0}$'}}1$.
\end{myAssumptionA}
\begin{myAssumptionA}
The expectation value of the logical variable can be predicted using the Born rule and the formal solution of the equation of motion, namely
\begin{align}
\big\langle \xvar{`A,\,$t$'}{} \big\rangle
& = {\mathrm{Tr}}\Big[\,\xop{`A,\,$t$'}{} \hat{\rho}(\mbox{`$\mathrm{A}_{0},t_{0}$'})\,\Big],
\label{eeBasic3}\\
\xop{`A,\,$t$'}{} 
& \equiv \hat{U}^{\dagger}(t)\,\xop{`A'}{}\,\hat{U}(t),
\label{eeBasic4}
\end{align}
where $\hat{\rho}(\mbox{`$\mathrm{A}_{0},t_{0}$'})$ is the density operator of the given ensemble and $\xop{`A,\,$t$'}{}$ is an operator corresponding to `A,\,$t$'.
For clarity, we consider a closed system and assume that the time evolution operator $\hat{U}(t)\,{\equiv}\,\exp({t}\hat{H}/{{\rm{i}}\hbar})$ is unitary and $\hat{\rho}(\mbox{`$\mathrm{A}_{0},t_{0}$'})$ is Hermitian.
\end{myAssumptionA}
\noindent
Under these assumptions, the expectation value of a logical variable represents the probability of `A,\,$t$' conditioned by `$\mathrm{A}_{0},t_{0}$', as can be seen from
\begin{align}
\big\langle \xvar{`A,\,$t$'}{} \big\rangle
\equiv \lim_{N\to\infty} \frac{\sum_{j}^{\mbox{\scriptsize `$\mathrm{A}_{0},t_{0}$'}} \xjvar{`A,\,$t$'}{}}{\sum_{j}^{\mbox{\scriptsize `$\mathrm{A}_{0},t_{0}$'}} 1} 
= \lim_{N\to\infty} \frac{\sum_{j}^{\mbox{\scriptsize `$\mathrm{A}_{0},t_{0}$'}\,\wedge\mbox{\scriptsize{`A,\,$t$'}}} 1}{\sum_{j}^{\mbox{\scriptsize `$\mathrm{A}_{0},t_{0}$'}} 1}.
\label{eeBasic6}
\end{align}
This shows that Assumptions A1--A3 define a proposition `A,\,$t$' whose probability is predictable by quantum theory.
Note that $\xop{`A,\,$t$'}{}$ is always Hermitian and positive-semidefinite since $0\,{\le}\,{\mathrm{Tr}}[\,\xop{`A,\,$t$'}{} \hat{\rho}(\mbox{`$\mathrm{A}_{0}$'})\,]\,{\le}\,1$ by Eq.\eqref{eeBasic6}.
It follows that $\xop{`A,\,$t$'}{}$ can be decomposed using the non-Hermitian operator $\zop{`A,\,$t$'}{}$ into the following form:
\begin{align}
\xop{`A,\,$t$'}{} = \zopd{`A,\,$t$'}{}\,\zop{`A,\,$t$'}{}.
\label{eeBasic7}
\end{align}
This means that Eqs.\eqref{eeBasic3}--\eqref{eeBasic4} can always be rewritten as follows~\cite{Breuer}:
\begin{align}
\big\langle \xvar{`A,\,$t$'}{} \big\rangle
& = {\mathrm{Tr}}\Big[\,\zop{`A,\,$t$'}{} \hat{\rho}(\mbox{`$\mathrm{A}_{0},t_{0}$'}) \zopd{`A,\,$t$'}{}\,\Big],
\label{eeBasic8}\\
\zop{`A,\,$t$'}{}
& \equiv \hat{U}^{\dagger}(t)\,\zop{`A'}{}\,\hat{U}(t).
\label{eeBasic9}
\end{align}
Here, the operators $\xop{`A,\,$t$'}{}$ and $\zop{`A,\,$t$'}{}$ play a central role in the theory presented below, so we will refer to them as the \emph{characteristic and partial characteristic operators} of `A,\,$t$', respectively.\footnote{In our theory, the most general form of the partial characteristic operator $\hat{z}$ is the product of the projection operators of propositions satisfying the compatibility condition, while in the Consistent Histories approach it is the product of the projection operators satisfying the consistency condition (i.e.,\,the chain operator of History)~\cite{Griffiths2001}. The characteristic operator $\hat{x}$ becomes significant only when considering general Boolean logic operations, as shown below.}
\par
Next, we assume that the logical variables further satisfy the laws of Boolean logic.
\begin{myAssumptionA}
Given a proposition `$\mathrm{A},\,t_{\mathrm{A}}$' and a proposition `$\mathrm{B},\,t_{\mathrm{B}}$', their negation ($\lnot$, NOT), conjunction ($\wedge$, AND), and disjunction ($\vee$, OR) are defined by
\begin{align}
\xjvar{$\lnot$`$\mathrm{A},\,t_{\mathrm{A}}$'}{}
& = 1 - \xjvar{`$\mathrm{A},\,t_{\mathrm{A}}$'}{}
,~~ {\forall}j,
\label{eeLogical1}\\
\xjvar{`$\mathrm{A},\,t_{\mathrm{A}}$'$\wedge$`$\mathrm{B},\,t_{\mathrm{B}}$'}{}
& = \xjvar{`$\mathrm{A},\,t_{\mathrm{A}}$'}{}\,\xjvar{`$\mathrm{B},\,t_{\mathrm{B}}$'}{}
,~~ {\forall}j,
\label{eeLogical2}\\
\xjvar{`$\mathrm{A},\,t_{\mathrm{A}}$'$\vee$`$\mathrm{B},\,t_{\mathrm{B}}$'}{}
& = \xjvar{`$\mathrm{A},\,t_{\mathrm{A}}$'}{}
+ \xjvar{`$\mathrm{B},\,t_{\mathrm{B}}$'}{}
- \xjvar{`$\mathrm{A},\,t_{\mathrm{A}}$'$\wedge$`$\mathrm{B},\,t_{\mathrm{B}}$'}{}
,~~ {\forall}j, \!\!\!\!\!\!\!\!
\label{eeLogical3}
\end{align}
where Eq.\eqref{eeLogical1} holds for `$\mathrm{B},\,t_{\mathrm{B}}$' as well as `$\mathrm{A},\,t_{\mathrm{A}}$'.
\end{myAssumptionA}
\noindent
Note that the combination of Assumptions A1--A4 is prohibited in the standard formulation.
In contrast, we introduce additional assumptions so that Assumptions A1 and A4 hold in the world of reality and Assumption A3 holds in the world of expectation.
Assuming that Assumption A3 also holds for the result of logical operations, we can immediately see that Assumptions A1--A4 are not closed by themselves.
More specifically, even if `$\mathrm{A},\,t_{\mathrm{A}}$' and `$\mathrm{B},\,t_{\mathrm{B}}$' satisfy Assumptions A1--A4, their conjunction and disjunction may not satisfy Assumption A3.
Taking the expectation value of Eqs.\eqref{eeLogical1}--\eqref{eeLogical3} and substituting Eq.\eqref{eeBasic3}, we see that Assumptions A1--A4 are consistent only if
\begin{align}
{\rm{Tr}}\big[\,\xop{$\lnot$`$\mathrm{A},\,t_{\mathrm{A}}$'}{}\,\hat{\rho}(\mbox{`$\mathrm{A}_{0},t_{0}$'})\,\big]
& = {\rm{Tr}}\big[\,\big(1 - \xop{`$\mathrm{A},\,t_{\mathrm{A}}$'}{}\big)\,\hat{\rho}(\mbox{`$\mathrm{A}_{0},t_{0}$'})\,\big],
\label{eeLogical6}\\
{\rm{Tr}}\big[\,\xop{`$\mathrm{A},\,t_{\mathrm{A}}$'$\wedge$`$\mathrm{B},\,t_{\mathrm{B}}$'}{}\,\hat{\rho}(\mbox{`$\mathrm{A}_{0},t_{0}$'})\,\big]
& = {\rm{Tr}}\big[\,\xop{`$\mathrm{B},\,t_{\mathrm{B}}$'$\wedge$`$\mathrm{A},\,t_{\mathrm{A}}$'}{}\,\hat{\rho}(\mbox{`$\mathrm{A}_{0},t_{0}$'})\,\big],
\label{eeLogical7}\\
{\rm{Tr}}\big[\,\xop{`$\mathrm{A},\,t_{\mathrm{A}}$'$\vee$`$\mathrm{B},\,t_{\mathrm{B}}$'}{}\,\hat{\rho}(\mbox{`$\mathrm{A}_{0},t_{0}$'})\,\big]
& = {\rm{Tr}}\big[\,\big(\xop{`$\mathrm{A},\,t_{\mathrm{A}}$'}{}
+ \xop{`$\mathrm{B},\,t_{\mathrm{B}}$'}{}
\nonumber\\
& ~~~~~~~~~~ 
- \xop{`$\mathrm{A},\,t_{\mathrm{A}}$'$\wedge$`$\mathrm{B},\,t_{\mathrm{B}}$'}{}\big)
\hat{\rho}(\mbox{`$\mathrm{A}_{0},t_{0}$'})\,\big].
\label{eeLogical8}
\end{align}
Focusing on the case $t_{\mathrm{A}}\,{\neq}\,t_{\mathrm{B}}$, these equations are satisfied if the characteristic operators are defined as
\begin{align}
\xop{$\lnot$`$\mathrm{A},\,t_{\mathrm{A}}$'}{}
& = \hat{1} - \xop{`$\mathrm{A},\,t_{\mathrm{A}}$'}{},
\label{eeLogical9}\\
\xop{`$\mathrm{A},\,t_{\mathrm{A}}$'$\wedge$`$\mathrm{B},\,t_{\mathrm{B}}$'}{}
& = \mathcal{T}::\xop{`$\mathrm{A},\,t_{\mathrm{A}}$'}{}\,\xop{`$\mathrm{B},\,t_{\mathrm{B}}$'}{}::,
\label{eeLogical10}\\
\xop{`$\mathrm{A},\,t_{\mathrm{A}}$'$\vee$`$\mathrm{B},\,t_{\mathrm{B}}$'}{} 
& = \xop{`$\mathrm{A},\,t_{\mathrm{A}}$'}{}
+ \xop{`$\mathrm{B},\,t_{\mathrm{B}}$'}{} - \xop{`$\mathrm{A},\,t_{\mathrm{A}}$'$\wedge$`$\mathrm{B},\,t_{\mathrm{B}}$'}{},
\label{eeLogical11}
\end{align}
where $\mathcal{T}$ and $::\,::$ are the time-ordering and normal-ordering symbols for partial characteristic operators, respectively.
It is easy to see that Eq.\eqref{eeLogical9} and Eq.\eqref{eeLogical11} are necessary for Eq.\eqref{eeLogical6} and Eq.\eqref{eeLogical8} to hold for any $\hat{\rho}(\mbox{`$\mathrm{A}_{0},t_{0}$'})$.
On the other hand, Eq.\eqref{eeLogical10} involves an operator product and must be derived from a heuristic consideration as described below.

\subsection{Derivation of the compatibility condition}
\label{sCompatibility}
To derive Eq.\eqref{eeLogical10}, we return to the definitions of characteristic and partial characteristic operators and note that the following relationship exists:
\begin{align}
\big\langle \xvar{`$\mathrm{A},\,t_{\mathrm{A}}$'$\wedge$`$\mathrm{B},\,t_{\mathrm{B}}$'}{} \big\rangle
&={\mathrm{Tr}}\Big[\,\xop{`$\mathrm{A},\,t_{\mathrm{A}}$'$\wedge$`$\mathrm{B},\,t_{\mathrm{B}}$'}{}\,\hat{\rho}(\mbox{`$\mathrm{A}_{0},t_{0}$'})\,\Big]
\label{eeLogical21}\\
&={\mathrm{Tr}}\Big[\,\zop{`$\mathrm{A},\,t_{\mathrm{A}}$'$\wedge$`$\mathrm{B},\,t_{\mathrm{B}}$'}{}\,\hat{\rho}(\mbox{`$\mathrm{A}_{0},t_{0}$'})\,\zopd{`$\mathrm{A},\,t_{\mathrm{A}}$'$\wedge$`$\mathrm{B},\,t_{\mathrm{B}}$'}{}\,\Big].
\label{eeLogical22}
\end{align}
As the simplest case, let us consider propositions `A,\,$t_{\mathrm{A}}$' and `B,\,$t_{\mathrm{B}}$' that describe the occupation of single quantum states $|\mbox{A}, t_{\mathrm{A}}\rangle$ and $|\mbox{B}, t_{\mathrm{B}}\rangle$, respectively. 
In this case, $\big\langle \xvar{`$\mathrm{A},\,t_{\mathrm{A}}$'$\wedge$`$\mathrm{B},\,t_{\mathrm{B}}$'}{} \big\rangle$ should physically correspond to the transition probability between $|\mbox{A}, t_{\mathrm{A}}\rangle$ and $|\mbox{B}, t_{\mathrm{B}}\rangle$~\cite{DiracQM}:
\begin{align}
\big\langle \xvar{`$\mathrm{A},\,t_{\mathrm{A}}$'$\wedge$`$\mathrm{B},\,t_{\mathrm{B}}$'}{} \big\rangle
= \big|\langle \mbox{A}, t_{\mathrm{A}}|\mbox{B}, t_{\mathrm{B}}\rangle\big|^{2}.
\label{eeLogical24}
\end{align}
Meanwhile, by the Born rule, we have
\begin{align}
\zop{`$\mathrm{A},\,t_{\mathrm{A}}$'}{}\,{=}\,|\mbox{A}, t_{\mathrm{A}}\rangle\langle \mbox{A}, t_{\mathrm{A}}|
,\quad
\zop{`$\mathrm{B},\,t_{\mathrm{B}}$'}{}\,{=}\,|\mbox{B}, t_{\mathrm{B}}\rangle \langle \mbox{B}, t_{\mathrm{B}}|.
\label{eeLogical23}
\end{align}
Comparing Eqs.\eqref{eeLogical22}--\eqref{eeLogical23}, we find that $\zop{`$\mathrm{A},\,t_{\mathrm{A}}$'$\wedge$`$\mathrm{B},\,t_{\mathrm{B}}$'}{}$ must be the product of $\zop{`$\mathrm{A},\,t_{\mathrm{A}}$'}{}$ and $\zop{`$\mathrm{B},\,t_{\mathrm{B}}$'}{}$~\cite[Chap.10]{Griffiths2001}.
It then follows from the formal solution of the Heisenberg equation (Eq.\eqref{eeBasic4}) that the partial characteristic operators must act on the density operator in ascending order of time as
\begin{align}
\zop{`$\mathrm{A},\,t_{\mathrm{A}}$'$\wedge$`$\mathrm{B},\,t_{\mathrm{B}}$'}{}
=\begin{cases}
\zop{`$\mathrm{A},\,t_{\mathrm{A}}$'}{}\,\zop{`$\mathrm{B},\,t_{\mathrm{B}}$'}{}, & t_{\mathrm{A}}>t_{\mathrm{B}},\\
\zop{`$\mathrm{B},\,t_{\mathrm{B}}$'}{}\,\zop{`$\mathrm{A},\,t_{\mathrm{A}}$'}{}, & t_{\mathrm{A}}<t_{\mathrm{B}}.\\
\end{cases}
\label{eeLogical25}
\end{align}
Substituting Eq.\eqref{eeLogical25} into Eq.\eqref{eeBasic7}, we obtain the case $t_{\mathrm{A}}\,{\neq}\,t_{\mathrm{B}}$ of Eq.\eqref{eeLogical10}:
\begin{align}
\xop{`$\mathrm{A},\,t_{\mathrm{A}}$'$\wedge$`$\mathrm{B},\,t_{\mathrm{B}}$'}{}
&= \zopd{`$\mathrm{A},\,t_{\mathrm{A}}$'$\wedge$`$\mathrm{B},\,t_{\mathrm{B}}$'}{}\,\zop{`$\mathrm{A},\,t_{\mathrm{A}}$'$\wedge$`$\mathrm{B},\,t_{\mathrm{B}}$'}{}
\label{eeLogical26}\\
&= \begin{cases}
\zopd{`$\mathrm{B},\,t_{\mathrm{B}}$'}{}\,\zopd{`$\mathrm{A},\,t_{\mathrm{A}}$'}{}\,\zop{`$\mathrm{A},\,t_{\mathrm{A}}$'}{}\,\zop{`$\mathrm{B},\,t_{\mathrm{B}}$'}{}, & t_{\mathrm{A}}>t_{\mathrm{B}},\\
\zopd{`$\mathrm{A},\,t_{\mathrm{A}}$'}{}\,\zopd{`$\mathrm{B},\,t_{\mathrm{B}}$'}{}\,\zop{`$\mathrm{B},\,t_{\mathrm{B}}$'}{}\,\zop{`$\mathrm{A},\,t_{\mathrm{A}}$'}{}, & t_{\mathrm{A}}<t_{\mathrm{B}}.
\end{cases}
\label{eeLogical27}
\end{align}
\par
Even in previous theories, it has been believed that Eq.\eqref{eeLogical25} holds in more general cases than assumed in Eq.\eqref{eeLogical23}.
For example, the Consistent Histories approach assumes that the chain operator of History can be defined when projection operators commute~\cite[Chap.10]{Griffiths2001}.
Additionally, for general fundamental excitations, quantum correlation functions~\cite[{\S}11.12]{Mandel} and many-body Green functions~\cite[{\S}3.7]{Fetter} are defined as similar time- and normally-ordered products of creation and annihilation operators.
The validity of such generalizations should be judged by whether they can explain the known experimental facts.
Thus, to represent integer functions that previous theories cannot describe, we relax the existing constraints.
We assume that Eq.\eqref{eeLogical25} holds as long as the partial characteristic operator of a single-time proposition is a projection operator.
\begin{myAssumptionA}
Boolean logic operations transform characteristic operators according to the following equations:
\begin{align}
\xop{$\lnot$`$\mathrm{A},\,t_{\mathrm{A}}$'}{}
& = \hat{1} - \xop{`$\mathrm{A},\,t_{\mathrm{A}}$'}{},
\label{eeComp1}\\
\xop{`$\mathrm{A},\,t_{\mathrm{A}}$'$\wedge$`$\mathrm{B},\,t_{\mathrm{B}}$'}{}
& = \mathcal{T}::\xop{`$\mathrm{A},\,t_{\mathrm{A}}$'}{}\,\xop{`$\mathrm{B},\,t_{\mathrm{B}}$'}{}::
\label{eeComp2}\\
\xop{`$\mathrm{A},\,t_{\mathrm{A}}$'$\vee$`$\mathrm{B},\,t_{\mathrm{B}}$'}{} ,
& = \xop{`$\mathrm{A},\,t_{\mathrm{A}}$'}{}
+ \xop{`$\mathrm{B},\,t_{\mathrm{B}}$'}{}
- \xop{`$\mathrm{A},\,t_{\mathrm{A}}$'$\wedge$`$\mathrm{B},\,t_{\mathrm{B}}$'}{}.
\label{eeComp3}
\end{align}
Here, $\mathcal{T}$ and $::\,::$ are the time-ordering and normal-ordering symbols for partial characteristic operators, respectively.
\end{myAssumptionA}
\begin{myAssumptionA}
The partial characteristic operator of a single-time proposition is the projection operator onto a set of orthonormalized quantum states:
\begin{align}
\zop{`A'}{} = \sum^{\mbox{\scriptsize `A'}}_{| a \rangle} | a \rangle\langle a |
,\quad \langle a | a' \rangle\,{=}\,\delta_{a,a'}.
\label{eeComp4}
\end{align}
Here, $\sum^{\mbox{\scriptsize `A'}}_{| a \rangle}$ denotes the sum over the quantum states such that proposition `A' is true.
See Section \ref{cReality} for a more specific definition of the set of quantum states.
\end{myAssumptionA}
\par
We can now see that Eq.\eqref{eeComp2} (or Eq.\eqref{eeLogical27}) takes different forms for $t_{\mathrm{A}}\!>\!t_{\mathrm{B}}$ and $t_{\mathrm{A}}\!<\!t_{\mathrm{B}}$, indicating that additional assumptions are necessary to guarantee the consistency of Assumptions A1--A6.
Substituting Eq.\eqref{eeLogical27} into Eq.\eqref{eeLogical21}, we find
\begin{align}
& \big\langle \xvar{`$\mathrm{A},\,t_{\mathrm{A}}$'$\wedge$`$\mathrm{B},\,t_{\mathrm{B}}$'}{} \big\rangle
\nonumber\\
& = \begin{cases}
\displaystyle {\rm{Tr}}\big[\,\zopd{`$\mathrm{B},\,t_{\mathrm{B}}$'}{}\,\zopd{`$\mathrm{A},\,t_{\mathrm{A}}$'}{}\,\zop{`$\mathrm{A},\,t_{\mathrm{A}}$'}{}\,\zop{`$\mathrm{B},\,t_{\mathrm{B}}$'}{}\,\hat{\rho}(\mbox{`$\mathrm{A}_{0},t_{0}$'})\,\big] & t_{\mathrm{A}}>t_{\mathrm{B}},\\
\displaystyle {\rm{Tr}}\big[\,\zopd{`$\mathrm{A},\,t_{\mathrm{A}}$'}{}\,\zopd{`$\mathrm{B},\,t_{\mathrm{B}}$'}{}\,\zop{`$\mathrm{B},\,t_{\mathrm{B}}$'}{}\,\zop{`$\mathrm{A},\,t_{\mathrm{A}}$'}{}\,\hat{\rho}(\mbox{`$\mathrm{A}_{0},t_{0}$'})\,\big] & t_{\mathrm{A}}<t_{\mathrm{B}}.
\end{cases}
\label{eeComp11}
\end{align}
For the probability of $\mbox{`$\mathrm{A},\,t_{\mathrm{A}}$'}\,{\wedge}\,\mbox{`$\mathrm{B},\,t_{\mathrm{B}}$'}$ to be uniquely determined in the limit $t_{\mathrm{A}}\,{\to}\,t$ and $t_{\mathrm{B}}\,{\to}\,t$, it is necessary to assume the following auxiliary condition:
\begin{align}
& {\rm{Tr}}\big[\,\zopd{`B,\,$t$'}{} \zopd{`A,\,$t$'}{} \zop{`A,\,$t$'}{} \zop{`B,\,$t$'}{}\,\hat{\rho}(\mbox{`$\mathrm{A}_{0},t_{0}$'})\,\big]
\nonumber\\
& = {\rm{Tr}}\big[\,\zopd{`A,\,$t$'}{} \zopd{`B,\,$t$'}{} \zop{`B,\,$t$'}{} \zop{`A,\,$t$'}{}\,\hat{\rho}(\mbox{`$\mathrm{A}_{0},t_{0}$'})\,\big]
\label{eeComp12}
\end{align}
Accordingly, we make the following assumptions to harmonize the laws of quantum theory and the laws of Boolean logic at $t_{\mathrm{A}}\,{=}\,t_{\mathrm{B}}$.
\begin{myAssumptionA}
Any meaningful propositions `A' and `B' must satisfy the condition
\begin{align}
& {\rm{Tr}}\big[\,\zopd{`B'}{} \zopd{`A'}{} \zop{`A'}{} \zop{`B'}{}\,\hat{\rho}(\,t\,|\mbox{`$\mathrm{A}_{0},t_{0}$'})\,\big]
\nonumber\\
& = {\rm{Tr}}\big[\,\zopd{`A'}{} \zopd{`B'}{} \zop{`B'}{} \zop{`A'}{}\,\hat{\rho}(\,t\,|\mbox{`$\mathrm{A}_{0},t_{0}$'})\,\big],
\label{eeComp15}
\end{align}
where $\hat{\rho}(\,t\,|\mbox{`$\mathrm{A}_{0},t_{0}$'})\,{\equiv}\,\hat{U}(t)\,\hat{\rho}(\mbox{`$\mathrm{A}_{0},t_{0}$'})\,\hat{U}^{\dagger}(t)$ is the density operator in the Schr\"{o}dinger picture.
\end{myAssumptionA}
\begin{myAssumptionA}
The time-ordering symbol $\mathcal{T}$ and the normal-ordering symbol $::\,::$ for the partial characteristic operator are defined by
\begin{align}
& \mathcal{T}::\xop{`$\mathrm{A},\,t_{\mathrm{A}}$'}{} \xop{`$\mathrm{B},\,t_{\mathrm{B}}$'}{}::\,
\nonumber\\
& \equiv \begin{cases}
\zopd{`$\mathrm{B},\,t_{\mathrm{B}}$'}{} \zopd{`$\mathrm{A},\,t_{\mathrm{A}}$'}{} \zop{`$\mathrm{A},\,t_{\mathrm{A}}$'}{} \zop{`$\mathrm{B},\,t_{\mathrm{B}}$'}{}, & t_{\mathrm{A}}>t_{\mathrm{B}},\phantom{\Big|}\\
\zopd{`$\mathrm{A},\,t_{\mathrm{A}}$'}{} \zopd{`$\mathrm{B},\,t_{\mathrm{B}}$'}{} \zop{`$\mathrm{B},\,t_{\mathrm{B}}$'}{} \zop{`$\mathrm{A},\,t_{\mathrm{A}}$'}{}, & t_{\mathrm{A}}<t_{\mathrm{B}},\phantom{\Big|}\\
\Big(\zopd{`$\mathrm{B},\,t_{\mathrm{B}}$'}{} \zopd{`$\mathrm{A},\,t_{\mathrm{A}}$'}{} \zop{`$\mathrm{A},\,t_{\mathrm{A}}$'}{} \zop{`$\mathrm{B},\,t_{\mathrm{B}}$'}{} & ~\\
~~+\zopd{`$\mathrm{A},\,t_{\mathrm{A}}$'}{} \zopd{`$\mathrm{B},\,t_{\mathrm{B}}$'}{} \zop{`$\mathrm{B},\,t_{\mathrm{B}}$'}{} \zop{`$\mathrm{A},\,t_{\mathrm{A}}$'}{}\Big)/2, & t_{\mathrm{A}}=t_{\mathrm{B}}.
\end{cases}
\label{eeComp19}
\end{align}
\end{myAssumptionA}
\noindent
For convenience of reference, we will refer to Eq.\eqref{eeComp15} in Assumption A7 as the \emph{compatibility condition} between proposition `$\mathrm{A},\,t_{\mathrm{A}}$' and proposition `$\mathrm{B},\,t_{\mathrm{B}}$'.
Note that it is necessary to assume the compatibility condition to justify Assumption A8.
Compatibility conditions for two or more propositions can be obtained in the same way (See {\S}\ref{sGeneralizations}).

\subsection{Joint and conditional probabilities}
\label{sJoint}
Before introducing the model of reality, let us summarize what can be said based on the minimal assumptions described above.
Under Assumptions A1--A8, the joint probability of `$\mathrm{A},\,t_{\mathrm{A}}$' and `$\mathrm{B},\,t_{\mathrm{B}}$' can be defined as follows:
\begin{align}
{\mathrm{Pr}}(\mbox{`$\mathrm{A},\,t_{\mathrm{A}}$'}\,{\wedge}\,\mbox{`$\mathrm{B},\,t_{\mathrm{B}}$'})
\equiv \big\langle \xvar{`$\mathrm{A},\,t_{\mathrm{A}}$'$\wedge$`$\mathrm{B},\,t_{\mathrm{B}}$'}{} \big\rangle
= {\rm{Tr}}\Big[\,\xop{`$\mathrm{A},\,t_{\mathrm{A}}$'$\wedge$`$\mathrm{B},\,t_{\mathrm{B}}$'}{}\,\hat{\rho}(\mbox{`$\mathrm{A}_{0},t_{0}$'})\,\Big].
\label{eeJoint1}
\end{align}
The conditional probability of `A,\,$t_{\mathrm{A}}$' under the information of `B,\,$t_{\mathrm{B}}$' is given by
\begin{align}
{\mathrm{Pr}}(\mbox{`$\mathrm{A},\,t_{\mathrm{A}}$'}|\mbox{`$\mathrm{B},\,t_{\mathrm{B}}$'})
& \equiv \frac{\langle\xvar{`$\mathrm{A},\,t_{\mathrm{A}}$'$\wedge$`$\mathrm{B},\,t_{\mathrm{B}}$'}{}\rangle}{\langle\xvar{`$\mathrm{B},\,t_{\mathrm{B}}$'}{}\rangle}
= \frac{{\rm{Tr}}\Big[\,\xop{`$\mathrm{A},\,t_{\mathrm{A}}$'$\wedge$`$\mathrm{B},\,t_{\mathrm{B}}$'}{}\,\hat{\rho}(\mbox{`$\mathrm{A}_{0},t_{0}$'})\,\Big]}{{\rm{Tr}}\Big[\,\xop{`$\mathrm{B},\,t_{\mathrm{B}}$'}{}\,\hat{\rho}(\mbox{`$\mathrm{A}_{0},t_{0}$'})\,\Big]},
\label{eeJoint2}
\end{align}
where we assumed $t_{\mathrm{A}}\,{>}\,t_{\mathrm{B}}$ for simplicity.
Using Assumption A5 (Eq.\eqref{eeComp2}) and Assumption A8 (Eq.\eqref{eeComp19}), this conditioinal probability can be rewitten in the form
\begin{align}
{\mathrm{Pr}}(\mbox{`$\mathrm{A},\,t_{\mathrm{A}}$'}|\mbox{`$\mathrm{B},\,t_{\mathrm{B}}$'})
= {\rm{Tr}}\Big[\,\hat{x}(\mbox{`A'},\,t_{\mathrm{A}})\,\hat{\rho}\big(\mbox{`$\mathrm{B},\,t_{\mathrm{B}}$'}\big)\,\Big]
,\quad t_{\mathrm{A}} > t_{\mathrm{B}},
\label{eeJoint3}
\end{align}
where the density operator conditioned by proposition `$\mathrm{B},\,t_{\mathrm{B}}$' is defined as
\begin{align}
\hat{\rho}\left(\mbox{`$\mathrm{B},\,t_{\mathrm{B}}$'} \right)
\equiv 
\frac{\zop{`$\mathrm{B},\,t_{\mathrm{B}}$'}{} \hat{\rho}(\mbox{`$\mathrm{A}_{0},t_{0}$'}) \zopd{`$\mathrm{B},\,t_{\mathrm{B}}$'}{}}{{\rm{Tr}}\left[\,\zop{`$\mathrm{B},\,t_{\mathrm{B}}$'}{} \hat{\rho}(\mbox{`$\mathrm{A}_{0},t_{0}$'}) \zopd{`$\mathrm{B},\,t_{\mathrm{B}}$'}{}\,\right]}
,\quad t_{\mathrm{A}} > t_{\mathrm{B}}.
\label{eeJoint3a}
\end{align}
Marginal probabilities can also be defined consistently by the laws of Boolean logic.
On the other hand, the appearance of the conjunctions ($\wedge$) in Eqs.\eqref{eeJoint1}--\eqref{eeJoint2} implies that the compatibility condition between `$\mathrm{A},\,t_{\mathrm{A}}$' and `$\mathrm{B},\,t_{\mathrm{B}}$' (Eq.\eqref{eeComp15}) must be satisfied for joint and conditional probabilities to be defined without contradiction.
\par
Notably, the above definition allows for joint probabilities to be defined for a broader range of propositions than in conventional theory.
In the standard formulation of quantum theory, Nelson's theorem states that a joint probability of two observables ${\mathcal A}$ and ${\mathcal B}$ can only be defined if and only if the corresponding operators are commutative with each other (i.e., $\hat{\mathcal A}\hat{\mathcal B}=\hat{\mathcal B}\hat{\mathcal A}$)~\cite[{\S}2.1.4]{Breuer}.
Correspondingly, the von Neumann-L\"{u}ders' theory~\cite{Lueders1951,Neumann} and the Consistent Histories approach~\cite[(10.44),{\S}4.6]{Griffiths2001} require the commutativity of projection operators for the existence of joint probabilities.
In contrast, our theory allows us to define joint probabilities even if partial characteristic operators do not commute.
To see this, let us rewrite the compatibility condition in the following form
\begin{align}
{\rm{Tr}}\Big[\,\Big(\zopd{`B'}{} \zopd{`A'}{} \zop{`A'}{} \zop{`B'}{} - \zopd{`A'}{} \zopd{`B'}{} \zop{`B'}{} \zop{`A'}{}\Big)\,\hat{\rho}(\,t\,|\mbox{`$\mathrm{A}_{0},t_{0}$'})\,\Big] = 0.
\label{eeJoint21}
\end{align}
Distinguishing the two cases where the compatibility condition holds, namely
\begin{align}
& \mbox{\bf Type-I:}
& \Big(\zop{`A'}{} \zop{`B'}{} \Big)^{\dagger} \zop{`A'}{} \zop{`B'}{}
= \Big(\zop{`B'}{} \zop{`A'}{} \Big)^{\dagger} \zop{`B'}{} \zop{`A'}{},
\label{eeJoint22}\\
& \mbox{\bf Type-II:}
& \Big(\zop{`A'}{} \zop{`B'}{} \Big)^{\dagger} \zop{`A'}{} \zop{`B'}{}
\neq \Big(\zop{`B'}{} \zop{`A'}{} \Big)^{\dagger} \zop{`B'}{} \zop{`A'}{},
\label{eeJoint23}
\end{align}
we see that conventional theories have only considered Type-I compatibility due to the requirement of the commutativity of projection operators (i.e., $\zop{`A'}{} \zop{`B'}{}\,{=}\,\zop{`B'}{}\zop{`A'}{}$).
As is well known, this requirement is based on the premise that the definition of joint probabilities should not be situation-dependent~\cite{Breuer}.
On the other hand, our theory abandons this implicit assumption, allowing Type-II compatibility to enable the situation-dependent definition of joint probabilities.
\par
It is important to note that Assumptions A1--A8 contain neither the consistency condition~\cite{Griffiths2001} nor the decoherence condition~\cite{Plenio1998,Zurek2003}.
This is because the compatibility condition guarantees the consistency of Boolean logic operations in a different way than in previous theories.
In our theory, it is no longer necessary to link the observation of individual events with the collapse of the density operator.
Thus, off-diagonal elements of the density operator vanish only by the intentional modification of the statistical ensemble (see Eq.\eqref{eeJoint3a}).
This is made possible by distinguishing between the world of reality, in which individual events occur, and the world of expectation, in which the laws of quantum theory holds.

\subsection{Deterministic relations between individual events}
\label{sDeterministic}
To see the important role played by the compatibility condition, it is crucial to identify the meaning of the statistical concepts used in the explanation.
The following definitions are classical but are explicitly written down for later use.
\begin{myDefinition*}[Independence]
Propositions `$\mathrm{A},\,t_{\mathrm{A}}$' and `$\mathrm{B},\,t_{\mathrm{B}}$' are \emph{independent} if the conditional probabilities equal the marginal probabilities, namely
\begin{align}
{\mathrm{Pr}}(\mbox{`$\mathrm{A},\,t_{\mathrm{A}}$'}|\mbox{`$\mathrm{B},\,t_{\mathrm{B}}$'})
& = {\mathrm{Pr}}(\mbox{`$\mathrm{A},\,t_{\mathrm{A}}$'}),
\label{eeJoint4a}\\
{\mathrm{Pr}}(\mbox{`$\mathrm{B},\,t_{\mathrm{B}}$'}|\mbox{`$\mathrm{A},\,t_{\mathrm{A}}$'})
& = {\mathrm{Pr}}(\mbox{`$\mathrm{B},\,t_{\mathrm{B}}$'}).
\label{eeJoint4}
\end{align}
This means that the joint probability of independent propositions can always be factorized into marginal probabilities so that
\begin{align}
{\mathrm{Pr}}(\mbox{`$\mathrm{A},\,t_{\mathrm{A}}$'}\,{\wedge}\,\mbox{`$\mathrm{B},\,t_{\mathrm{B}}$'})={\mathrm{Pr}}(\mbox{`$\mathrm{A},\,t_{\mathrm{A}}$'})\,{\mathrm{Pr}}(\mbox{`$\mathrm{B},\,t_{\mathrm{B}}$'}).
\label{eeJoint5}
\end{align}
\end{myDefinition*}
\begin{myDefinition*}[Exclusivity]
Propositions `$\mathrm{A},\,t_{\mathrm{A}}$' and `$\mathrm{B},\,t_{\mathrm{B}}$' are \emph{exclusive} if their joint probability is zero, such that
\begin{align}
{\mathrm{Pr}}(\mbox{`$\mathrm{A},\,t_{\mathrm{A}}$'}\,{\wedge}\,\mbox{`$\mathrm{B},\,t_{\mathrm{B}}$'})
\equiv \big\langle\xvar{`$\mathrm{A},\,t_{\mathrm{A}}$'$\wedge$`$\mathrm{B},\,t_{\mathrm{B}}$'}{}\big\rangle
= 0.
\label{eeJoint6}
\end{align}
This implies that the conjunction of exclusive propositions is always false:
\begin{align}
x^{(j)}(\mbox{`$\mathrm{A},\,t_{\mathrm{A}}$'}\,{\wedge}\,\mbox{`$\mathrm{B},\,t_{\mathrm{B}}$'})=0.
\label{eeJoint7}
\end{align}
It then follows from Eq.\eqref{eeJoint7} and the laws of Boolean logic (Assumption A4) that the logical variables of mutually exclusive propositions are mutually additive:
\begin{align}
\xjvar{`$\mathrm{A},\,t_{\mathrm{A}}$'$\vee$`$\mathrm{B},\,t_{\mathrm{B}}$'}{}
= \xjvar{`$\mathrm{A},\,t_{\mathrm{A}}$'}{} + \xjvar{`$\mathrm{B},\,t_{\mathrm{B}}$'}{}.
\label{eeJoint9}
\end{align}
\end{myDefinition*}
\begin{myDefinition*}[Deterministic Relations]
A \emph{deterministic relation} exists between propositions `$\mathrm{A},\,t_{\mathrm{A}}$' and `$\mathrm{B},\,t_{\mathrm{B}}$' when their truth values always concide as
\begin{align}
\xjvar{`$\mathrm{A},\,t_{\mathrm{A}}$'}{} = \xjvar{`$\mathrm{B},\,t_{\mathrm{B}}$'}{}
,\quad \forall{j}.
\label{eeJoint11}
\end{align}
By the definition of conditional probabilities (Eq.\eqref{eeBasic6}), this equation is equivalent to
\begin{align}
1 = {\mathrm{Pr}}(\mbox{`$\mathrm{A},\,t_{\mathrm{A}}$'}|\mbox{`$\mathrm{B},\,t_{\mathrm{B}}$'}) = {\mathrm{Pr}}(\mbox{`$\mathrm{B},\,t_{\mathrm{B}}$'}|\mbox{`$\mathrm{A},\,t_{\mathrm{A}}$'}),
\label{eeJoint12}
\end{align}
which can be further rewritten using Eq.\eqref{eeJoint2} as
\begin{align}
\langle \xvar{`$\mathrm{A},\,t_{\mathrm{A}}$'$\wedge$`$\mathrm{B},\,t_{\mathrm{B}}$'}{} \rangle
= \langle \xvar{`$\mathrm{A},\,t_{\mathrm{A}}$'}{} \rangle
= \langle \xvar{`$\mathrm{B},\,t_{\mathrm{B}}$'}{} \rangle.
\label{eeJoint13}
\end{align}
\end{myDefinition*}
\noindent
According to the above definition, a deterministic relation between `$\mathrm{A},\,t_{\mathrm{A}}$' and `$\mathrm{B},\,t_{\mathrm{B}}$' should be expressed in the form of Eqs.\eqref{eeJoint11}--\eqref{eeJoint13}.
On the other hand, von Neumann-L\"{u}ders' theory~\cite{Lueders1951} and the Consistent Histories approach~\cite{Griffiths2001} could not define the left-hand side of Eq.\eqref{eeJoint13} when projection operators are not commutative (i.e., $\zop{`A'}{}\zop{`B'}{}\,{\neq}\,$ $\zop{`B'}{}\zop{`A'}{}$).
Consequently, the only relation that could be safely expressed in conventional theories was the equality between expectation values, namely
\begin{align}
\big\langle \xvar{`$\mathrm{A},\,t_{\mathrm{A}}$'}{} \big\rangle = \big\langle \xvar{`$\mathrm{B},\,t_{\mathrm{B}}$'}{} \big\rangle.
\label{eeJoint14}
\end{align}
This means that the deterministic relation in Eq.\eqref{eeJoint11} may represent a law of reality that cannot be expressed in the standard formulation.
Note that the compatibility condition between `$\mathrm{A},\,t_{\mathrm{A}}$' and `$\mathrm{B},\,t_{\mathrm{B}}$' must hold for the deterministic relation to exist.
\par
Since our logical variables take only two values (0 or 1), it is necessary to check the implications of applying approximations to propositions.
To this end, consider an approximate proposition `$\underline{\mathrm{B}}$' replacing original proposition `B' and let `TN' and `FP' be the true-negative and false-positive type confusions due to the replacement, respectively.
Then, the influence of erroneous determination can be incorporated by rewriting Eq.\eqref{eeJoint11} as follows:
\begin{align}
\xjvar{`$\mathrm{A},\,t_{\mathrm{A}}$'}{}
& = \xjvar{$\big(\mbox{$\lnot$`TN'$\wedge$`$\underline{\mathrm{B}},\,t_{\mathrm{B}}$'}\big)\,{\vee}\, 
\big(\mbox{`FP'$\wedge\lnot$`$\underline{\mathrm{B}},\,t_{\mathrm{B}}$'}\big)$}{Big}
,\quad \forall{j}
\label{eeDeterministic10}\\
& = \xjvar{$\lnot$`TN'}{} \, \xjvar{`$\underline{\mathrm{B}},\,t_{\mathrm{B}}$'}{} 
+ \xjvar{`FP'}{} \, \xjvar{$\lnot$`$\underline{\mathrm{B}},\,t_{\mathrm{B}}$'}{}
,\quad \forall{j}.
\label{eeDeterministic11}
\end{align}
Assuming that the probabilities of errorneous determinations are $p_{\mathrm{TN}}$ and $p_{\mathrm{FP}}$ and that `TN' and `FP' are independent with any other propositions, we have $\zop{`TN'}{}\,{=}\,\sqrt{p_{\mathrm{TN}}}\,\hat{1}$, $\zop{`FP'}{}\,{=}\,\sqrt{p_{\mathrm{FP}}}\,\hat{1}$, so that
\begin{align}
\langle \xvar{`$\underline{\mathrm{B}},\,t_{\mathrm{B}}$'$\wedge\lnot$`TN'}{} \rangle 
& \cong \langle \xvar{`$\underline{\mathrm{B}},\,t_{\mathrm{B}}$'}{} \rangle \langle \xvar{$\lnot$`TN'}{} \rangle,
\label{eeDeterministic12}\\
\langle \xvar{`$\underline{\mathrm{B}},\,t_{\mathrm{B}}$'$\wedge$`FP'}{} \rangle 
& \cong \langle \xvar{`$\underline{\mathrm{B}},\,t_{\mathrm{B}}$'}{} \rangle \langle \xvar{`FP'}{} \rangle.
\label{eeDeterministic13}
\end{align}
The use of conjunction can be justified because the compatibility condition between `$\mathrm{A},\,t_{\mathrm{A}}$' and `$\underline{\mathrm{B}},\,t_{\mathrm{B}}$' will give the same conclusion as the compatibility condition between `$\mathrm{A},\,t_{\mathrm{A}}$' and `$\mathrm{B},\,t_{\mathrm{B}}$'.
Multiplying $\xjvar{`$\mathrm{A},\,t_{\mathrm{A}}$'}{}$ or $\xjvar{`$\underline{\mathrm{B}},\,t_{\mathrm{B}}$'}{}$ to the both side of Eq.\eqref{eeDeterministic11}, taking the expectation value, and substituting Eqs.\eqref{eeDeterministic12}--\eqref{eeDeterministic13} into the resulting equations, we get
\begin{align}
\langle \xvar{`$\mathrm{A},\,t_{\mathrm{A}}$'$\wedge$`$\underline{\mathrm{B}},\,t_{\mathrm{B}}$'}{} \rangle
& = \frac{ 1 - \langle \xvar{`FP'}{} \rangle }{ \langle \xvar{$\lnot$`TN'}{} \rangle - \langle \xvar{`FP'}{} \rangle } \, \langle \xvar{`$\mathrm{A},\,t_{\mathrm{A}}$'}{} \rangle
\label{eeDeterministic14}\\
& = \langle \xvar{$\lnot$`TN'}{} \rangle \, \langle \xvar{`$\underline{\mathrm{B}},\,t_{\mathrm{B}}$'}{} \rangle.
\label{eeDeterministic15}
\end{align}
When the probabilities of erroneous determinations are sufficiently small so that $\langle \xvar{$\lnot$`TN'}{} \rangle\,{\cong}\,1$, $\langle \xvar{`FP'}{} \rangle\,{\cong}\,0$, this equation yields
\begin{align}
\langle \xvar{`$\mathrm{A},\,t_{\mathrm{A}}$'$\wedge$`$\underline{\mathrm{B}},\,t_{\mathrm{B}}$'}{} \rangle
\cong \langle \xvar{`$\mathrm{A},\,t_{\mathrm{A}}$'}{} \rangle
\cong \langle \xvar{`$\underline{\mathrm{B}},\,t_{\mathrm{B}}$'}{} \rangle.
\label{eeDeterministic17}
\end{align}
Thus, an approximate proposition can be reasonably defined if erroneous determinations are independent of the proposition under consideration and if numerical errors are allowed in evaluating joint and conditional probabilities.
To avoid complications, propositions `TN' and `FP' will be omitted as much as possible in the following.

\section{Model of reality}
\label{cReality}
\subsection{Overview}
\label{sReality0}
In the previous section, we introduced a two-layer theoretical framework where Boolean logical variables describe individual trials and the density operator describes statistical properties.
This dual structure aligns with Einstein's perspective on quantum theory~\cite{EPR,Einstein1949}.
On the other hand, subsequent experiments have observed quantum jumps between stationary states in both atomic systems~\cite{QuantumJump1,QuantumJump2,QuantumJump3} and radiation fields~\cite{Haroche1,Haroche2}.
This indicates the necessity of adopting a stationary-state model of reality~\cite{Bohr1913,Dirac1927}, not the point-particle model by Einstein~\cite{EPR,Einstein1949}.
Historically, however, Bohr overemphasized the fact that his atomic model renders the light quantum hypothesis superfluous~\cite{Bohr1913,Bohr1924}, denying the possibility that quantum theory is a two-layer theory~\cite{Bohr1950}.
Therefore, we reconcile Einstein's and Bohr's claims by using the logical variables described above to represent stationary state occupation.
The list of assumptions has already been presented in Table \ref{tAssumptions}, so this section describes the details of individual assumptions.

\subsection{Definition of stationary states}
\label{sReality1}
First, we make the following assumptions to define stationary states in the non-relativistic QED sense~\cite{Mandel,Cohen,Messiah}.
\begin{myAssumptionB}[Definition of the target system]
The Hamiltonian of the target system is given by
\begin{align}
\hat{H}^{\hat{\mathsf{W}}} 
& = {\hat{\mathsf{W}}}^{\dagger} \left(\hat{H}_{\mathrm{A}} + \hat{H}_{\mathrm{F}} + \hat{H}_{\mathrm{I}}\right) {\hat{\mathsf{W}}}
\label{eeModel11}\\
\hat{H}_{\mathrm{A}}
& = \sum_{k}^{K}\frac{1}{2m_{k}}\|{\hat{\bm p}}_{k}\|^{2}
+\frac{1}{2}\sum_{k}^{K}\sum_{k'}^{k'{\neq}k}\frac{e_{k}e_{k'}}{4\pi\varepsilon_{0}}\frac{1}{\left\|\hat{\bm r}_{k}-\hat{\bm r}_{k'}\right\|},
\label{eeModel12}\\
\hat{H}_{\mathrm{F}}
& = \iiint\frac{\varepsilon_{0}}{2}\left[\|\hat{\bm E}_{\mathrm{T}}({\bm x})\|^{2}+{c^{2}}\|\hat{\bm B}({\bm x})\|^{2}\right]d{\bm x},
\label{eeModel13}\\
\hat{H}_{\mathrm{I}}
& = -\sum^{K}_{k}\frac{e_{k}}{m_{k}} \hat{\bm A}_{\mathrm{T}}(\hat{\bm r}_{k})\cdot{\hat{\bm p}}_{k} 
+ \sum^{K}_{k}\frac{e_{k}^{2}}{2m_{k}}\left\|{\hat{\bm A}_{\mathrm{T}}}(\hat{\bm r}_{k})\right\|^{2}
+ \sum_{k}^{K}\frac{g_{k}e_{k}}{2m_{k}}{\hat{\bm s}}_{k}\cdot\hat{\bm B}(\hat{\bm r}_{k}),
\label{eeModel14}
\end{align}
where $\hat{H}_{\mathrm{A}}$, $\hat{H}_{\mathrm{F}}$, and $\hat{H}_{\mathrm{I}}$ represent the Hamiltonian of the atomic system, radiation field, and interaction, respectively.
$\hat{\bm r}_{k}$ and $\hat{\bm p}_{k}$ ($1\,{\leqq}\,k\,{\leqq}\,K$) are the position and momentum operators of the $k$-th charged particle, and $\hat{\bm A}_{\mathrm{T}}({\bm x})$ and $\hat{\bm E}_{\mathrm{T}}({\bm x})$ are the transverse components of the vector potential and electric field ($0\,{=}\,{\bm \nabla}\cdot\hat{\bm A}_{\mathrm{T}}({\bm x})\,{=}\,{\bm \nabla}\cdot\hat{\bm E}_{\mathrm{T}}({\bm x})$), respectively.
The commutation relations are given by
\begin{align}
& \Big[\hat{r}_{k\xi},\,\hat{p}_{k'\xi'}\Big]_{-} = {\mathrm{i}}\hbar\,\delta_{k k'}\delta_{\xi \xi'}
,\quad 1\,{\leqq}\,k, k'\,{\leqq}\,K
,\quad \xi, \xi'\,{\in}\,\{x,y,z\},
\label{eeModel15}\\
& \Big[{\hat{E}_{{\mathrm{T}}\xi}}({\bm x}, t),\,{\hat{A}_{{\mathrm{T}}\xi'}}({\bm x}', t)\Big]_{-}=\frac{{\mathrm{i}}\hbar}{\varepsilon_{0}}\,\delta^{{\mathrm{T}}}_{\xi \xi'}({\bm x}-{\bm x}')
,\quad \xi, \xi'\,{\in}\,\{x,y,z\},
\label{eeModel16}
\end{align}
with $\delta^{{\mathrm{T}}}_{\xi \xi'}({\bm x}-{\bm x}')$ as the transverse delta function~\cite[Eq.(10.8-15)]{Mandel}.
The $\hat{\mathsf{W}}$ in Eq.\eqref{eeModel11} is a time-independent unitary operator, introduced to take into account the multipolar Hamiltonian~\cite[{\S}14.1]{Mandel} and the field Hamiltonian with displaced vacuum.
See the tables in Appendix \ref{sAppNotation} for the meaning of other symbols.\footnote{For clarity, we consider a closed system with periodic boundary conditions imposed on the surface of a large volume $V$. Thus, the transverse delta function can be expressed as
\begin{align}
\delta^{{\mathrm{T}}}_{\xi \xi'}({\bm x}-{\bm x}')
= \iiint \left(\delta_{\xi \xi'}-\frac{k_{\xi}k_{\xi'}}{\|{\bm k}\|^{2}} \right) {\mathrm{e}}^{{\mathrm{i}}{\bm k}\cdot({\bm x}-{\bm x}')}\,\frac{d{\bm k}}{(2\pi)^{3}}
= \frac{1}{V} \sum_{{\bm k}} \left(\delta_{\xi \xi'}-\frac{k_{\xi}k_{\xi'}}{\|{\bm k}\|^{2}} \right) {\mathrm{e}}^{{\mathrm{i}}{\bm k}\cdot({\bm x}-{\bm x}')}
\label{eeModel17}
\end{align}
Note also that the formal expressions $\hat{\bm B}(\hat{\bm r}_{k})$ and $\hat{\bm A}_{\mathrm{T}}(\hat{\bm r}_{k})$ can be evaluated in the position representation by letting $\hat{\bm r}_{k}\,{\to}\,{\bm r}_{k}$ and $\hat{\bm p}_{k}\,{\to}\,{-}{\mathrm{i}}\hbar{\bm \nabla}_{k}$. The superscript of the Hamiltonian $\hat{H}^{\hat{\mathsf{W}}}$ indicates that the time-independent unitary transformation $\hat{\mathsf{W}}$ was considered.}
\end{myAssumptionB}
\begin{myAssumptionB}[Definition of stationary states]
The stationary states are defined as the orthonormalized eigenstates of the non-interacting Hamiltonian
\begin{align}
\hat{H}^{\hat{\mathsf{U}}}_{0} 
\equiv {\hat{\mathsf{U}}}^{\dagger} \left(\hat{H}_{\mathrm{A}} + \hat{H}_{\mathrm{F}} \right) {\hat{\mathsf{U}}}
\label{eeModel22}
\end{align}
by the eigenequation
\begin{align}
\hat{H}^{\hat{\mathsf{U}}}_{0}|\phi\rangle = E(\phi)|\phi\rangle
,\quad
\langle\phi|\phi'\rangle=\delta_{\phi, \phi'}.
\label{eeModel21}
\end{align}
Here, $\phi$ denotes a complete set of quantum numbers, $E(\phi)$ is the corresponding eigenvalue, and $\delta_{\phi \phi'}$ is a Kronecker delta or a delta function, depending on the type of stationary states~\cite{Messiah}.
$\hat{\mathsf{U}}$ is a time-independent unitary operator, introduced to include coherent states, or more generally, quantum states outside of Hilbert space.
$\hat{\mathsf{U}}$ may include ${\hat{\mathsf{W}}}$ when ${\hat{\mathsf{W}}}\,{\neq}\,\hat{1}$.
\end{myAssumptionB}
\begin{myAssumptionB}[Definition of the set of statoinary states]
The set of orthonormalized stationary states satisfying the given condition ${\mathrm{C}}$ is denoted by\footnote{~For example, the stationary states of a free electron are labeled by $\phi\,{=}\,(p_{x},p_{y},p_{z},\sigma_{\xi})$, where $(p_{x},p_{y},p_{z})\,{\in}\,{\mathbb{R}}^{3}$ is a momentum vector and $\sigma_{\xi}\,{\in}\,\{+1/2,-1/2\}$ is a spin in the $\xi$ direction. Then, a set of stationary states having a particular eigenenergy ${E}$ can be expressed as $\stateset{E}\,{\equiv}\,\{\,|\phi\rangle\,|\,\phi\,{=}$ $\,(p_{E}\cos{\theta},\,p_{E}\sin{\theta}\cos{\phi},\,p_{E}\sin{\theta}\sin{\phi},\sigma_{\xi}),\,p_{E}\,{\equiv}\,\sqrt{2m_{\mathrm{e}}{E}},\,\theta\,{\in}\,[0,\pi),\,\phi\,{\in}\,[0,2\pi),\sigma_{\xi}\,{\in}\,\{+1/2,-1/2\}\}$.}
\begin{align}
\stateset{\mathrm{C}}_{\hat{\mathsf{U}}}
\equiv \left\{\,\forall |\phi\rangle\,\Big|
~\hat{H}^{\hat{\mathsf{U}}}_{0}|\phi\rangle\,{=}\,E(\phi)|\phi\rangle,
~\langle \phi | \phi' \rangle\,{=}\,\delta_{\phi,\phi'},
~\mbox{$|\phi\rangle$ satisfies $\mathrm{C}$\,} \right\}.
\label{eeModel31}
\end{align}
In particular, given a complete set of stationary states, 
\begin{align}
\stateset{\mathrm{E}}_{\hat{\mathsf{U}}}
\equiv \left\{\,\forall |\phi\rangle\,\Big|
~\hat{H}^{\hat{\mathsf{U}}}_{0}|\phi\rangle\,{=}\,E(\phi)|\phi\rangle
,~\langle \phi | \phi' \rangle\,{=}\,\delta_{\phi,\phi'}
\right\},
\label{eeModel32}
\end{align}
the sets of stationary states that can be expressed in the form
\begin{align}
\stateset{\mathrm{C}^{\mu}}_{\hat{\mathsf{U}}}
\equiv \Big\{\,\forall |\phi\rangle\,\Big|\,|\phi\rangle\,{\in}\,\stateset{\mathrm{E}}_{\hat{\mathsf{U}}},
\,\mbox{$|\phi\rangle$ satisfies $\mathrm{C}^{\,\mu}$\,} 
\Big\}
,\quad \mu \in \{1,\ldots,M\}
\label{eeModel33}
\end{align}
are termed as \emph{mutually orthogonalized sets}.
To save symbols, we use the same symbol $\stateset{\mathrm{C}}_{\hat{\mathsf{U}}}$ to represent the subspace of the Hilbert space spanned by set $\stateset{\mathrm{C}}_{\hat{\mathsf{U}}}$.
\end{myAssumptionB}
\noindent
Because of the time-independent unitary transformation $\hat{\mathsf{U}}$ in Eq.\eqref{eeModel22}, the stationary states defined by the above assumptions include a wider range of quantum states than the traditional definition~\cite[Chap.2]{Messiah}.
However, except in Section \ref{cDiscussion}, the rest of this paper will not discuss the case ${\hat{\mathsf{U}}}\,{\neq}\,\hat{1}$ or ${\hat{\mathsf{W}}}\,{\neq}\,\hat{1}$.
Hence, we tentatively use the conventional definition given below:
\begin{align}
\hat{H}_{0} 
& \equiv \hat{H}_{\mathrm{A}} + \hat{H}_{\mathrm{F}},
\label{eeModel35}\\
\hat{H}_{0}|\phi\rangle 
& = E(\phi)|\phi\rangle
,\quad
\langle\phi|\phi'\rangle=\delta_{\phi \phi'},
\label{eeModel36}\\
\stateset{\mathrm{C}}
& \equiv \left\{\,\forall |\phi\rangle\,\Big|
~\hat{H}_{0}|\phi\rangle\,{=}\,E(\phi)|\phi\rangle,
~\langle \phi | \phi' \rangle\,{=}\,\delta_{\phi,\phi'},
~\mbox{$|\phi\rangle$ satisfies $\mathrm{C}$\,} 
\right\}.
\label{eeModel37}
\end{align}

\subsection{Laws of stationary state occupation}
\label{sReality2}
Next, following the discussion in Section \ref{cLogical}, we define Boolean logical variables that describe stationary state occupation.
\begin{myAssumptionB}[Definition of elementary propositions]
An elementary proposition is defined as a proposition that represents the occupation of stationary states at a single instant of time.
\begin{align}
\mbox{`$\occupation{\state{\phi}{}}{},t$'} 
& \equiv \mbox{`A stationary state $|\phi\rangle$ is occupied at time $t$'}
\phantom{\Big|}
\label{eeModel41}\\
\mbox{`$\occupation{\stateset{{\mathrm{C}}}}{},t$'} 
& \equiv \mbox{`The stationary states occupied at time $t$ is contained in}
\nonumber\\
& ~~~~~\mbox{subspace $\stateset{\mathrm{C}}$'}
\label{eeModel42}
\end{align}
\end{myAssumptionB}
\begin{myAssumptionB}[Definition of logical variables]
The logical variable describing an elementary proposition is defined by
\begin{align}
\xjvar{`$\occupation{\state{\phi}{}}{},t$'}{big}
& \equiv \begin{cases}
  1, & \mbox{`$\occupation{\state{\phi}{}}{},t$' is True at time $t$}\\
  0, & \mbox{`$\occupation{\state{\phi}{}}{},t$' is False at time $t$}
\end{cases}
,\quad j \in {{\mathbb{N}}},
\label{eeModel51}\\
\xjvar{`$\occupation{\stateset{{\mathrm{C}}}}{},t$'}{big}
& \equiv \begin{cases}
  1, & \mbox{`$\occupation{\stateset{{\mathrm{C}}}}{},t$' is True at time $t$}\\
  0, & \mbox{`$\occupation{\stateset{{\mathrm{C}}}}{},t$' is False at time $t$}
\end{cases}
,\quad j \in {{\mathbb{N}}},
\label{eeModel52}
\end{align}
where $j$ is the trial number and $\mathbb{N}$ denotes natural numbers.
\end{myAssumptionB}
\begin{myAssumptionB}[Laws of Boolean logic]
Boolean logic operations on elementary propositions are defined by
\begin{align}
\xjvar{$\lnot$`$\occupation{\stateset{{\mathrm{C}}_{1}}}{},t_{1}$'}{big}
& = 1 - \xjvar{`$\occupation{\stateset{{\mathrm{C}}_{1}}}{},t_{1}$'}{big},
\label{eeModel61}\\
\xjvar{`$\occupation{\stateset{{\mathrm{C}}_{1}}}{},t_{1}$'$\wedge$`$\occupation{\stateset{{\mathrm{C}}_{2}}}{},t_{2}$'}{big}
& = \xjvar{`$\occupation{\stateset{{\mathrm{C}}_{1}}}{},t_{1}$'}{big}\,\xjvar{`$\occupation{\stateset{{\mathrm{C}}_{2}}}{},t_{2}$'}{big},
\label{eeModel62}\\
\xjvar{`$\occupation{\stateset{{\mathrm{C}}_{1}}}{},t_{1}$'$\vee$`$\occupation{\stateset{{\mathrm{C}}_{2}}}{},t_{2}$'}{big}
& = \xjvar{`$\occupation{\stateset{{\mathrm{C}}_{1}}}{},t_{1}$'}{big} + \xjvar{`$\occupation{\stateset{{\mathrm{C}}_{2}}}{},t_{2}$'}{big}
\nonumber\\
&~~~~ - \xjvar{`$\occupation{\stateset{{\mathrm{C}}_{1}}}{},t_{1}$'$\wedge$`$\occupation{\stateset{{\mathrm{C}}_{2}}}{},t_{2}$'}{big},
\label{eeModel63}
\end{align}
where the symbols $\lnot$, $\wedge$, and $\vee$ denote negation (NOT), conjunction (AND), and disjunction (OR), respectively.
These relations shall hold for elementary propositions defined at different times ($t_{1}\,{\neq}\,t_{2}$).
\end{myAssumptionB}
\begin{myAssumptionB}[Exclusivity]
For any orthogonal stationary states $|\phi_{1}\rangle$ and $|\phi_{2}\rangle$ that satisfy $\langle\phi_{1}|\phi_{2}\rangle\,{=}\,\delta_{\phi_{1},\phi_{2}}$, 
\begin{align}
& \xjvar{`$\occupation{\state{\phi_{1}}{}}{},t$'$\wedge$`$\occupation{\state{\phi_{2}}{}}{},t$'}{big} = 0
\label{eeModel71}
\end{align}
for $\phi_{1}\,{\neq}\,\phi_{2}$ and 
\begin{align}
& \xjvar{`$\occupation{\state{\phi_{1}}{}}{},t$'$\wedge$`$\occupation{\state{\phi_{2}}{}}{},t$'}{big} 
= \xjvar{`$\occupation{\state{\phi_{1}}{}}{},t$'}{big} 
= \xjvar{`$\occupation{\state{\phi_{2}}{}}{},t$'}{big} 
\label{eeModel72}
\end{align}
for $\phi_{1}\,{=}\,\phi_{2}$.
\end{myAssumptionB}
\begin{myAssumptionB}[Definition of compound propositions]
A compound proposition is defined as the conjunction of elementary propositions `${\mathrm{A}}_{i},t_{i}$' ($1\,{\leqq}\,i\,{\leqq}\,I$) specified at different times:
\begin{align}
\mbox{`$\mathrm{A},\{t\}$'}
\equiv \bigwedge^{I}_{i=1} \mbox{`${\mathrm{A}}_{i},t_{i}$'}
,\quad I \in {\mathbb{N}}.
\label{eeModel81}
\end{align}
Here, $\{t\}$ denotes the set of times $t_{1},\ldots,t_{I}$.
For simplicity, the elementary propositions are assumed to be numbered in ascending order of time ($t_{i}\,{<}\,t_{i+1}$).
By definition, elementary propositions can always be expressed as
\begin{align}
\mbox{`${\mathrm{A}}_{i},t_{i}$'}
= \mbox{`$\occupation{\stateset{{\mathrm{C}}_{i}}}{},t_{i}$'},
\label{eeModel82}
\end{align}
but the sets $\{\!\!\{{\mathrm{C}}_{i}\}\!\!\}$ are not necessarily orthogonalized with each other.\footnote{~Writing explicitly, $\langle\phi_{i'}|\phi_{i''}\rangle$ is always 0 if $|\phi_{i'}\rangle$ and $|\phi''\rangle$ are both in $\stateset{{\mathrm{C}}_{i}}$, but not necessarily if $|\phi_{i'}\rangle$ is in $\stateset{{\mathrm{C}}_{i}}$ while $|\phi_{i''}\rangle$ is not.}
\end{myAssumptionB}
\noindent
It is worth noting that the exclusivity assumption (Assumption B7) is necessary to maintain the consistency of the entire set of assumptions (See {\S}\ref{sRelevant}) and must be distinguished from the Pauli exclusion principle.
Accepting the exclusivity assumption, we have 
\begin{align}
\mbox{`$\occupation{\stateset{{\mathrm{C}}}}{},t$'} 
= \bigvee_{|\phi\rangle}^{\stateset{\mathrm{C}}} \mbox{`$\occupation{\state{\phi}{}}{},t$'} 
\label{eeModel90}
\end{align}
where $\bigvee_{|\phi\rangle}^{\stateset{\mathrm{C}}}$ stands for the conjunction over all stationary states in $\stateset{\mathrm{C}}$.

\subsection{Laws of quantum theory}
\label{sReality3}
Now, the laws of quantum theory must be restated to allow for the calculation of expectation values.
In conventional theories, the probabilities of compound propositions can be calculated only if all sets $\stateset{{\mathrm{C}}_{i}}$ ($1\,{\leqq}\,i\,{\leqq}\,I$) are orthogonalized with each other~\cite{Griffiths2001}.
In contrast, we assume that the probabilities of compound propositions can be calculated as long as the compatibility condition is satisfied, regardless of whether the sets $\stateset{{\mathrm{C}}_{i}}$ are mutually orthogonalized.
Specific assumptions are as follows:
\begin{myAssumptionB}[Definition of expectation values]
The expectation value of the logical variable is defined by
\begin{align}
\big\langle \xvar{`A,$\{t\}$'}{} \big\rangle
\equiv 
\frac{\sum_{j}^{\mbox{\scriptsize `${\mathrm{A}}_{0},t_{0}$'}} \xjvar{`A,$\{t\}$'}{}}
{\sum_{j}^{\mbox{\scriptsize `${\mathrm{A}}_{0},t_{0}$'}} \,1},
\label{eeModel91}
\end{align}
where the sums on the right-hand side shall be taken over the ensemble specified by the proposition `$\mathrm{A}_{0},t_{0}$' and the density operator $\hat{\rho}(\mbox{`${\mathrm{A}}_{0},t_{0}$'})$.
For brevity, we omit the limit symbol $\lim_{N\to\infty}$ because the timing of taking the statistical limit is obvious in this paper.
\end{myAssumptionB}
\begin{myAssumptionB}[Born rule (1)]
The probability of a compound proposition `A,$\{t\}$ can be calculated by the trace formula of quantum theory, namely
\begin{align}
\big\langle \xvar{`A,$\{t\}$'}{} \big\rangle
= {\mathrm{Tr}}\Big[\,\xop{`A,$\{t\}$'}{}\,\hat{\rho}(\mbox{`${\mathrm{A}}_{0},t_{0}$'})\,\Big].
\label{eeModel101}
\end{align}
Here, the density operator $\hat{\rho}(\mbox{`${\mathrm{A}}_{0},t_{0}$'})$ is Hermitian and the characteristic operator $\xop{`A,$\{t\}$'}{}$ can be expressed as 
\begin{align}
\xop{`A,$\{t\}$'}{}
= \zopd{`A,$\{t\}$'}{} \zop{`A,$\{t\}$'}{}
\label{eeModel102}
\end{align}
using a partial characteristic operator $\zop{`A,$\{t\}$'}{}$.
\end{myAssumptionB}
\begin{myAssumptionB}[Born rule (2)]
The partial characteristic operator of an elementary proposition is a projection operator onto a stationary state or a subspace of the Hilbert space spanned by a set of stationary states:
\begin{align}
\zop{`$\occupation{\state{\phi}{}}{}$'}{big}
& = |\phi\rangle\langle\phi|,
\label{eeModel131}\\
\zop{`$\occupation{\stateset{{\mathrm{C}}}}{}$'}{big}
& = \sum_{|\phi\rangle}^{\stateset{\mathrm{C}}} |\phi\rangle\langle\phi|.
\label{eeModel132}
\end{align}
\end{myAssumptionB}
\begin{myAssumptionB}[Conversion rules]
Boolean logic operations transform the characteristic operators of elementary propositions as follows:
\begin{align}
\xop{$\lnot$`$\occupation{\stateset{{\mathrm{C}}_{1}}}{},t_{1}$'}{big}
& = \hat{1} - \xop{`$\occupation{\stateset{{\mathrm{C}}_{1}}}{},t_{1}$'}{big},
\label{eeModel111}\\
\xop{`$\occupation{\stateset{{\mathrm{C}}_{1}}}{},t_{1}$'$\wedge$`$\occupation{\stateset{{\mathrm{C}}_{2}}}{},t_{2}$'}{big}
& = \mathcal{T}::\xop{`$\occupation{\stateset{{\mathrm{C}}_{1}}}{},t_{1}$'}{big}\,\xop{`$\occupation{\stateset{{\mathrm{C}}_{2}}}{},t_{2}$'}{big}::,
\label{eeModel112}\\
\xop{`$\occupation{\stateset{{\mathrm{C}}_{1}}}{},t_{1}$'$\vee$`$\occupation{\stateset{{\mathrm{C}}_{2}}}{},t_{2}$'}{big}
& = \xop{`$\occupation{\stateset{{\mathrm{C}}_{1}}}{},t_{1}$'}{big} + \xop{`$\occupation{\stateset{{\mathrm{C}}_{2}}}{},t_{2}$'}{big}
\nonumber\\
&
~~~ - \xop{`$\occupation{\stateset{{\mathrm{C}}_{1}}}{},t_{1}$'$\wedge$`$\occupation{\stateset{{\mathrm{C}}_{2}}}{},t_{2}$'}{big},
\label{eeModel113}
\end{align}
where the time-ordering and normal-ordering symbols are defined by
\begin{align}
& \mathcal{T}::\xop{`$\occupation{\stateset{{\mathrm{C}}_{1}}}{},t_{1}$'}{big}\,\xop{`$\occupation{\stateset{{\mathrm{C}}_{2}}}{},t_{2}$'}{big}::
\label{eeModel114}\\
& \equiv \begin{cases}
\zopd{`$\occupation{\stateset{{\mathrm{C}}_{2}}}{},t_{2}$'}{big} \zopd{`$\occupation{\stateset{{\mathrm{C}}_{1}}}{},t_{1}$'}{big} \zop{`$\occupation{\stateset{{\mathrm{C}}_{1}}}{},t_{1}$'}{big} \zop{`$\occupation{\stateset{{\mathrm{C}}_{2}}}{},t_{2}$'}{big}, & \!\!\!\!\!\!\!\!\!\!\!\!\!\!\!\! t_{1}>t_{2},\phantom{\Big|}\\
\zopd{`$\occupation{\stateset{{\mathrm{C}}_{1}}}{},t_{1}$'}{big} \zopd{`$\occupation{\stateset{{\mathrm{C}}_{2}}}{},t_{2}$'}{big} \zop{`$\occupation{\stateset{{\mathrm{C}}_{2}}}{},t_{2}$'}{big} \zop{`$\occupation{\stateset{{\mathrm{C}}_{1}}}{},t_{1}$'}{big}, & \!\!\!\!\!\!\!\!\!\!\!\!\!\!\!\! t_{1}<t_{2},\phantom{\Big|}\\
\Big( \zopd{`$\occupation{\stateset{{\mathrm{C}}_{2}}}{},t_{2}$'}{big} \zopd{`$\occupation{\stateset{{\mathrm{C}}_{1}}}{},t_{1}$'}{big} \zop{`$\occupation{\stateset{{\mathrm{C}}_{1}}}{},t_{1}$'}{big} \zop{`$\occupation{\stateset{{\mathrm{C}}_{2}}}{},t_{2}$'}{big} & ~\\
~ + \zopd{`$\occupation{\stateset{{\mathrm{C}}_{1}}}{},t_{1}$'}{big} \zopd{`$\occupation{\stateset{{\mathrm{C}}_{2}}}{},t_{2}$'}{big} \zop{`$\occupation{\stateset{{\mathrm{C}}_{2}}}{},t_{2}$'}{big} \zop{`$\occupation{\stateset{{\mathrm{C}}_{1}}}{},t_{1}$'}{big} \Big)/2, & \\
 & \!\!\!\!\!\!\!\!\!\!\!\!\!\!\!\! t_{1}=t_{2}.
\end{cases}\!\!\!\!\!\!\!\!\!\!\!\!\!\!\!\!\!\!\!\!
\nonumber
\end{align}
\end{myAssumptionB}
\begin{myAssumptionB}[Equation of motion]
The partial characteristic operator of an elementary proposition evolves in time according to
\begin{align}
\zop{`$\occupation{\state{\phi}{}}{},t$'}{big}
& = \hat{U}^{\dagger}(t)\,\zop{`$\occupation{\state{\phi}{}}{}$'}{big}\,\hat{U}(t),
\label{eeModel121}\\
\zop{`$\occupation{\stateset{{\mathrm{C}}}}{},t$'}{big}
& = \hat{U}^{\dagger}(t)\,\zop{`$\occupation{\stateset{{\mathrm{C}}}}{}$'}{big}\,\hat{U}(t),
\label{eeModel122}
\end{align}
where the time evolution operator $\hat{U}(t)$ is defined by
\begin{align}
\hat{U}(t) \equiv \exp\left(\frac{t}{{\rm{i}}\hbar}\hat{H}\right).
\label{eeModel123}
\end{align}
\end{myAssumptionB}
\begin{myAssumptionB}[Comatibility condition]
The elementary propositions `${\mathrm{A}}_{i},t_{i}$' \\($1\,{\leqq}\,{i}\,{\leqq}\,I$), used to define a compound proposition `$\mathrm{A},\{t\}$', shall be chosen to satisfy the compatibility conditions, namely
\begin{align}
& {\rm{Tr}}\Big[\,\hat{K}_{i_{1},\ldots,i_{\mathcal{I}}}(t)\,\hat{\rho}(\mbox{`$\mathrm{A}_{0},t_{0}$'})\,\hat{K}_{i_{1},\ldots,i_{\mathcal{I}}}(t){}^{\dagger}\,\Big]
\nonumber\\
& = {\rm{Tr}}\Big[\,\left({\mathcal{P}}\hat{K}_{i_{1},\ldots,i_{\mathcal{I}}}(t)\right) \hat{\rho}(\mbox{`$\mathrm{A}_{0},t_{0}$'}) \left({\mathcal{P}}\hat{K}_{i_{1},\ldots,i_{\mathcal{I}}}(t)\right)^{\dagger} \Big]
,\quad 2\,{\leqq}\,\mathcal{I}\,{\leqq}\,I,
\label{eeModel141}
\end{align}
where the chain operator $\hat{K}_{i_{1},\ldots,i_{\mathcal{I}}}(t)$ for $i_{1}\,{<}\,{\ldots}\,{<}\,i_{\mathcal{I}}$ is defined as
\begin{align}
\hat{K}_{i_{1},\ldots,i_{\mathcal{I}}}(t) \equiv 
\lim_{t_{i_{1}}\,{\to}\,t} \cdots \lim_{t_{i_{\mathcal{I}}}\,{\to}\,t} 
\zop{`${\mathrm{A}}_{i_{1}},t_{i_{1}}$'}{} \cdots \zop{`${\mathrm{A}}_{i_{\mathcal{I}}},t_{i_{\mathcal{I}}}$'}{}
,\quad 2\,{\leqq}\,\mathcal{I}\,{\leqq}\,I,
\label{eeModel142}
\end{align}
and ${\mathcal{P}}\hat{K}_{i_{1},\ldots,i_{\mathcal{I}}}(t)$ is an operator obtained by rearranging the partial characteristic operators constituting $\hat{K}_{i_{1},\ldots,i_{\mathcal{I}}}(t)$ in an arbitrary order.
\end{myAssumptionB}
\noindent
Since the compatibility condition depends only on a single time variable, it can also be expressed as
\begin{align}
& {\rm{Tr}}\Big[\,\hat{K}_{i_{1},\ldots,i_{\mathcal{I}}}\,\hat{\rho}(\,t\,|\mbox{`$\mathrm{A}_{0},t_{0}$'})\,\hat{K}_{i_{1},\ldots,i_{\mathcal{I}}}{}^{\dagger}\,\Big]
\nonumber\\
& = {\rm{Tr}}\Big[\,\left({\mathcal{P}}\hat{K}_{i_{1},\ldots,i_{\mathcal{I}}}\right)\,\hat{\rho}(\,t\,|\mbox{`$\mathrm{A}_{0},t_{0}$'})\,\left({\mathcal{P}}\hat{K}_{i_{1},\ldots,i_{\mathcal{I}}}\right)^{\dagger} \Big]
,\quad 2\,{\leqq}\,\mathcal{I}\,{\leqq}\,I
\label{eeModel143}
\end{align}
where the density operator in the Schr\"{o}dinger picture and the time-independent chain operator are defined as follows:
\begin{align}
\hat{\rho}(\,t\,|\mbox{`${\mathrm{A}}_{0},t_{0}$'})
& \equiv \hat{U}(t)\,\hat{\rho}(\mbox{`${\mathrm{A}}_{0},t_{0}$'})\,\hat{U}^{\dagger}(t),
\label{eeModel144}\\
\hat{K}_{i_{1},\ldots,i_{\mathcal{I}}} 
& \equiv \zop{`${\mathrm{A}}_{i_{1}}$'}{} \cdots \zop{`${\mathrm{A}}_{i_{\mathcal{I}}}$'}{}.
\label{eeModel145}
\end{align}

\subsection{Role of compatibility conditions}
\label{sReality4}
To elucidate the physical meaning of the aforementioned assumptions, it is beneficial to introduce the following additional assumption:
\setcounter{myAssumptionB}{-1}
\begin{myAssumptionB}[Model of reality]
In the complete description based on the Hamiltonians of the entire system, only one stationary state $|\Omega^{(j)}(t)\rangle$ is occupied at any given moment:
\begin{align}
\xjvar{`$\occupation{\state{\phi}{}}{},t$'}{big}\,{=}\,\delta_{\phi,\Omega^{(j)}(t)}
,\quad
{\forall}j.
\label{eeRole1}
\end{align}
Correspondingly, a logical variable describing the occupation of the set $\stateset{{\mathrm{C}}_{\Omega}}$ containing $|\Omega^{(j)}(t)\rangle$ always has a definite value 1.
\begin{align}
\xjvar{`$\occupation{\stateset{{\mathrm{C}}_{\Omega}}}{},t$'}{big} 
= 1 
\quad \mbox{if~~~$\Omega^{(j)}(t)\,{\in}\,\stateset{{\mathrm{C}}_{\Omega}}$}
,\quad {\forall} j.
\label{eeRole2}
\end{align}
\end{myAssumptionB}
\noindent
When Eq.\eqref{eeRole1} is combined with Assumptions B1--B14, $|\Omega^{(j)}(t)\rangle$ undergoes stochastic evolution over time.
Consequently, our theory exhibits non-deterministic behavior in the sharp representation of Eq.\eqref{eeRole1}.
Conversely, in the non-sharp representation of Eq.\eqref{eeRole2}, deterministic relations between logical variables can emerge when the set $\stateset{{\mathrm{C}}_{\Omega}}$ is sufficiently large to encompass the variation of $|\Omega^{(j)}(t)\rangle$.
This characteristic allows our theory to exhibit deterministic behavior when employing a non-sharp representation.
\par
To highlight the distinction between the compatibility condition and the consistency condition, we can rewrite the compatibility condition as
\begin{align}
& {\rm{Tr}}\Bigg[\,\Bigg( \hat{K}_{i_{1},\ldots,i_{\mathcal{I}}}^{\dagger}\,\hat{K}_{i_{1},\ldots,i_{\mathcal{I}}}
- \left({\mathcal{P}}\hat{K}_{i_{1},\ldots,i_{\mathcal{I}}}\right)^{\dagger}\left({\mathcal{P}}\hat{K}_{i_{1},\ldots,i_{\mathcal{I}}}\right) \bigg)
\,\hat{\rho}(\,t\,|\mbox{`${\mathrm{A}}_{0},t_{0}$'})\,\Bigg] = 0
,~~ 2\,{\leqq}\,\mathcal{I}\,{\leqq}\,I
\label{eeRole3}
\end{align}
and redefine the Type-I and Type-II compatibility by the following conditions:
\begin{align}
\mbox{\bf Type-I:}
& ~~~ \hat{K}_{i_{1},\ldots,i_{\mathcal{I}}}^{\dagger}\,\hat{K}_{i_{1},\ldots,i_{\mathcal{I}}}
= \left({\mathcal{P}}\hat{K}_{i_{1},\ldots,i_{\mathcal{I}}}\right)^{\dagger} \left({\mathcal{P}}\hat{K}_{i_{1},\ldots,i_{\mathcal{I}}}\right)
,~~ 2\,{\leqq}\,\mathcal{I}\,{\leqq}\,I,
\label{eeRole4}\\
\mbox{\bf Type-II:}
& ~~~ \hat{K}_{i_{1},\ldots,i_{\mathcal{I}}}^{\dagger}\,\hat{K}_{i_{1},\ldots,i_{\mathcal{I}}}
\neq \left({\mathcal{P}}\hat{K}_{i_{1},\ldots,i_{\mathcal{I}}}\right)^{\dagger} \left({\mathcal{P}}\hat{K}_{i_{1},\ldots,i_{\mathcal{I}}} \right)
,~~ 2\,{\leqq}\,\mathcal{I}\,{\leqq}\,I.
\label{eeRole5}
\end{align}
Given that ${\mathcal{P}}\hat{K}_{i_{1},\ldots,i_{\mathcal{I}}}$ represents an arbitrary permutation of $\hat{K}_{i_{1},\ldots,i_{\mathcal{I}}}$, it becomes evident that any theory demanding commutativity of partial characteristic operators permits only Type-I compatibility.
In particular, the Consistent Histories approach states that any meaningful History must satisfy the consistency condition:
\begin{align}
{\rm{Tr}}\left[\,\hat{K}({\mathrm{A}}_{1},t_{1};\ldots;{\mathrm{A}}_{I},t_{I})\,\hat{\rho}(\mbox{`${\mathrm{A}}_{0},t_{0}$'})\,\hat{K}({\mathrm{A}}'_{1},t_{1};\ldots;{\mathrm{A}}'_{I},t_{I})^{\dagger}\,\right]
= 0.
\label{eeRole6}
\end{align}
Practically, this condition is satisfied only when projection operators are commutative with each other
~\cite[Chap.10]{Griffiths2001}.
Thus, the Consistent Histories approach and the von Neumann-L\"{u}ders theory only consider the Type-I compatibility.
On the other hand, assuming a compatibility condition instead of a consistency condition would allow the Type-II compatibility to be considered, albeit at the cost of permitting a situation-dependent definition of joint probabilities.
In this sense, the compatibility condition (Eq.\eqref{eeRole3}) is a generalization of the consistency condition of the Consistent Histories approach.
\par
It is crucial to acknowledge the distinct ways in which compatibility conditions are satisfied in sharp and non-sharp representations.
In the sharp representation of Eq.\eqref{eeRole1}, the second-order compatibility condition between  elementary proposition takes the following form:
\begin{align}
0 & = {\rm{Tr}}\left[\,\Big( \zopd{`$\occupation{\state{\phi_{1}}{}}{}$'}{big} \zopd{`$\occupation{\state{\phi_{2}}{}}{}$'}{big} \zop{`$\occupation{\state{\phi_{2}}{}}{}$'}{big} \zop{`$\occupation{\state{\phi_{1}}{}}{}$'}{big} \right.
\nonumber\\
& ~~~~~~~~~ - \left. \zopd{`$\occupation{\state{\phi_{2}}{}}{}$'}{big} \zopd{`$\occupation{\state{\phi_{1}}{}}{}$'}{big} \zop{`$\occupation{\state{\phi_{1}}{}}{}$'}{big} \zop{`$\occupation{\state{\phi_{2}}{}}{}$'}{big} \Big)\,\hat{\rho}(\,t\,|\mbox{`${\mathrm{A}}_{0},t_{0}$'})\,\right]
\nonumber\\
& = \left|\langle \phi_{1} | \phi_{2} \rangle\right|^{2} 
\Big( \langle \phi_{1} |\,\hat{\rho}(\,t\,|\mbox{`${\mathrm{A}}_{0},t_{0}$'})\,| \phi_{1} \rangle - \langle \phi_{2} |\,\hat{\rho}(\,t\,|\mbox{`${\mathrm{A}}_{0},t_{0}$'})\,| \phi_{2} \rangle \Big).
\label{eeRole7}
\end{align}
This condition is satisfied only if the stationary states $|\phi_{1}\rangle$ and $|\phi_{2}\rangle$ are orthogonal with each other (apart from coincidence).
Thus, the Type-I compatibility is sufficient when the representation is sharp.
On the other hand, in the non-sharp representation of Eq.\eqref{eeRole2}, the second-order compatibility condition has the following form:
\begin{align}
0 & = {\rm{Tr}}\left[\,\Big( \zopd{`$\occupation{\stateset{{\mathrm{C}}_{1}}}{}$'}{big} \zopd{`$\occupation{\stateset{{\mathrm{C}}_{2}}}{}$'}{big} \zop{`$\occupation{\stateset{{\mathrm{C}}_{2}}}{}$'}{big} \zop{`$\occupation{\stateset{{\mathrm{C}}_{1}}}{}$'}{big} \right.
\nonumber\\
& ~~~~~~~~~ - \left. \zopd{`$\occupation{\stateset{{\mathrm{C}}_{2}}}{}$'}{big} \zopd{`$\occupation{\stateset{{\mathrm{C}}_{1}}}{}$'}{big} \zop{`$\occupation{\stateset{{\mathrm{C}}_{1}}}{}$'}{big} \zop{`$\occupation{\stateset{{\mathrm{C}}_{2}}}{}$'}{big} \Big)\,\hat{\rho}(\,t\,|\mbox{`${\mathrm{A}}_{0},t_{0}$'})\,\right]
\nonumber\\
& = \sum^{\stateset{{\mathrm{C}}_{1}}}_{|\phi_{1}\rangle} \sum^{\stateset{{\mathrm{C}}_{2}}}_{|\phi_{2}\rangle}
\left|\langle \phi_{1} | \phi_{2} \rangle\right|^{2} 
\Big( \langle \phi_{1} |\,\hat{\rho}(\,t\,|\mbox{`${\mathrm{A}}_{0},t_{0}$'})\,| \phi_{1} \rangle - \langle \phi_{2} |\,\hat{\rho}(\,t\,|\mbox{`${\mathrm{A}}_{0},t_{0}$'})\,| \phi_{2} \rangle \Big).
\label{eeRole8}
\end{align}
At first glance, it is not obvious that this expression can be satisfied for reasons other than the commutativity of partial characteristic operators~\cite{Griffiths2001}.
Nevertheless, we can show that the compatibility condition always holds among elementary propositions satisfying Eq.\eqref{eeRole2} (i.e., $\xjvar{`$\occupation{\stateset{{\mathrm{C}}_{\Omega}}}{},t$'}{}\,{=}\,1$).
The proof is simple and is shown in {\S}\ref{sAppendix100}.
In this respect, the compatibility condition is a condition that determines the validity of approximate representations of reality.

\section{Key features of our model}
\label{cMeasurement}
Next, we summarize the key properties derived from the above assumptions.
First, we illustrate how to describe degenerate systems and composite systems, demonstrating that deterministic relations are established between entangled subsystems.
Second, we define the instantaneous values of physical quantities in individual trials, showing that the compatibility condition limits the possibility of arithmetic operations between instantaneous values.
Third, we formulate a classical approximation for the instantaneous values of physical quantities, deriving the complementarity principle and classical equations of motion. 
Fourth, by focusing on entangled spin systems, we show how our theory avoids the contradiction caused by the Bell and CHSH inequalities.

\subsection{Deterministic relations between entangled subsystems}
\subsubsection{Representation of degenerate system}
\label{sDegeneracy}
Let us first look at a simple example to see how the compatibility condition adequately captures the constraints imposed on degenerate systems.
For this purpose, consider a spin-1/2 system with zero magnetic field and let $|{\uparrow}_{\xi}\rangle$ and $|{\downarrow}_{\xi}\rangle$ be two stationary states with $\xi$ (${\in}\{x,y,z\}$) as the measurement direction.
In this example, the stationary states cannot be uniquely determined due to degeneracy.
However, if we define elementary propositions by
\begin{align}
\mbox{`$\occupation{\state{{\uparrow}_{\xi}}{}}{}$'} 
& \equiv \mbox{`$|{\uparrow}_{\xi}\rangle$ is occupied'}
,\quad \xi \in \{x,y,z\}
\label{eeSimpleEx2}\\
\mbox{`$\occupation{\state{{\downarrow}_{\xi}}{}}{}$'} 
& \equiv \mbox{`$|{\downarrow}_{\xi}\rangle$ is occupied'}
,\quad \xi \in \{x,y,z\}
\label{eeSimpleEx3}
\end{align}
and use a concrete form of partial characteristic operators, namely
\begin{align}
& \zop{`$\occupation{\state{{\uparrow}_{\xi}}{}}{}$'}{} = | {\uparrow}_{\xi} \rangle\langle {\uparrow}_{\xi} |
,\quad \zop{`$\occupation{\state{{\downarrow}_{\xi}}{}}{}$'}{} = | {\downarrow}_{\xi} \rangle\langle {\downarrow}_{\xi} |
,\quad \xi \in \{x,y,z\},
\label{eeSimpleEx4}
\end{align}
we can determine whether or not the given set of propositions satisfies the compatibility condition.
It is easy to see that the compatibility conditions hold among\footnote{~The compatibility condition always holds between `$\occupation{\stateset{\mathsf{E}}}{}$' and any proposition if $\zop{`$\occupation{\stateset{\mathsf{E}}}{}$'}{}\,{=}\,\hat{1}$. This is why Eq.\eqref{eeSimpleEx6} and Eq.\eqref{eeSimpleEx7} contain `$\occupation{\state{{\uparrow}_{x}}{}}{}$'${\vee}$`$\occupation{\state{{\downarrow}_{x}}{}}{}$' and `$\occupation{\state{{\uparrow}_{y}}{}}{}$'${\vee}$`$\occupation{\state{{\downarrow}_{y}}{}}{}$'. Note also that $x^{(j)}(\mbox{`$\occupation{\state{{\uparrow}_{x}}{}}{}$'}{\vee}\mbox{`$\occupation{\state{{\downarrow}_{x}}{}}{}$'})\,{=}\,1$ and $x^{(j)}(\mbox{`$\occupation{\state{{\uparrow}_{y}}{}}{}$'}{\vee}\mbox{`$\occupation{\state{{\downarrow}_{y}}{}}{}$'})\,{=}\,1$.}
\begin{align}
& \big\{\,
  \mbox{`$\occupation{\state{{\uparrow}_{x}}{}}{}$'},\,
  \mbox{`$\occupation{\state{{\downarrow}_{x}}{}}{}$'},\, 
  \mbox{`$\occupation{\state{{\uparrow}_{x}}{}}{}$'${\vee}$`$\occupation{\state{{\downarrow}_{x}}{}}{}$'},\,
  \mbox{`$\occupation{\state{{\uparrow}_{y}}{}}{}$'${\vee}$`$\occupation{\state{{\downarrow}_{y}}{}}{}$'}
\,\big\}
\label{eeSimpleEx6}
\end{align}
and among
\begin{align}
& \big\{\,
  \mbox{`$\occupation{\state{{\uparrow}_{y}}{}}{}$'},\,
  \mbox{`$\occupation{\state{{\downarrow}_{y}}{}}{}$'},\,
  \mbox{`$\occupation{\state{{\uparrow}_{x}}{}}{}$'${\vee}$`$\occupation{\state{{\downarrow}_{x}}{}}{}$'},\,
  \mbox{`$\occupation{\state{{\uparrow}_{y}}{}}{}$'${\vee}$`$\occupation{\state{{\downarrow}_{y}}{}}{}$'}
\,\big\},
\label{eeSimpleEx7}
\end{align}
but not for the set that involves propositions with different measurement directions, e.g.,
\begin{align}
\big\{\,
  \mbox{`$\occupation{\state{{\uparrow}_{x}}{}}{}$'}, 
  \mbox{`$\occupation{\state{{\uparrow}_{y}}{}}{}$'}
\,\big\}
\quad\mbox{and}\quad
\big\{\,
  \mbox{`$\occupation{\state{{\uparrow}_{x}}{}}{}$'}, 
  \mbox{`$\occupation{\state{{\downarrow}_{y}}{}}{}$'}
\,\big\}.
\label{eeSimpleEx8}
\end{align}
Therefore, there is no conjunction and no joint probability for propositions describing different measurement directions.
This suggests that the instantaneous value of a particular component of the spin angular momentum must be defined using only propositions with the same measurement direction.
From a comparison with the quantum theoretical formulas for their expectation values, the only possibility is
\begin{align}
S^{(j)}_{\xi}(t) 
& = \frac{\hbar}{2} x^{(j)}\Big(\mbox{`o(${\uparrow}_{\xi}$),\,$t$'}\Big) 
- \frac{\hbar}{2} x^{(j)}\Big(\mbox{`o(${\downarrow}_{\xi}$),\,$t$'}\Big)
,\quad \xi \in \{x,y,z\}.
\label{eeSimpleEx11}
\end{align}
Similarly, simultaneous determination of the instantaneous values of $S^{(j)}_{x}(t)$ and $S^{(j)}_{y}(t)$ is impossible due to the lack of a conjunction of underlying propositions.
This is exactly what is expected for a spin-1/2 system.
\par
Incidentally, if $\stateset{{E}_{0}}$ is the set of all stationary states with a given energy ${E}_{0}$ and if $|{E}_{1},\chi\rangle$ is another stationary state with eigenenergy ${E}_{1}$ and a degeneracy index $\chi$, then the compatibility condition always holds between `$\occupation{\stateset{{E}_{0}}}{}$' and `$\occupation{\state{{E}_{1},\chi}{}}{}$'.
For example, the second-order compatibility condition takes the following form:
\begin{align}
& {\rm{Tr}}\left[\,\zopd{`$\occupation{\stateset{{E}_{0}}}{}$'}{} \zopd{`$\occupation{\state{{E}_{1},\chi}{}}{}$'}{} \zop{`$\occupation{\state{{E}_{1},\chi}{}}{}$'}{} \zop{`$\occupation{\stateset{{E}_{0}}}{}$'}{}\,\hat{\rho}(\,t\,|\mbox{`${\mathrm{A}}_{0},t_{0}$'})\,\right]
\nonumber\\
& = {\rm{Tr}}\left[\,\zopd{`$\occupation{\state{{E}_{1},\chi}{}}{}$'}{} \zopd{`$\occupation{\stateset{{E}_{0}}}{}$'}{} \zop{`$\occupation{\stateset{{E}_{0}}}{}$'}{} \zop{`$\occupation{\state{{E}_{1},\chi}{}}{}$'}{}\,\hat{\rho}(\,t\,|\mbox{`${\mathrm{A}}_{0},t_{0}$'})\,\right].
\label{eeSimpleEx0}
\end{align}
The proof is self-explanatory due to the orthogonality of stationary states (Assumption B2).
This result is useful when calculating the light detection probability (cf.\,Supplementary Material B and Ref.~\cite{Theory2}).

\subsubsection{Representation of non-interacting subsystems}
\label{sSubsystem}
Next, we define the logical variables of subsystems for a system consisting of non-interacting subsystems.
To do this, suppose that the non-interacting Hamiltonian can be approximated by the effective expression
\begin{align}
\hat{H}_{0} = \sum_{l=1}^{L} \hat{H}_{0}^{(l)},
\label{eeSubSystems1}
\end{align}
and that the stationary state $|\phi^{(l)}\rangle^{\!l}$ of the $l$-th subsystem is defined by
\begin{align}
\hat{H}_{0}^{(l)}|\phi^{(l)}\rangle^{\!l} = {E}^{(l)}(\phi^{(l)})\,|\phi^{(l)}\rangle^{\!l}
,\quad {}^{l}\!\langle{\phi}^{(l)}|\phi'^{(l)}\rangle^{\!l} = \delta_{{\phi}^{(l)},{\phi'}^{(l)}}
,\quad
1 \leqq l \leqq L.
\label{eeSubSystems2}
\end{align}
Here, $\hat{H}_{0}^{(l)}$ is the Hamiltoian of the $l$-th subsystem, and $\phi^{(l)}$ denotes the quantum number of the $l$-th subsystem.
Then, the stationary state $|\phi\rangle$ of the entire system can be expressed as a tensor product, namely
\begin{align}
|\phi\rangle 
= |\phi^{(1)}\ldots\phi^{(L)}\rangle
\equiv |\phi^{(1)}\rangle^{\!1}{\otimes}\cdots{\otimes}|\phi^{(L)}\rangle^{\!L},
\label{eeSubSystems3}
\end{align}
and from Eq.\eqref{eeSubSystems3} and the Born rule (Assumption B11), we obtain
\begin{align}
\zop{`$\occupation{\state{\phi}{}}{}$'}{big}
& = \prod^{L}_{l=1} \zop{$\mbox{`$\occupation{\state{\phi^{(l)}}{l}}{}$'}$}{big}
,\quad
\zop{$\mbox{`$\occupation{\state{\phi^{(l)}}{l}}{}$'}$}{big}
= | \phi^{(l)} \rangle^{\!l}{}^{l}\!\langle \phi^{(l)} |,
\label{eeSubSystems5}\\
\xop{`$\occupation{\state{\phi}{}}{}$'}{big}
& = \prod^{L}_{l=1} \xop{$\mbox{`$\occupation{\state{\phi^{(l)}}{l}}{}$'}$}{big}
,\quad
\xop{$\mbox{`$\occupation{\state{\phi^{(l)}}{l}}{}$'}$}{big} 
= | \phi^{(l)} \rangle^{\!l}{}^{l}\!\langle \phi^{(l)} |.
\label{eeSubSystems6}
\end{align}
Here, `$\occupation{\state{\phi^{(l)}}{l}}{}$' represents the proposition `A stationary state $|\phi^{(l)}\rangle^{\!l}$ of the $l$-th subsystem is occupied'.
It then follows from the commutativity of $\zop{`$\occupation{\state{\phi^{(l)}}{l}}{}$'}{}$ that compatibility conditions are satisfied between propositions describing different subsystems.
Accordingly, Eqs.\eqref{eeSubSystems5}--\eqref{eeSubSystems6} can be rewritten using the conjunction of $L$ propositioins as
\begin{align}
\zop{`$\occupation{\state{\phi}{}}{}$'}{big}
= \zop{$\displaystyle \bigwedge^{L}_{l=1} \mbox{`$\occupation{\state{\phi^{(l)}}{l}}{}$'}$}{leftright}
,\quad
\xop{`$\occupation{\state{\phi}{}}{}$'}{big}
= \xop{$\displaystyle \bigwedge^{L}_{l=1} \mbox{`$\occupation{\state{\phi^{(l)}}{l}}{}$'}$}{leftright}.
\label{eeSubSystems6a}
\end{align}
Then, the simple lemma presented in Supplementary Material A guarantees that for non-interacting subsystems, the logical variable of the entire system can always be factored into the logical variables of the subsystems as follows:
\begin{align}
\xjvar{`$\occupation{\state{\phi}{}}{},t$'}{big}
= \xjvar{$\displaystyle \bigwedge^{L}_{l=1} \mbox{`$\occupation{\state{\phi^{(l)}}{l}}{},t$'}$}{leftright}
= \prod^{L}_{l=1} \xjvar{$\mbox{`$\occupation{\state{\phi^{(l)}}{l}}{},t$'}$}{big}.
\label{eeSubSystems8}
\end{align}
On the other hand, the expectation value $\langle \xvar{`$\occupation{\state{\phi}{}}{},t$'}{} \rangle$ of the entire system is not necessarily factorized into the expectation value $\langle \xvar{$\mbox{`$\occupation{\state{\phi^{(l)}}{l}}{},t$'}$}{} \rangle$ of the subsystems.
This results from quantum entanglement, which prevents propositions of different subsystems from becoming independent ({\S}\ref{sJoint}) due to interaction at an earlier time.

\subsubsection{Deterministic relations in entangled subsystems}
\label{sEntanglement}
To see this concretely, let us see how entanglement is described using the logic variables of subsystems.
For simplicity, consider a system consisting of two subsystems and assume that the interaction is negligible for $t\,{\geqq}\,t_{0}$, but not necessarily for $t\,{<}\,t_{0}$.
Additionally, we assume that for $t\,{\geqq}\,t_{0}$, each subsystem has degenerate stationary states $\{|\phi^{(1)}_{\chi_{1}}\rangle\,|\,\chi_{1}\,{\in}\,{\mathsf{K}}_{1}\}$ and $\{|\phi^{(2)}_{\chi_{2}}\rangle\,|\,\chi_{2}\,{\in}\,{\mathsf{K}}_{2}\}$, with $\chi_{1}$ and $\chi_{2}$ as degenerate indices, and that the system is entangled at least at $t\,{=}\,t_{0}$.
In this case, the stationary state of the entire system is generally of the form
\begin{align}
|\phi\rangle 
= \sum_{\chi_{1}} \sum_{\chi_{2}} c_{\chi_{1},\chi_{2}} |\phi^{(1)}_{\chi_{1}}\rangle^{\!1}{\otimes}|\phi^{(2)}_{\chi_{2}}\rangle^{\!2}
,\quad
\sum_{\chi_{1}} \sum_{\chi_{2}} |c_{\chi_{1},\chi_{2}}|^{2} = 1,
\label{eeEntangled0}
\end{align}
but by the assumption of entanglement, the density operator of the entire system has the following form:
\begin{align}
& \hat{\rho}(\,t\,|\,\mbox{`$\mathrm{entangled},t_{0}$'})
= \big|\phi_{\mathrm{entangled}}\big\rangle\big\langle\phi_{\mathrm{entangled}}\big| 
,\quad t > t_{0},
\label{eeEntangled2}
\end{align}
where $|\phi_{\mathrm{entangled}}\rangle$ is a stationary state defined by
\begin{align}
|\phi_{\mathrm{entangled}}\rangle
\equiv \sum_{\chi} c_{\chi} |\phi^{(1)}_{\chi}\rangle^{\!1}{\otimes}|\phi^{(2)}_{\chi}\rangle^{\!2}
,\quad
\sum_{\chi} |c_{\chi}|^{2} = 1.
\label{eeEntangled1}
\end{align}
\par
Under these assumptions, the logical variables of subsystems can be defined consistently for $t\,{\geqq}t_{0}$ because there is no interaction.
Using the specific form of partial characteristic operators, namely
\begin{align}
& \zop{$\mbox{`$\occupation{\state{\phi^{(1)}_{\chi_{1}}}{1}}{}$'}$}{big} 
= |\phi^{(1)}_{\chi_{1}}\rangle^{\!1}{}^{1}\!\langle\phi^{(1)}_{\chi_{1}}|
,\quad
\zop{$\mbox{`$\occupation{\state{\phi^{(2)}_{\chi_{2}}}{2}}{}$'}$}{big} 
= |\phi^{(2)}_{\chi_{2}}\rangle^{\!2}{}^{2}\!\langle\phi^{(2)}_{\chi_{2}}|,
\label{eeEntangled3}
\end{align}
we can show that the compatibility condition is satisfied between propositions `$\occupation{\state{\phi^{(1)}_{\chi_{1}}}{1}}{}$' and `$\occupation{\state{\phi^{(2)}_{\chi_{2}}}{2}}{}$'.
Furthermore, the joint probability of `$\occupation{\state{\phi^{(1)}_{\chi_{1}}}{1}}{}$' and `$\occupation{\state{\phi^{(2)}_{\chi_{2}}}{2}}{}$' can be calculated using Eq.\eqref{eeEntangled2} and Eq.\eqref{eeEntangled3} as
\begin{align}
& \Big\langle x\Big({\mbox{`$\occupation{\state{\phi^{(1)}_{\chi_{1}}}{1}}{},t$'}}{\wedge}{\mbox{`$\occupation{\state{\phi^{(2)}_{\chi_{2}}}{2}}{},t$'}}\Big) \Big\rangle
\nonumber\\
& = \mathrm{Tr}\Big[\,::\hat{x}\Big({\mbox{`$\occupation{\state{\phi^{(1)}_{\chi_{1}}}{1}}{}$'}}\Big)\,\hat{x}\Big({\mbox{`$\occupation{\state{\phi^{(2)}_{\chi_{2}}}{2}}{}$'}}\Big)::\,\hat{\rho}(\,t\,|\mbox{`${\mathrm{A}}_{0},t_{0}$'})\,\Big]
\label{eeEntangled4}\\
& = {}^{1}\!\langle\phi^{(1)}_{\chi_{1}}| {}^{2}\!\langle\phi^{(2)}_{\chi_{2}}| 
\Bigg(\sum_{\chi} c_{\chi} |\phi^{(1)}_{\chi}\rangle_{1}{\otimes}|\phi^{(2)}_{\chi}\rangle_{2}\Bigg)
\Bigg(\sum_{\chi'} c^{*}_{\chi'} {}^{1}\!\langle\phi^{(1)}_{\chi'}|{\otimes}{}^{2}\!\langle\phi^{(2)}_{\chi'}|\Bigg)
|\phi^{(1)}_{\chi_{1}}\rangle^{\!1}|\phi^{(2)}_{\chi_{2}}\rangle^{\!2}
\label{eeEntangled5}\\
& = \big| c_{\chi} \big|^{2} \delta_{\chi,\chi_{1}} \delta_{\chi,\chi_{2}}.
\label{eeEntangled6}
\end{align}
The merginal probabilities can similarly be computed as
\begin{align}
\Big\langle x\Big({\mbox{`$\occupation{\state{\phi^{(l)}_{\chi_{l}}}{l}}{},t$'}}\Big) \Big\rangle
= \mathrm{Tr}\Big[\,\hat{x}\Big({\mbox{`$\occupation{\state{\phi^{(l)}_{\chi_{l}}}{l}}{}$'}}\Big)\,\hat{\rho}(\,t\,|\mbox{`${\mathrm{A}}_{0},t_{0}$'})\,\Big]
& = \big| c_{\chi_{l}} \big|^{2}
,\quad
l \in \{1, 2\}.
\label{eeEntangled8}
\end{align}
Combining Eq.\eqref{eeEntangled6} and Eq.\eqref{eeEntangled8}, we obtain
\begin{align}
\Big\langle x\Big({\mbox{`$\occupation{\state{\phi^{(1)}_{\chi_{1}}}{1}}{},t$'}}{\wedge}{\mbox{`$\occupation{\state{\phi^{(2)}_{\chi_{2}}}{2}}{},t$'}_{2}}\Big) \Big\rangle
& = \Big\langle x\Big({\mbox{`$\occupation{\state{\phi^{(1)}_{\chi_{1}}}{1}}{},t$'}}\Big) \Big\rangle \delta_{\chi_{1},\chi_{2}}
\nonumber\\
& = \Big\langle x\Big({\mbox{`$\occupation{\state{\phi^{(2)}_{\chi_{2}}}{2}}{},t$'}_{2}}\Big) \Big\rangle \delta_{\chi_{1},\chi_{2}},
\label{eeEntangled9}
\end{align}
so that for $\chi_{1}\,{=}\,\chi_{2}$,
\begin{align}
\xjvar{`$\occupation{\state{\phi^{(1)}_{\chi}}{1}}{}$'}{leftright}
= \xjvar{`$\occupation{\state{\phi^{(2)}_{\chi}}{2}}{}$'}{leftright} 
,\quad \forall \chi
,\quad \forall j.
\label{eeEntangled10}
\end{align}
\par
In the standard formulation of quantum theory, deterministic relations between entangled subsystems have conventionally been explained based on the collapse of density operators or state vectors~\cite{Cat}.
By contrast, the above calculation shows that the collapse of state vectors is unnecessary to establish deterministic relations in entangled systems.
This conclusion was reached because our theory distinguished between the world of reality and the world of expectation, decoupling the time evolution of the density operator from the occurrence of individual events.
The emergence of the deterministic relations is due to the specific form of the density operator in Eq.\eqref{eeEntangled2}.
If the interaction is non-negligible and the stationary states of subsystems are not well-defined, a more accurate description based on the stationary state of the entire system is required.

\subsection{Instantaneous values of physical quantities}
\label{sMeasurement1}
\subsubsection{Basic definitions}
\label{sInstantaneous0}
Our model of reality (Assumption B0) requires that instantaneous values of physical quantities in individual trials be defined based on the occupation of stationary states. 
In other words, for a physical quantity $F$ to have a definite value in each stationary state, the corresponding operator $\hat{F}$ must commute with the non-interacting Hamiltonian $\hat{H}_{0}$ (Eq.\eqref{eeModel35}). 
Therefore, we begin by considering physical quantities $F$ where the operator $\hat{F}$ satisfies $[\hat{F},\hat{H}_{0}]_{-}\,{=}\,0$, and demonstrate that its instantaneous value can be defined rationally in both statistical and quantum-theoretical sense.
In this case, we can choose stationary state $|\phi\rangle$ as a simultaneous eigenstate of $\hat{F}$ and $\hat{H}_{0}$, and express the spectral decomposition~\cite{Breuer} of $\hat{F}$ and $\hat{H}_{0}$ as
\begin{align}
\hat{F} = \sum_{|\phi\rangle} {F}(\phi)\,|\phi\rangle\langle\phi|
,\quad
\hat{H}_{0} = \sum_{|\phi\rangle} E(\phi)\,|\phi\rangle\langle\phi|
,\quad
\hat{1} = \sum_{|\phi\rangle} |\phi\rangle\langle\phi|.
\label{eeInstantaneous1}
\end{align}
Here, $F(\phi)\,{=}\,\langle\phi|\hat{F}|\phi\rangle$ and $E(\phi)\,{=}\,\langle\phi|\hat{H}_{0}|\phi\rangle$ due to the orthogonality of the stationary states (Assumption B2).
The range notation is omitted for the sums over the complete set of stationary states.
If we define the instantaneous value of $F$ as
\begin{align}
F^{(j)}(t) \equiv \sum_{|\phi\rangle} F(\phi)\,\xjvar{`$\occupation{\state{\phi}{}}{},t$'}{big}
,\quad F(\phi)\,{\equiv}\,\langle\phi|\hat{F}|\phi\rangle
,\quad \forall j,
\label{eeInstantaneous2}
\end{align}
its expectation value satisfies
\begin{align}
\langle F(t) \rangle 
= \sum_{|\phi\rangle} F(\phi)\,\big\langle \xvar{`$\occupation{\state{\phi}{}}{},t$'}{big} \big\rangle,
\label{eeInstantaneous2a}
\end{align}
while Assumptions B1--B14 yield
\begin{align}
\big\langle \xvar{`$\occupation{\state{\phi}{}}{},t$'}{big} \big\rangle
& \equiv \frac{\sum_{j}^{\mbox{\scriptsize `${\mathrm{A}}_{0},t_{0}$'}} \xjvar{`$\occupation{\state{\phi}{}}{},t$'}{big}}{\sum_{j}^{\mbox{`\scriptsize ${\mathrm{A}}_{0},t_{0}$'}} 1}
= \mathrm{Tr}\left[\,\xop{`$\occupation{\state{\phi}{}}{},t$'}{big} \hat{\rho}(\,t\,|\mbox{`${\mathrm{A}}_{0},t_{0}$'})\,\right],
\label{eeInstantaneous3}\\
\xop{`$\occupation{\state{\phi}{}}{},t$'}{big} 
& = |\phi\rangle\langle\phi|.
\label{eeInstantaneous3a}
\end{align}
Substituting Eqs.\eqref{eeInstantaneous3}--\eqref{eeInstantaneous3a} into Eq.\eqref{eeInstantaneous2a} and using the first equation of Eq.\eqref{eeInstantaneous1}, we find that the expectation value of $F$, defined in the statistical sense, can be correctly calculated using the trace formula of quantum theory:
\begin{align}
\langle F(t) \rangle 
& \equiv \frac{\sum_{j}^{\mbox{\scriptsize `${\mathrm{A}}_{0},t_{0}$'}} F^{(j)}(t)}{\sum_{j}^{\mbox{`\scriptsize ${\mathrm{A}}_{0},t_{0}$'}} 1}
= \mathrm{Tr}\left[\,\hat{F}\hat{\rho}(\,t\,|\mbox{`${\mathrm{A}}_{0},t_{0}$'})\,\right].
\label{eeInstantaneous4}
\end{align}
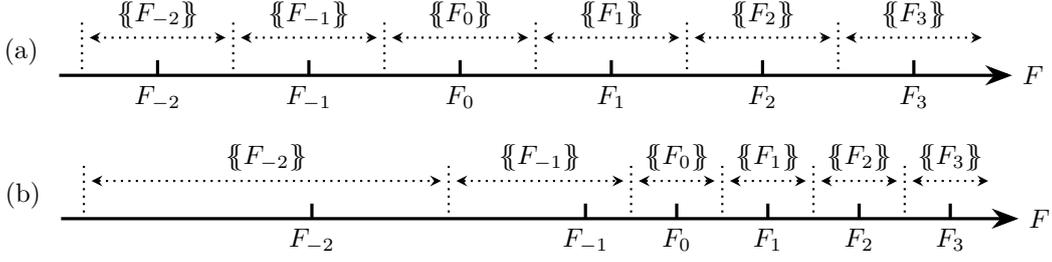
\begin{figure}[bt]
\centering
\begin{tikzpicture}
  \tikzmath{
    \UD   = 1.99; 
    \UNR  = 3;    
    \UNL  = 2;    
    \UNLL = 1;    
    \LDL  = 1.2;  
    \LDR  = 1.2;  
    \LNR  = 3;    
    \LNL  = 2;    
    \LNLL = 1;
    \X = -2.85;
    \Y = 1.9;
  }
  \draw(-\UD*\UNL-\UD/2-0.8 +\X, 0+\Y )node[above]{\small (a)};
  \draw [arrows={-Stealth[scale=1.2]},very thick] (-\UD*\UNL-\UD/2-0.3 +\X, 0+\Y )--(\UD*\UNR+\UD/2+0.3 +\X, 0+\Y );
  \draw(\UD*\UNR+\UD/2+0.3 +\X, 0+\Y )node[right]{\small ${F}$};
  \foreach \k in {-\UNL,-\UNLL,...,\UNR}{
    \draw[very thick] (\k*\UD +\X,0.2 +\Y )--($(0 +\X,0+\Y)!(\k*\UD +\X,0.2+\Y)!(1 +\X,0+\Y)$);
    \draw[thick,dotted](\UD*\k-\UD*0.5 +\X, 0.7+\Y )--(\UD*\k-\UD*0.5 +\X, 0+\Y );
    \draw(\UD*\k +\X, 0+\Y )node[below]{\small ${F}_{\k}$};
    \draw [<->,>=stealth,thick,dotted] (\UD*\k-\UD*0.5+0.1 +\X, 0.5+\Y )--(\UD*\k+\UD*0.5-0.1 +\X, 0.5+\Y );
    \draw(\UD*\k +\X, 0.5+\Y )node[above]{\small $\{\!\!\{{F}_{\k}\}\!\!\}$};
  }
  \draw(-\LDL*\LNL*\LNL-\LDL*\LNL-\LDL/2-0.8,0)node[above]{\small (b)};
  \draw [arrows={-Stealth[scale=1.2]},very thick] (-\LDL*\LNL*\LNL-\LDL*\LNL-\LDL/2-0.3,0)--(\LDR*\LNR+\LDR*0.5+0.3,0);
  \draw(\LDR*\LNR+\LDR*0.5+0.3, 0 )node[right]{\small ${F}$};
  \foreach \k in {-\LNL,-\LNLL,...,-1}{
    \draw[very thick] (-\k*\k*\LDL,0.2)--($(0,0)!(-\k*\k*\LDL,0.2)!(1,0)$);
    \draw[thick,dotted](-\LDL*\k*\k+\LDL*\k-\LDL/2,0.7)--(-\LDL*\k*\k+\LDL*\k-\LDL/2,0);
    \draw(-\LDL*\k*\k,0)node[below]{\small ${F}_{\k}$};
    \draw [<->,>=stealth,thick,dotted] (-\LDL*\k*\k+\LDL*\k-\LDL/2+0.1,0.5)--(-\LDL*\k*\k-\LDL*\k-\LDL/2-0.1,0.5);
    \draw(-\LDL*\k*\k-\LDL/2,0.5)node[above]{\small $\{\!\!\{{F}_{\k}\}\!\!\}$};
  }
  \foreach \k in {0,1,...,\LNR}{
    \draw[very thick] (\k*\LDR,0.2)--($(0,0)!(\k*\LDR,0.2)!(1,0)$);
    \draw[thick,dotted](\LDR*\k-\LDR*0.5,0.7)--(\LDR*\k-\LDR*0.5,0);
    \draw(\LDR*\k,0)node[below]{\small ${F}_{\k}$};
    \draw [<->,>=stealth,thick,dotted] (\LDR*\k-\LDR*0.5+0.1,0.5)--(\LDR*\k+\LDR*0.5-0.1,0.5);
    \draw(\LDR*\k,0.5)node[above]{\small $\{\!\!\{{F}_{\k}\}\!\!\}$};
  }
\end{tikzpicture}
\caption{Discretization grids for $F$ with (a) uniform and (b) non-uniform grid intervals.}
\label{figNumberLine}
\end{figure}
%
\subsubsection{Definition of instantaneous values: $\big[\hat{F},\hat{H}_{0}\big]_{-}\,{=}\,0$}
\label{sInstantaneous1}
To define arithmetic operations between instantaneous values of two or more physical quantities, it is useful to introduce a non-sharp representation that does not distinguish between stationary states with the same value of $F$.
\begin{myAssumptionC}[Set of stationary states with specified value and precision of a physical quantity]
For an operator $\hat{F}$ that satisfies
\begin{align}
\big[\hat{F},\hat{H}_{0}\big]_{-} = 0,
\label{eeInstantaneous10}
\end{align}
the set of stationary states with physical quantity $F$ having value ${F}_{i}$ is defined by 
\begin{align}
& \stateset{{F}_{i}}
\equiv \bigg\{\,\forall |\phi\rangle\,\bigg|\,\frac{{F}_{i-1}+{F}_{i}}{2}\,{\leqq}\,F(\phi)\,{<}\,\frac{{F}_{i}+{F}_{i+1}}{2}
,~\hat{F}|\phi\rangle\,{=}\,F(\phi)|\phi\rangle,
\nonumber\\
& ~~~~~~~~~~~~~~~~~~~~~
\hat{H}_{0}|\phi\rangle\,{=}\,E(\phi)|\phi\rangle
,~\langle\phi|\phi'\rangle = \delta_{\phi,\phi'}
\,\bigg\}
,\quad
i \in {\mathbb{Z}},
\label{eeInstantaneous11}
\end{align}
where the discretization grid of $F$ with interval $\{\delta{F}_{i}\,|\,i\,{\in}\,{\mathbb{Z}}\}$ is defined by
\begin{align}
{F}_{i+1} \equiv {F}_{i} + \delta{F}_{i}
,\quad F_{0} \equiv 0
,\quad i \in {\mathbb{Z}}
\label{eeInstantaneous12}
\end{align}
(Fig.\ref{figNumberLine}).
The interval of the discretization grids, abbreviated as $\delta{F}$, is hereafter referred to as the precision of proposition representation for $F$.
To save symbol, the subspace spanned by set $\stateset{{F}_{i}}$ is also represented by the same symbol $\stateset{{F}_{i}}$.
\end{myAssumptionC}
\noindent
According to this definition, proposition `$\occupation{\stateset{{F}_{i}}}{leftright},t$' can alternatively be expressed as `Physical quantity $F$ has a value $F_{i}$ with precision $\delta{F}$, at time $t$'.
The definition of instantaneous values given in Eq.\eqref{eeInstantaneous2} can then be generalized as follows.
\begin{myAssumptionC}[Instantaneous value of physical quantity]
For an operator $\hat{F}$ that satisfies
\begin{align}
\big[\hat{F},\hat{H}_{0}\big]_{-} = 0,
\label{eeInstantaneous13}
\end{align}
the instantaneous value of $F$ in the $j$-th trial is defined by
\begin{align}
{F}^{(j)}(t) 
\equiv \sum_{i} {F}_{i}\,\xjvar{`$\occupation{\stateset{{F}_{i}}}{leftright},t$'}{big}
,\quad \forall j.
\label{eeInstantaneous14}
\end{align}
\end{myAssumptionC}
\noindent
Using this definition, the expectation value of $F$, defined statistically in Eq.\eqref{eeInstantaneous4}, can be correctly calculated as
\begin{align}
\big\langle {F}(t) \big\rangle
\cong \sum_{i} {F}_{i}\,\big\langle \xvar{`$\occupation{\stateset{{F}_{i}}}{leftright},t$'}{big} \big\rangle
= \mathrm{Tr}\left[\,\hat{F}\hat{\rho}(\,t\,|\mbox{`${\mathrm{A}}_{0},t_{0}$'})\,\right]
\label{eeInstantaneous16}
\end{align}
within an approximation that assigns a value $F_{i}$ to all stationary states in $\stateset{{F}_{i}}$:
\begin{align}
\langle\phi|\hat{F}|\phi\rangle\,{\cong}\,F_{i}
,\quad \forall |\phi\rangle \in \stateset{{F}_{i}}.
\label{eeInstantaneous15}
\end{align}
It is worth noting that Eq.\eqref{eeInstantaneous14} and Eq.\eqref{eeInstantaneous16} exactly match Eq.\eqref{eeInstantaneous2} and Eq.\eqref{eeInstantaneous4} when the grid spacings $\{\delta{F}_{i}\,|\,i\,{\in}\,{\mathbb{Z}}\}$ approach zero in the continuous part of the eigenvalue spectrum and equals the spectral spacings in the discrete part.
This means that Definition C2 includes the simple case of Eq.\eqref{eeInstantaneous2}.
\begin{figure}[t]
\centering
\includegraphics{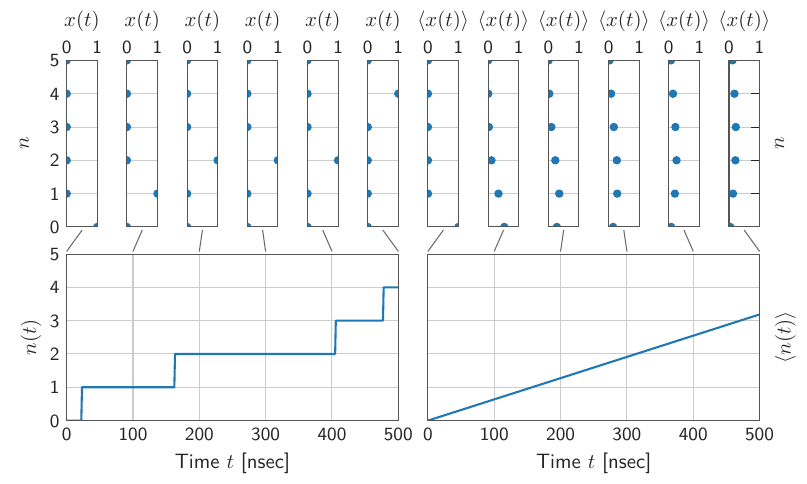}
\caption{The count number $n(t)$ of a photodetector~\cite{Theory2}. Lower left: the instantaneous value of $n(t)$ in an individual trial. Lower right: the expectation value of $n(t)$. Upper left: the instantaneous values of logical variables $x^{(j)}(\mbox{`o$(n), t$'})$. Upper right: the expectation values of logical variables $\langle x(\mbox{`o$(n), t$'})\rangle$. To clarify the logical connection with Ref.\cite{Theory2}, the same figure in Ref.\cite{Theory2} is used as is. See Ref.\cite{Theory2} for details of the calculations.}
\label{figPhysical1}
\end{figure}

\subsubsection{Arithmetic operations between instantaneous values: $0\,{=}\,\big[\hat{F},\hat{H}_{0}\big]_{-}\,{=}\,\big[\hat{G},\hat{H}_{0}\big]_{-}$}
\label{sArith1}
According to Definition C2, specifying the instantaneous value $F$ is equivalent to specifying the instantaneous values of the following logical variables:
\begin{align}
\left\{\,\xjvar{`$\occupation{\stateset{{F}_{i}}}{leftright},t$'}{big}\,\Big|\,i \in {\mathbb{Z}}\,\right\}
\label{eeArith1-1}
\end{align}
(Fig.\ref{figPhysical1}, Upper left).
Note that at any given moment, only one of these logical variables will have a value of 1.
The probability distribution of $F$ can also be represented by the expectation values of these logical variables, namely
\begin{align}
\left\{\,\big\langle \xvar{`$\occupation{\stateset{{F}_{i}}}{leftright},t$'}{big} \big\rangle\,\Big|\,i \in {\mathbb{Z}}\,\right\}
\label{eeArith1-2}
\end{align}
(Fig.\ref{figPhysical1}, Upper right).
Accordingly, under definitions C1 and C2, both the instantaneous value and probability distribution of $F$ can be fully represented by proposition $\big\{\mbox{`$\occupation{\stateset{{F}_{i}}}{leftright},t$'}$ $\,\big|\,i \in {\mathbb{Z}}\,\big\}$ alone.
It then follows that arithmetic operations between the instantaneous values of $F$ and $G$ should also be represented by propositions, specifically the result of logical operations on propositions $\left\{\mbox{`$\occupation{\stateset{{F}_{i}}}{leftright},t$'}\,\big|\,i \in {\mathbb{Z}}\,\right\}$ and $\left\{\mbox{`$\occupation{\stateset{{G}_{i}}}{leftright},t$'}\,\big|\,i \in {\mathbb{Z}}\,\right\}$.
For later use, we provide the definitions of addition and multiplication of $F$ and $G$:
\begin{myAssumptionC}[Arithmetic operations between instantaneous values]
Let $\hat{F}$ and $\hat{G}$ be operators satisfying
\begin{align}
0\,{=}\,\big[\hat{F},\hat{H}_{0}\big]_{-}\,{=}\,\big[\hat{G},\hat{H}_{0}\big]_{-},
\label{eeArith1-3}
\end{align}
and let the compatibility conditions be satisfied for all combinations of propositions `$\occupation{\stateset{{F}_{i}}}{leftright},t$' and `$\occupation{\stateset{{G}_{i'}}}{leftright},t$' ($i\,{\in}\,{\mathbb{Z}}$, $i'\,{\in}\,{\mathbb{Z}}$).
Then, the addition and multiplication of $F$ and $G$ can be defined as
\begin{align}
({F}+{G})^{(j)}(t)
& \equiv \sum_{i} \sum_{i'} \Big({F}_{i}+{G}_{i'}\Big)\,
\xjvar{`$\occupation{\stateset{{F}_{i}}}{leftright},t$'${\wedge}$`$\occupation{\stateset{{G}_{i'}}}{leftright},t$'}{big}
,\quad \forall j,
\label{eeArith1-4}\\
({F} {G})^{(j)}(t)
& \equiv \sum_{i} \sum_{i'} \Big({F}_{i}{G}_{i'}\Big)\,
\xjvar{`$\occupation{\stateset{{F}_{i}}}{leftright},t$'${\wedge}$`$\occupation{\stateset{{G}_{i'}}}{leftright},t$'}{big}
,\quad \forall j.
\label{eeArith1-5}
\end{align}
\end{myAssumptionC}
\noindent
The compatibility conditions between propositions `$\occupation{\stateset{{F}_{i}}}{leftright}$' and `$\occupation{\stateset{{G}_{i'}}}{leftright}$' are required because the conjunction ($\wedge$) must exist for the right-hand sides of Eqs.\eqref{eeArith1-4}--\eqref{eeArith1-5} to be defined.
When these conditions are satisfied, Eqs.\eqref{eeArith1-4}--\eqref{eeArith1-5} can be simplified using the laws of Boolean logic (Assumption B6) as
\begin{align}
({F}+{G})^{(j)}(t)
& = {F}^{(j)}(t) + {G}^{(j)}(t)
,\quad \forall j,
\label{eeArith1-7}\\
({F} {G})^{(j)}(t)
& = {F}^{(j)}(t)\,{G}^{(j)}(t)
,\quad \forall j.
\label{eeArith1-8}
\end{align}
The extension to more general functional operations follows naturally from these definitions.
Moreover, the condition for defining the joint probability of $F$ and $G$, $\big\langle x\big(\mbox{`$\occupation{\stateset{{F}_{i}}}{leftright},t$'}$ $\mbox{${\wedge}$`$\occupation{\stateset{{G}_{i'}}}{leftright},t$'}\big) \big\rangle$, is identical to that in Definition C3.
\par
It is important to note that, according to Definition C3, $[\hat{F},\hat{G}]_{-}\,{\neq}\,0$ does not necessarily preclude the existence of a joint probability of $F$ and $G$.
More precisely, if $[\hat{F},\hat{G}]_{-}\,{\neq}\,0$, there is no stationary state to which simultaneous values of $F$ and $G$ can be assigned.
However, if $0\,{=}\,[\hat{F},\hat{H}_{0}]_{-}\,{=}\,[\hat{G},\hat{H}_{0}]_{-}$, there exist a set of stationary states, $\stateset{{F}_{i}}$, having a definite value of $F$, and a set of stationary states, $\stateset{{G}_{i'}}$, having a definite value of $G$.
The compatibility conditions between `$\occupation{\stateset{{F}_{i}}}{leftright},t$' and `$\occupation{\stateset{{G}_{i'}}}{leftright},t$' are then satisfied if
\begin{align}
1 
= \langle \xvar{`$\occupation{\stateset{{F}_{i}}}{leftright},t$'}{big} \rangle
= \langle \xvar{`$\occupation{\stateset{{G}_{i'}}}{leftright},t$'}{big} \rangle
,\quad \forall j.
\label{eeArith1-9}
\end{align}
This means that in our theory, the arithmetic operations and joint probabilities of $F$ and $G$ may exist even when $\big[\hat{F},\hat{G}\big]_{-}\,{\neq}\,0$.
This is because, our theory allows for Type-II compatibility through the compatibility condition, whereas conventional theory prohibits Type-II compatibility through the orthogonality or consistency conditions (See {\S}\ref{sJoint} or {\S}\ref{sReality4}).

\subsubsection{Definition of instantaneous values:~$\big[\hat{F},\hat{H}_{0}\big]_{-}\,{\neq}\,0$ or $\big[\hat{G},\hat{H}_{0}\big]_{-}\,{\neq}\,0$}
\label{sMeasGeneral}
For general physical quantities $F$ where the operator $\hat{F}$ satisfies the condition $[\hat{F},\hat{H}_{0}]_{-}\,{\neq}\,0$, no simultaneous eigenstates of $F$ and $\hat{H}_{0}$ exist.
Consequently, according to our model of reality, it is impossible to define instantaneous values for such physical quantities $F$.
Concrete examples of such quantities include the position ${\bm r}$ of a charged particle, and the real and imaginary parts of the modal amplitude $\tilde{a}_{{\bm k}\lambda}$ of a radiation field.
Experimentally, however, instead of these quantities, instantaneous values of different physical quantities $\mathcal{F}$ are measured in systems that include additional measurement instruments. 
For instance, in the measurement of a charged particle's position~\cite{Compton1925}, the instantaneous value of $F\,{=}\,{\bm r}$ is estimated from the instantaneous value of the position of water droplets $\mathcal{F}\,{=}\,{\mathcal{R}}$ in a cloud chamber, using the following empirical formula:
\begin{align}
{\bm r}^{(j)}(t) & = {\mathcal{R}}^{(j)}(t).
\label{eeMeasGeneral1}
\end{align}
If a magnetic field is applied during measurement, the instantaneous value of the momentum $G\,{=}\,{\bm p}$ of the charged particle, ${\bm p}^{(j)}(t)\,{=}\,{\mathcal{P}}^{(j)}(t)$, can be estimated from the radius of curvature of the same water droplet trajectory used for position measurement. 
Similarly, in the measurement of radiation field mode amplitudes~\cite{Mandel}, the instantaneous values of the real part $F\,{=}\,{\mathrm{Re}}[\tilde{\alpha}]$ and imaginary part $G\,{=}\,{\mathrm{Im}}[\tilde{\alpha}_{\lambda}]$ are determined based on the count numbers ${\mathcal{N}}_{1}$ and ${\mathcal{N}}_{2}$ of two photodetectors constituting a heterodyne photodetector~\cite{Lvovsky2008,Kikuchi2016}, using the following empirical formulas:
\begin{align}
{\mathrm{Re}}[\tilde{\alpha}^{(j)}_{{\bm k}\lambda}(t)]
= {\mathrm{Const.}} \frac{{\mathcal{N}}_{1}^{(j)}(t) + {\mathcal{N}}_{2}^{(j)}(t)}{2}
,\quad
{\mathrm{Im}}[\tilde{\alpha}^{(j)}_{{\bm k}\lambda}(t)]
= {\mathrm{Const.}} \frac{{\mathcal{N}}_{1}^{(j)}(t) - {\mathcal{N}}_{2}^{(j)}(t)}{2{\mathrm{i}}}.
\label{eeMeasGeneral2}
\end{align}
In other words, in experiments, the instantaneous values of unmeasurable physical quantities $F$ are determined based on the instantaneous values of physical quantities $\mathcal{F}$ in an extended system with an additional measurement instrument.
These empirical formulas can be generally expressed using measurement propositions as follows:
\begin{align}
\xjvar{`$\occupation{\stateset{{F}_{i}}}{leftright},t$'}{big}
& = \xjvar{`$\occupation{\stateset{{\mathcal{F}}_{i}}}{leftright},t$'}{big}
,\quad \forall j
,\quad \forall i,
\label{eeMeasGeneral3}\\
\xjvar{`$\occupation{\stateset{{G}_{i}}}{leftright},t$'}{big}
& = \xjvar{`$\occupation{\stateset{{\mathcal{G}}_{i}}}{leftright},t$'}{big}
,\quad \forall j
,\quad \forall i.
\label{eeMeasGeneral4}
\end{align}
For this reason, we next consider defining the instantaneous values of physical quantities $F$ and $G$ based on the instantaneous values of physical quantities $\mathcal{F}$ and $\mathcal{G}$ using these deterministic relations.
\par
To this end, we first clarify the definition of instantaneous values for $\mathcal{F}$ and $\mathcal{G}$ in the extended system. 
Generally, we consider cases where the operators for $F$ and $G$ do not commute with the non-interacting Hamiltonian $\hat{H}_{0}$ while $\mathcal{F}$ and $\mathcal{G}$ can be measured by adding measurement instruments.
It then follows from the assumptions that
\begin{align}
\big[\hat{F},\hat{H}_{0}\big]_{-} \neq 0
\quad\mbox{or}\quad
\big[\hat{G},\hat{H}_{0}\big]_{-} \neq 0
\label{eeMeasGeneral5}
\end{align}
and
\begin{align}
\big[\hat{\mathcal{F}},\hat{\mathcal{H}}_{0,\mathcal{F}}\big]_{-} = 0
\quad\mbox{and}\quad
\big[\hat{\mathcal{G}},\hat{\mathcal{H}}_{0,\mathcal{G}}\big]_{-} = 0.
\label{eeMeasGeneral6}
\end{align}
Note that there is no guarantee that $\mathcal{F}$ and $\mathcal{G}$ can be measured using common measurement instrument, so we have distinguished between the non-interacting Hamiltonian $\hat{\mathcal{H}}_{0,\mathcal{F}}$ describing the measurement of $\mathcal{F}$ and $\hat{\mathcal{H}}_{0,\mathcal{G}}$ describing the measurement of $\mathcal{G}$. 
In general, when the instantaneous values of $\mathcal{F}$ and $\mathcal{G}$ cannot be determined in a single trial (i.e., using common measurement instrument), it is impossible to use common non-interacting Hamiltonians, total Hamiltonians, stationary states, and density operators for the measurements of $\mathcal{F}$ and $\mathcal{G}$ ($\hat{\mathcal{H}}_{0,\mathcal{F}}\,{\neq}\,\hat{\mathcal{H}}_{0,\mathcal{G}}$). 
In such cases, the measurements of $\mathcal{F}$ and $\mathcal{G}$ cannot be represented by a common density operator, and consequently, the compatibility conditions between $\mathcal{F}$ and $\mathcal{G}$ do not hold. 
On the other hand, when the instantaneous values of $\mathcal{F}$ and $\mathcal{G}$ can be determined in a single trial, it is possible to use common non-interacting Hamiltonians, total Hamiltonians, stationary states, and density operators for the measurements of $\mathcal{F}$ and $\mathcal{G}$ ($\hat{\mathcal{H}}_{0,\mathcal{F}}\,{=}\,\hat{\mathcal{H}}_{0,\mathcal{G}}\,{\equiv}\,\hat{\mathcal{H}}_{0}$), allowing for the following definition:
\begin{myAssumptionC}[Instantaneous values of physical quantities in extended systems (1)]
When physical quantities $\mathcal{F}$ and $\mathcal{G}$ can be measured simultaneously in a single trial in the extended system with additional measurement instruments, it is possible to use common non-interacting Hamiltonians, total Hamiltonians, stationary states, and density operators for the measurements of $\mathcal{F}$ and $\mathcal{G}$, and the following relationship holds:
\begin{align}
0 
= \big[\hat{\mathcal{F}},\hat{\mathcal{G}}\big]_{-}
= \big[\hat{\mathcal{F}},\hat{\mathcal{H}}_{0}\big]_{-}
= \big[\hat{\mathcal{G}},\hat{\mathcal{H}}_{0}\big]_{-}.
\label{eeMeasGeneral7}
\end{align}
In this case, the instantaneous values of $\mathcal{F}$ and $\mathcal{G}$ can be defined by the following equations:
\begin{align}
{\mathcal{F}}^{(j)}(t)
= \sum_{i} {\mathcal{F}}_{i}\,\xjvar{`$\occupation{\stateset{{\mathcal{F}}_{i}}}{leftright},t$'}{big}
,\quad \forall j,
\label{eeMeasGeneral8}\\
{\mathcal{G}}^{(j)}(t)
= \sum_{i} {\mathcal{G}}_{i}\,\xjvar{`$\occupation{\stateset{{\mathcal{G}}_{i}}}{leftright},t$'}{big}
,\quad \forall j.
\label{eeMeasGeneral9}
\end{align}
In this case, compatibility conditions between `$\occupation{\stateset{{\mathcal{F}}_{i}}}{leftright},t$' and `$\occupation{\stateset{{\mathcal{G}}_{i'}}}{leftright},t$' hold by assumption, allowing for the definition of arithmetic operations and joint probabilities for instantaneous values ${\mathcal{F}}^{(j)}(t)$ and ${\mathcal{G}}^{(j)}(t)$.
\end{myAssumptionC}
\begin{myAssumptionC}[Instantaneous values of physical quantities in extended systems (2)]
When physical quantities $\mathcal{F}$ and $\mathcal{G}$ cannot be measured simultaneously in a single trial, due to differences in measurement instruments, it is impossible to use common non-interacting Hamiltonians, total Hamiltonians, stationary states, and density operators for the measurements of $\mathcal{F}$ and $\mathcal{G}$, meaning that
\begin{align}
\big[\hat{\mathcal{F}},\hat{\mathcal{H}}_{0,\mathcal{F}}\big]_{-} = 0
& \quad\mbox{and}\quad
\big[\hat{\mathcal{G}},\hat{\mathcal{H}}_{0,\mathcal{G}}\big]_{-} = 0
\label{eeMeasGeneral11}
\end{align}
with
\begin{align}
& \hat{\mathcal{H}}_{0,\mathcal{F}} \neq \hat{\mathcal{H}}_{0,\mathcal{G}}.
\label{eeMeasGeneral12}
\end{align}
In this case, the instantaneous values of $\mathcal{F}$ and $\mathcal{G}$ can still be defined by the following equations:
\begin{align}
{\mathcal{F}}^{(j)}(t)
= \sum_{i} {\mathcal{F}}_{i}\,\xjvar{`$\occupation{\stateset{{\mathcal{F}}_{i}}}{leftright},t$'}{big}
,\quad \forall j,
\label{eeMeasGeneral13}\\
{\mathcal{G}}^{(j)}(t)
= \sum_{i} {\mathcal{G}}_{i}\,\xjvar{`$\occupation{\stateset{{\mathcal{G}}_{i}}}{leftright},t$'}{big}
,\quad \forall j.
\label{eeMeasGeneral14}
\end{align}
However, because the compatibility conditions between `$\occupation{\stateset{{\mathcal{F}}_{i}}}{leftright},t$' and `$\occupation{\stateset{{\mathcal{G}}_{i'}}}{leftright},t$' do not hold, arithmetic operations and joint probabilities for instantaneous values $\mathcal{F}^{(j)}(t)$ and $\mathcal{G}^{(j)}(t)$ cannot be defined.
\end{myAssumptionC}
%
\subsubsection{Physical quantities as proxy representations}
\label{sProxy}
We now define the instantaneous values of physical quantities $F$ and $G$ in the original system using the deterministic relations in Eqs.\eqref{eeMeasGeneral3}--\eqref{eeMeasGeneral4}.
Note that these equations merely indicate that the instantaneous values of $F$ and $G$ in the original system are proxy representations of the instantaneous values of $\mathcal{F}$ and $\mathcal{G}$ in the extended system. 
Crucially, for this type of proxy representations to faithfully represent the behavior of the original quantities in both individual trials and expectations, the truth values, probability distributions, and joint probabilities of the measurement propositions must all coincide (cf.\,{\S}\ref{sArith1}). 
Therefore, we define physical quantities as proxy representations as follows:
\begin{myAssumptionC}[Complete proxy representation]
When the instantaneous values of physical quantities $\mathcal{F}$ and $\mathcal{G}$ can be measured simultaneously in a single trial in the extended system including measurement instruments (Definition C4), we define propositions `$\occupation{\stateset{{F}_{i}}}{leftright},t$' and `$\occupation{\stateset{{G}_{i}}}{leftright},t$' in the idealized system without measurement instruments to be complete proxy representations of propositions `$\occupation{\stateset{{\mathcal{F}}_{i}}}{leftright},t$' and `$\occupation{\stateset{{\mathcal{G}}_{i}}}{leftright},t$' in the extended system including measurement instruments if the following relations hold:\\
(Equivalence of truth values)
\begin{align}
\xjvar{`$\occupation{\stateset{{F}_{i}}}{leftright},t$'}{big}
& = \xjvar{`$\occupation{\stateset{{\mathcal{F}}_{i}}}{leftright},t$'}{big}
,\quad \forall j
,\quad \forall i
\label{eeProxy1}\\
\xjvar{`$\occupation{\stateset{{G}_{i}}}{leftright},t$'}{big}
& = \xjvar{`$\occupation{\stateset{{\mathcal{G}}_{i}}}{leftright},t$'}{big}
,\quad \forall j
,\quad \forall i
\label{eeProxy2}
\end{align}
(Equivalence of probability distributions)
\begin{align}
& {\rm{Tr}}\Big[\,\zop{`$\occupation{\stateset{{F}_{i}}}{leftright}$'}{leftright}\,\hat{\rho}(\,t\,|\mbox{`$\mathrm{A}_{0},t_{0}$'})\,\zopd{`$\occupation{\stateset{{F}_{i}}}{leftright}$'}{leftright}\,\Big]
\nonumber\\
& = {\rm{Tr}}\Big[\,\zop{`$\occupation{\stateset{{\mathcal{F}}_{i}}}{leftright}$'}{leftright}\,\hat{\wp}(\,t\,|\mbox{`$\mathrm{A}_{0},t_{0}$'})\,\zopd{`$\occupation{\stateset{{\mathcal{F}}_{i}}}{leftright}$'}{leftright}\,\Big]
,\quad \forall i
\label{eeProxy3}\\
& {\rm{Tr}}\Big[\,\zop{`$\occupation{\stateset{{G}_{i}}}{leftright}$'}{leftright} \hat{\rho}(\,t\,|\mbox{`$\mathrm{A}_{0},t_{0}$'}) \zopd{`$\occupation{\stateset{{G}_{i}}}{leftright}$'}{leftright}\,\Big]
\nonumber\\
& = {\rm{Tr}}\Big[\,\zop{`$\occupation{\stateset{{\mathcal{G}}_{i}}}{leftright}$'}{leftright} \hat{\wp}(\,t\,|\mbox{`$\mathrm{A}_{0},t_{0}$'}) \zopd{`$\occupation{\stateset{{\mathcal{G}}_{i}}}{leftright}$'}{leftright}\,\Big]
,\quad \forall i
\label{eeProxy4}
\end{align}
(Equivalence of joint probabilities in the simultaneous limit)
\begin{align}
& {\rm{Tr}}\Big[\,\zop{`$\occupation{\stateset{{F}_{i}}}{leftright}$'}{leftright}\,\zop{`$\occupation{\stateset{{G}_{i'}}}{leftright}$'}{leftright}\,\hat{\rho}(\,t\,|\mbox{`$\mathrm{A}_{0},t_{0}$'})\,\zopd{`$\occupation{\stateset{{G}_{i'}}}{leftright}$'}{leftright}\,\zopd{`$\occupation{\stateset{{F}_{i}}}{leftright}$'}{leftright}\,\Big]
\nonumber\\
& = {\rm{Tr}}\Big[\,\zop{`$\occupation{\stateset{{\mathcal{F}}_{i}}}{leftright}$'}{leftright}\,\zop{`$\occupation{\stateset{{\mathcal{G}}_{i'}}}{leftright}$'}{leftright}\,\hat{\wp}(\,t\,|\mbox{`$\mathrm{A}_{0},t_{0}$'})\,\zopd{`$\occupation{\stateset{{\mathcal{G}}_{i'}}}{leftright}$'}{leftright}\,\zopd{`$\occupation{\stateset{{\mathcal{F}}_{i}}}{leftright}$'}{leftright}\,\Big],
\nonumber\\
& \quad \forall i,~\forall i'
\label{eeProxy5}\\
& {\rm{Tr}}\Big[\,\zop{`$\occupation{\stateset{{G}_{i}}}{leftright}$'}{leftright}\,\zop{`$\occupation{\stateset{{F}_{i'}}}{leftright}$'}{leftright}\,\hat{\rho}(\,t\,|\mbox{`$\mathrm{A}_{0},t_{0}$'})\,\zopd{`$\occupation{\stateset{{F}_{i'}}}{leftright}$'}{leftright}\,\zopd{`$\occupation{\stateset{{G}_{i}}}{leftright}$'}{leftright}\,\Big]
\nonumber\\
& = {\rm{Tr}}\Big[\,\zop{`$\occupation{\stateset{{\mathcal{G}}_{i}}}{leftright}$'}{leftright}\,\zop{`$\occupation{\stateset{{\mathcal{F}}_{i'}}}{leftright}$'}{leftright}\,\hat{\wp}(\,t\,|\mbox{`$\mathrm{A}_{0},t_{0}$'})\,\zopd{`$\occupation{\stateset{{\mathcal{F}}_{i'}}}{leftright}$'}{leftright}\,\zopd{`$\occupation{\stateset{{\mathcal{G}}_{i}}}{leftright}$'}{leftright}\,\Big],
\nonumber\\
& \quad \forall i,~\forall i'
\label{eeProxy6}
\end{align}
For simplicity of notation, we assume that the indices $i$ and $i'$ of the sets have been adjusted so that the deterministic relations in Eqs.\eqref{eeProxy1}--\eqref{eeProxy2} hold.
When these conditions are satisfied, $F$ and $G$ are expressed as complete proxy representations of $\mathcal{F}$ and $\mathcal{G}$.
\end{myAssumptionC}
\begin{myAssumptionC}[Incomplete proxy representation]
When the instantaneous values of physical quantities $\mathcal{F}$ and $\mathcal{G}$ cannot be measured simultaneously in a single trial in the extended system including measurement instruments (Definition C5), we define propositions `$\occupation{\stateset{{F}_{i}}}{leftright},t$' and `$\occupation{\stateset{{G}_{i}}}{leftright},t$' in the idealized system without measurement instruments to be incomplete proxy representations of propositions `$\occupation{\stateset{{\mathcal{F}}_{i}}}{leftright},t$' and `$\occupation{\stateset{{\mathcal{G}}_{i}}}{leftright},t$' in the extended system including measurement instruments if the following equations hold:\\
(Equivalence of truth values)
\begin{align}
\xjvar{`$\occupation{\stateset{{F}_{i}}}{leftright},t$'}{big}
& = \xjvar{`$\occupation{\stateset{{\mathcal{F}}_{i}}}{leftright},t$'}{big}
,\quad \forall j
,\quad \forall i
\label{eeProxy11}\\
\xjvar{`$\occupation{\stateset{{G}_{i}}}{leftright},t$'}{big}
& = \xjvar{`$\occupation{\stateset{{\mathcal{G}}_{i}}}{leftright},t$'}{big}
,\quad \forall j
,\quad \forall i
\label{eeProxy12}
\end{align}
(Equivalence of probability distributions)
\begin{align}
& {\rm{Tr}}\Big[\,\zop{`$\occupation{\stateset{{F}_{i}}}{leftright}$'}{leftright}\,\hat{\rho}(\,t\,|\mbox{`$\mathrm{A}_{0},t_{0}$'})\,\zopd{`$\occupation{\stateset{{F}_{i}}}{leftright}$'}{leftright}\,\Big]
\nonumber\\
& = {\rm{Tr}}\Big[\,\zop{`$\occupation{\stateset{{\mathcal{F}}_{i}}}{leftright}$'}{leftright}\,\hat{\wp}_{\mathcal{F}}(\,t\,|\mbox{`$\mathrm{A}_{0},t_{0}$'})\,\zopd{`$\occupation{\stateset{{\mathcal{F}}_{i}}}{leftright}$'}{leftright}\,\Big]
,\quad \forall i
\label{eeProxy13}\\
& {\rm{Tr}}\Big[\,\zop{`$\occupation{\stateset{{G}_{i}}}{leftright}$'}{leftright} \hat{\rho}(\,t\,|\mbox{`$\mathrm{A}_{0},t_{0}$'}) \zopd{`$\occupation{\stateset{{G}_{i}}}{leftright}$'}{leftright}\,\Big]
\nonumber\\
& = {\rm{Tr}}\Big[\,\zop{`$\occupation{\stateset{{\mathcal{G}}_{i}}}{leftright}$'}{leftright} \hat{\wp}_{\mathcal{G}}(\,t\,|\mbox{`$\mathrm{A}_{0},t_{0}$'}) \zopd{`$\occupation{\stateset{{\mathcal{G}}_{i}}}{leftright}$'}{leftright}\,\Big]
,\quad \forall i
\label{eeProxy14}
\end{align}
Here, considering the differences in measurement instruments, we denote the density operator describing the measurement of $\mathcal{F}$ in the extended system as $\hat{\wp}_{\mathcal{F}}(\,t\,|\mbox{`$\mathrm{A}_{0},t_{0}$'})$, and the density operator describing the measurement of $\mathcal{G}$ as $\hat{\wp}_{\mathcal{G}}(\,t\,|\mbox{`$\mathrm{A}_{0},t_{0}$'})$. 
When these conditions are satisfied, $F$ and $G$ are expressed as incomplete proxy representations of $\mathcal{F}$ and $\mathcal{G}$.
\end{myAssumptionC}
\begin{myAssumptionC}[Instantaneous values of physical quantities as proxy representations]
When propositions `$\occupation{\stateset{{F}_{i}}}{leftright},t$' and `$\occupation{\stateset{{G}_{i}}}{leftright},t$' are complete proxy representations of propositions `$\occupation{\stateset{{\mathcal{F}}_{i}}}{leftright},t$' and `$\occupation{\stateset{{\mathcal{G}}_{i}}}{leftright},t$', we define the instantaneous values of physical quantities $F$ and $G$ by the following equations:
\begin{align}
{F}^{(j)}(t) 
= \sum_{i} {F}_{i}\,\xjvar{`$\occupation{\stateset{{F}_{i}}}{leftright},t$'}{big}
,\quad \forall j,
\label{eeProxy21}\\
{G}^{(j)}(t) 
= \sum_{i} {G}_{i}\,\xjvar{`$\occupation{\stateset{{G}_{i}}}{leftright},t$'}{big}
,\quad \forall j.
\label{eeProxy22}
\end{align}
\end{myAssumptionC}

\subsubsection{Differences from conventional theory}
\label{sContextuality}
With the definitions provided above, arithmetic operations and joint probabilities between instantaneous values of $F$ and $G$ can be defined not when $[\hat{F},\hat{G}]_{-}\,{=}\,0$, but when the instantaneous values of $\mathcal{F}$ and $\mathcal{G}$ can be measured in a single trial. 
Specifically, when the instantaneous values of $\mathcal{F}$ and $\mathcal{G}$ can be measured in a single trial, the compatibility conditions between $\mathcal{F}$ and $\mathcal{G}$ hold according to Definition C4, and the compatibility conditions between $F$ and $G$ hold according to Definition C6. 
Therefore, in this case, arithmetic operations and joint probabilities can be defined between the instantaneous values of both $\mathcal{F}$ and $\mathcal{G}$, and $F$ and $G$. 
Conversely, when the instantaneous values of $\mathcal{F}$ and $\mathcal{G}$ cannot be measured in a single trial, the compatibility conditions between $\mathcal{F}$ and $\mathcal{G}$ do not hold according to Definition C5, and the compatibility conditions between $F$ and $G$ do not hold according to Definition C7. 
Consequently, in this case, the instantaneous values of $\mathcal{F}$, $\mathcal{G}$, $F$, and $G$ can be defined according to Definitions C5 and C7, but arithmetic operations and joint probabilities between the instantaneous values of $\mathcal{F}$ and $\mathcal{G}$, or $F$ and $G$, cannot be defined.
\par
It is important to note that this conclusion differs from conventional theory.
To see this, we focus on two typical cases that satisfy the following conditions:
\begin{align}
\big[\hat{F},\hat{G}\big]_{-} \neq 0.
\label{eeContextuality1}
\end{align}
In this case, Nelson's theorem in the standard formulation states that the joint probability of $F$ and $G$ is not definable.
In contrast, our theory predicts that the joint probabilities of $\mathcal{F}$ and $\mathcal{G}$, and $F$ and $G$, are definable if the instantaneous values of $\mathcal{F}$ and $\mathcal{G}$ can be measure in a single trial and if $F$ and $G$ are proxy representations of $\mathcal{F}$ and $\mathcal{G}$.
For example, in the measurement of the position and momentum of charged particles, it is possible to estimate the functions $\mathcal{F}\,{=}\,{\mathcal{R}}$ and $\mathcal{G}\,{=}\,{\mathcal{P}}$ of the arrangement of water droplets in a single trial.
Moreover, ${F}\,{=}\,{\bm r}$ and ${G}\,{=}\,{\bm p}$ become proxy representations of $\mathcal{F}\,{=}\,{\mathcal{R}}$ and $\mathcal{G}\,{=}\,{\mathcal{P}}$ if we adjust the precision $\delta{F}$ and $\delta{G}$ such that the compatibility conditions between `$\occupation{\stateset{{F}_{i}}}{leftright},t$' and `$\occupation{\stateset{{G}_{i}}}{leftright},t$' are satisfied (Eqs.\eqref{eeProxy5}--\eqref{eeProxy6}).
To this end, we have only to require
\begin{align}
\exists i,~\exists i',\quad  
1 
= \langle \xvar{`$\occupation{\stateset{{F}_{i}}}{leftright},t$'}{big} \rangle
= \langle \xvar{`$\occupation{\stateset{{G}_{i'}}}{leftright},t$'}{big} \rangle.
\label{eeContextuality2}
\end{align}
This correctly explains the experimental fact that the instantaneous values of position ${F}\,{=}\,{\bm r}$ and momentum ${G}\,{=}\,{\bm p}$ can be obtained simultaneously in a single trial when the measurement precision is sufficiently coarse compared to the uncertainty limit.
\par
As another example, in the measurement of the $x$ and $y$ components of the spin angular momentum, measurements provide the instantaneous values of the following physical quantities:
\begin{align}
{\mathcal{F}}
& = \frac{\hbar}{2} \Big( \xvar{`$\mbox{detected}_{{\uparrow}_{x}}$'}{big} - \xvar{`$\mbox{detected}_{{\downarrow}_{x}}$'}{big} \Big)
= \hbar \left( \xvar{`$\mbox{detected}_{{\uparrow}_{x}}$'}{big} - \frac{1}{2} \right),
\label{eeContextuality3}\\
{\mathcal{G}}
& = \frac{\hbar}{2} \Big( \xvar{`$\mbox{detected}_{{\uparrow}_{y}}$'}{big} - \xvar{`$\mbox{detected}_{{\downarrow}_{y}}$'}{big} \Big)
= \hbar \left( \xvar{`$\mbox{detected}_{{\uparrow}_{y}}$'}{big} - \frac{1}{2} \right),
\label{eeContextuality4}
\end{align}
where logical variables $\{ \xvar{`o(${\uparrow}_{x}$),\,$t$'}{}, \xvar{`o(${\downarrow}_{x}$),\,$t$'}{} \}$ and $\{ \xvar{`o(${\uparrow}_{y}$),\,$t$'}{}, \xvar{`o(${\downarrow}_{y}$),\,$t$'}{} \}$ represent the detection signals from two separate Stern-Gerlach experiments with different magnetic field directions (Eq.\eqref{eeSimpleEx11}).
Then, operators $\hat{\mathcal{F}}$ and $\hat{\mathcal{G}}$ satisfy $\big[\hat{\mathcal{F}},\hat{\mathcal{G}}\big]_{-}\,{=}\,0$, as can be confirmed by the first-order perturbation analysis of particle detectors.
This indicates that, according to conventional theory, the joint probability of $\mathcal{F}$ and $\mathcal{G}$ should be definable. 
In this case, however, the experiments are conducted separately, so the joint probability of $\mathcal{F}$ and $\mathcal{G}$ is meaningless in practice, and as a result, the effects of measurement contextuality need to be carefully considered.
In our theory, on the other hand, when the instantaneous values of $\mathcal{F}$ and $\mathcal{G}$ cannot be obtained in a single trial, their joint probability cannot be defined, eliminating the need for special commentary on the experimental facts of measurement contextuality.

\subsection{Classical approximations and classical equations of motion}
\label{sClassical}
\subsubsection{Classical approximations for instantaneous values of physical quantities}
\label{sClassical1}
To formulate a classical approximation for instantaneous values of physical quantities, we approximate the instantaneous value using a single logical variable as
\begin{align}
F^{(j)}(t) 
\cong F^{(j)}_{0}\,\xjvar{`$\occupation{\stateset{{F}^{(j)}_{0}}}{},t$'}{leftright}
,\quad \forall j
\label{eeMeasClassical1}
\end{align}
and adjust the precision $\delta{F}$ of proposition `$\occupation{\stateset{{F}^{(j)}_{0}}}{},t$' to satisfy the following condition:
\begin{align}
\xjvar{`$\occupation{\stateset{{F}^{(j)}_{0}}}{},t$'}{leftright} = 1
,\quad \forall j.
\label{eeMeasClassical2}
\end{align}
Since the compatibility conditions are always satisfied among propositions of this kind ({\S}\ref{sAppendix100}), the classical approximation allows arithmetic operations between any physical quantities.
At the same time, however, taking the expectation value of Eq.\eqref{eeMeasClassical1}, we obtain
\begin{align}
F^{(j)}(t) = F^{(j)}_{0} = \langle F(t)\rangle,
\label{eeMeasClassical3}
\end{align}
meaning that the classical approximation requires the ensemble definition and the grid center to be continuously updated so that the expectation value must be equal to the instantaneous value.
\par
For this reason, we denote the proposition 'The instantaneous value of $F$ at time $t$ is $F^{(j)}(t)$' by the symbol '$F^{(j)}(t)$' and define the conditional expectation value conditioned by proposition `$F^{(j)}(t)$' as
\begin{align}
\Big\langle\,F(t)\,\Big|\!\Big|\,\mbox{`$F^{(j)}(t)$'} \Big\rangle 
\equiv \frac{\sum_{j'}^{\mbox{\scriptsize `$F^{(j)}(t)$'}} F^{(j')}(t)}{\sum_{j'}^{\mbox{\scriptsize `$F^{(j)}(t)$'}} 1}
= F^{(j)}(t) 
,\quad \forall j.
\label{eeMeasClassical5}
\end{align}
It then follows from Assumptions B1--B14 and the definition of conditional probabilities (Eq.\eqref{eeJoint3a}) that the density operator $\hat{\rho}(t|\text{'$F^{(j)}(t)$'})$ in the classical approximation satisfies
\begin{align}
& \Big\langle\,F(t)\,\Big|\!\Big|\,\mbox{`$F^{(j)}(t)$'}\Big\rangle 
= {\mathrm{Tr}}\Big[\,\hat{F}\,\hat{\rho}\big(\,t\,\big|\,\mbox{`$F^{(j)}(t)$'}\big)\,\Big],
\label{eeMeasClassical6}\\
& \hat{\rho}\big(\,t\,\big|\,\mbox{`$F^{(j)}(t)$'}\,\big)
\equiv \frac{\hat{z}\big(\,\mbox{`$F^{(j)}(t)$'}\big)\,\hat{\rho}(\,t\,|\mbox{`${\mathrm{A}}_{0},t_{0}$'})\,\hat{z}^{\dagger}\big(\,\mbox{`$F^{(j)}(t)$'}\big)}
{{\rm{Tr}}\Big[\,\hat{z}(\,\mbox{`$F^{(j)}(t)$'})\,\hat{\rho}(\,t\,|\mbox{`${\mathrm{A}}_{0},t_{0}$'})\,\hat{z}^{\dagger}(\,\mbox{`$F^{(j)}(t)$'})\,\Big]}.
\label{eeMeasClassical7}
\end{align}
Here, the projection operator $\hat{z}\big(\,\mbox{`$F^{(j)}(t)$'}\big)$ on the right-hand side of Eq.\eqref{eeMeasClassical7} is defined by
\begin{align}
& \zop{`$F^{(j)}(t)$'}{big}
= \sum^{\stateset{F^{(j)}(t)}}_{|\varphi\rangle}
|\varphi\rangle\langle\varphi|,
\label{eeMeasClassical8}\\
& \stateset{F^{(j)}(t)} 
\equiv \bigg\{\,\forall |\varphi\rangle\,\bigg|\,
F^{(j)}(t)-\frac{\delta{F}_{i-1}}{2}\,{\leqq}\,F(\varphi)\,{<}\,F^{(j)}(t)+\frac{\delta{F}_{i}}{2}
,~\hat{F}|\varphi\rangle = F(\varphi)|\varphi\rangle,
\nonumber\\
& ~~~~~~~~~~~~~~~~~~~~~~~~~~~~
\langle\varphi|\varphi'\rangle = \delta_{\varphi,\varphi'}\,\bigg\},
\label{eeMeasClassical9}
\end{align}
with the corresponding Heisenberg operator given by
\begin{align}
\zop{`$F^{(j)}(t),t'$'}{big} = \hat{U}^{\dagger}(t'-t)\,\hat{z}\big(\,\mbox{`$F^{(j)}(t)$'}\big)\,\hat{U}(t'-t)
,\quad t' \geqq t.
\label{eeMeasClassical10}
\end{align}
Using Eqs.\eqref{eeMeasClassical7}--\eqref{eeMeasClassical10}, we can see that the density operator and the Heisenberg operator satisfy
\begin{align}
\hat{\rho}\big(\,t\,\big|\,\mbox{`$F^{(j)}(t)$'}\,\big)
& = \zop{`$F^{(j)}(t)$'}{big}
\hat{\rho}\big(\,t\,\big|\,\mbox{`$F^{(j)}(t)$'}\,\big)
\zop{`$F^{(j)}(t)$'}{big},
\label{eeMeasClassical11}\\
\hat{F}(t)
& = \hat{z}\big(\,\mbox{`$F^{(j)}(t)$'}\big)\,\hat{F}(t)\,\hat{z}\big(\,\mbox{`$F^{(j)}(t)$'}\big).
\label{eeMeasClassical12}
\end{align}
These equations show that, in the classical approximation, we need to keep the precision $\delta{F}$ of proposition representation low enough so that projection operator $\zop{\text{'$F^{(j)}(t)$'}}{big}$ does not affect the time evolution of $\hat{\rho}\big(\,t\,\big|\,\mbox{`$F^{(j)}(t)$'}\,\big)$ and $\hat{F}(t)$. 
It should also be noted that in the classical approximation, we have
\begin{align}
& \xjvar{`$\occupation{\stateset{{\mathrm{C}}^{(j)}_{\Omega}(t)}}{},t$'}{big} 
= \mathrm{Tr}\Big[\,\xop{`$\occupation{\stateset{{\mathrm{C}}^{(j)}_{\Omega}(t)}}{}$'}{big} \hat{\rho}\big(\,t\,\big|\,\mbox{`$F^{(j)}(t)$'}\,\big)\,\Big]
= 1
,\quad \forall j,
\label{eeMeasClassical13}
\end{align}
where $|\Omega^{(j)}(t)\rangle$ is the stationary state of the entire system occupied at time $t$.
$\stateset{{\mathrm{C}}^{(j)}_{\Omega}(t)}$ is the set of stationary states containing $|\Omega^{(j)}(t)\rangle$.
Combining Eq.\eqref{eeMeasClassical11} and Eq.\eqref{eeMeasClassical13}, we find that the characteristic operator in the Heisenberg picture satisfies
\begin{align}
\xop{`$\occupation{\stateset{{\mathrm{C}}^{(j)}_{\Omega}(t)}}{},t$'}{big} 
& = \hat{x}\big(\,\mbox{`$F^{(j)}(t)$'}\big)\,\xop{`$\occupation{\stateset{{\mathrm{C}}^{(j)}_{\Omega}(t)}}{},t$'}{big} \,\hat{x}\big(\,\mbox{`$F^{(j)}(t)$'}\big)
\label{eeMeasClassical14}\\
& = \hat{x}\big(\,\mbox{`$\occupation{\stateset{{\mathrm{C}}^{(j)}_{\Omega}(t)}}{},t$'$\wedge$`$F^{(j)}(t)$'}\big).
\label{eeMeasClassical15}
\end{align}

\subsubsection{Complementarity principle}
\label{sClassical3}
To derive the classical equations of motion, we need to specify the simultaneous values of position and momentum.
While this was not possible under the standard formulation, with the compatibility condition, it becomes possible by lowering the precision of proposition representation.
According to Eq.\eqref{eeMeasClassical15}, the classical approximation requires that proposition `$\occupation{\stateset{{\mathrm{C}}^{(j)}_{\Omega}(t)}}{}$' satisfy
\begin{align}
\mbox{`$\occupation{\stateset{{\mathrm{C}}^{(j)}_{\Omega}(t)}}{}$'}
& = \mbox{`$\occupation{\stateset{{\mathrm{C}}^{(j)}_{\Omega}(t)}}{}$'$\wedge$`${\bm r}^{(j)}(t)$'$\wedge$`${\bm p}^{(j)}(t)$'}
\label{eeMeasClassical21}\\
& = \mbox{`$\occupation{\stateset{{\mathrm{C}}^{(j)}_{\Omega}(t)}\,{\cap}\,\stateset{{\bm r}^{(j)}(t)}\,{\cap}\,\stateset{{\bm p}^{(j)}(t)}}{}$'},
\label{eeMeasClassical22}
\end{align}
which can be achieved by setting the precision of position $\delta{\bm r}$ and the precision of momentum $\delta{\bm p}$ so that
\begin{align}
\stateset{{\mathrm{C}}^{(j)}_{\Omega}(t)} \subseteq \stateset{{\bm r}^{(j)}(t)} \cap \stateset{{\bm p}^{(j)}(t)}.
\label{eeMeasClassical24}
\end{align}
If we define the variance of a physical quantity $F$ as
\begin{align}
{\mathrm{Var}}_{F}(t) \equiv \mathrm{Tr}\left[\left( \hat{F} - \Big\langle{F}(t)\Big|\!\Big|\mbox{`${\bm r}^{(j)}(t)$'$\wedge$`${\bm p}^{(j)}(t)$'}\Big\rangle\right)^{2} \hat{\rho}\big(\,t\,\big|\,\mbox{`${\bm r}^{(j)}(t)$'$\wedge$`${\bm p}^{(j)}(t)$'}\,\big)\right],
\label{eeComplementarity4}
\end{align}
Eq.\eqref{eeMeasClassical24} should hold if
\begin{align}
\delta{r}_{\xi} \gg \sqrt{{\mathrm{Var}}_{r_{\xi}}(t)}
\quad\mbox{and}\quad
\delta{p}_{\xi} \gg \sqrt{{\mathrm{Var}}_{p_{\xi}}(t)}
,\quad \xi \in \{x,y,z\}.
\label{eeComplementarity6}
\end{align}
Meanwhile, the Robertson inequality~\cite{Robertson} in quantum theory guarantees that
\begin{align}
\sqrt{{\mathrm{Var}}_{F}(t)}
\sqrt{{\mathrm{Var}}_{G}(t)}
\geqq \frac{1}{2}\,\Bigg| {\mathrm{Tr}}\left[\,\left[\hat{F},\hat{G}\right]_{-}\hat{\rho}\big(\,t\,\big|\,\mbox{`${\bm r}^{(j)}(t)$'$\wedge$`${\bm p}^{(j)}(t)$'}\,\big)\right] \Bigg|,
\label{eeComplementarity7}
\end{align}
indicating that the classical approximation is valid when
\begin{align}
\delta{r}_{\xi}\,\delta{p}_{\xi} 
\gg \frac{1}{2}\,\Bigg|\,{\mathrm{Tr}}\left[\,\left[\hat{r}_{\xi},\hat{p}_{\xi}\right]_{-}\hat{\rho}\big(\,t\,\big|\,\mbox{`${\bm r}^{(j)}(t)$'$\wedge$`${\bm p}^{(j)}(t)$'}\,\big)\right] \Bigg|
= \frac{\hbar}{2}
,\quad \xi \in \{x,y,z\}
\label{eeComplementarity8}
\end{align}
Eq.\eqref{eeComplementarity8} presents the precision limit of classical approximation in individual trials, an explicit expression of the complementarity principle~\cite{Messiah}.
Note, however, that we have not considered the possibility of squeezing~\cite[Chap.21]{Mandel} here.

\subsubsection{Classical equations of motion}
\label{sEqMotion}
Next, we consider cases where instantaneous values of canonical variables and canonical momenta can be defined at closely spaced time intervals based on the classical approximation.
The density operator $\hat{\rho}\big(\,t\,\big|\,\mbox{`${\bm r}^{(j)}(t)$'$\wedge$`${\bm p}^{(j)}(t)$'}\,\big)$ in the classical approximation is discontinuously updated by Eq.\eqref{eeMeasClassical7} at the moment new measurement values are obtained, and it changes continuously according to the quantum equation of motion between measurements. 
However, the projection operator used to update the ensemble is non-sharp, as can be seen from Eqs.\eqref{eeMeasClassical11}--\eqref{eeMeasClassical12}, so the time evolution of the density operator is not affected by the ensemble update operation. 
Consequently, the density operator in the classical approximation changes according to the quantum equation of motion at all times, and according to Ehrenfest's theorem~\cite[Chap.6]{Messiah}\cite{BornHeisenberg1925}, the following classical equations of motion hold:
\begin{align}
& \frac{d}{dt}q^{(j)}_{l}(t)
= + \Bigg\langle \frac{\partial \hat{H}}{\partial \hat{p}_{l}} \,\Bigg|\!\Bigg|\,\bigwedge^{L}_{l=1}\mbox{`${q}^{(j)}(t)$'$\wedge$`${p}^{(j)}(t)$'}\Bigg\rangle
,\quad
1\,{\leqq}\,l\,{\leqq}\,L
,\quad \forall j,
\label{eeEqMotion1}\\
& \frac{d}{dt}p^{(j)}_{l}(t)
= - \Bigg\langle \frac{\partial \hat{H}}{\partial \hat{q}_{l}} \,\Bigg|\!\Bigg|\,\bigwedge^{L}_{l=1}\mbox{`${q}^{(j)}(t)$'$\wedge$`${p}^{(j)}(t)$'}\Bigg\rangle
,\quad
1\,{\leqq}\,l\,{\leqq}\,L
,\quad \forall j.
\label{eeEqMotion2}
\end{align}
Writing these equations explicitly for the position ${\bm r}_{k}(t)$ and momentum ${\bm p}_{k}(t)$ of charged particles ($1\,{\leqq}\,k\,{\leqq}\,K$) gives the classical equations of motion for charged particles:
\begin{align}
{\bm v}^{(j)}_{k}(t)
& = \frac{d}{dt}{\bm r}^{(j)}_{k}(t)
,\quad \forall j,
\label{eeEqMotion3}\\
m_{k}\frac{d}{dt}{\bm v}^{(j)}_{k}(t)
& = {\bm f}^{(j)}_{\mathrm{Lorentz},k}(t) + {\bm f}^{(j)}_{\mathrm{spin},k}(t)
,\quad \forall j.
\label{eeEqMotion4}
\end{align}
Here, we defined the velocity of the $k$-th charged particle as
\begin{align}
\hat{\bm v}_{k} 
& \equiv \frac{1}{m_{k}}\left(\hat{\bm p}_{k} - e_{k}{\hat{\bm A}_{\mathrm{T}}}(\hat{\bm r}_{k})\right),
\label{eeEqMotion5}\\
{\bm v}^{(j)}_{k}(t)
& \equiv \frac{1}{m_{k}}\left({\bm p}^{(j)}_{k}(t) - e_{k}{{\bm A}^{(j)}_{\mathrm{T}}}({\bm r}^{(j)}_{k}(t),t)\right),
\label{eeEqMotion6}
\end{align}
and defined the Lorentz force and spin force acting on the $k$-th charged particle by the following equations:
\begin{align}
{\bm f}^{(j)}_{\mathrm{Lorentz},k}(t) 
& = {\rm{Tr}}\Bigg[\,{e_{k}} \Bigg(
  {\hat{\bm E}}_{\mathrm{T}}(\hat{\bm r}_{k})
  - \sum^{K}_{k'=1} \frac{e_{k'}}{4\pi\varepsilon_{0}}\frac{\hat{\bm r}_{k} - \hat{\bm r}_{k'}}{\left\|\hat{\bm r}_{k}-\hat{\bm r}_{k'}\right\|^{3}}
  + \hat{\bm v}_{k} \cdot {\hat{\bm B}}(\hat{\bm r}_{k})\Bigg)
  \hat{\rho}\big(\,t\,\big|\,\mbox{`$\mathrm{history}^{(j)}_{t}$'}\,\bigg)\,\Bigg],
\label{eeEqMotion7}\\
{\bm f}^{(j)}_{\mathrm{spin},k}(t) 
& = {\rm{Tr}}\Bigg[\,\Bigg(\frac{g_{k}|e_{k}|}{2m_{k}}
\frac{\partial}{\partial\hat{\bm r}_{k}}\left({\hat{\bm s}}_{k}\cdot\hat{\bm B}(\hat{\bm r}_{k})\right)\Bigg)\,\hat{\rho}\big(\,t\,\big|\,\mbox{`$\mathrm{history}^{(j)}_{t}$'}\,\big)\,\Bigg].
\label{eeEqMotion8}
\end{align}
If we set the time of the first measurement (i.e., the time when water droplets first form in the cloud chamber) as $t\,{=}\,0$, then for $t\,{>}\,0$, the classical equations of motion hold due to continuous measurements (i.e., generation of localized stationary states), so there is no ``spooky action at a distance.'' 
On the other hand, for $t\,{<}\,0$, charged particles may be in an entangled state, and at $t\,{=}\,0$, deterministic relation may be formed between the positions and momenta of different charged particles (Supplementary Material B).
This change at $t=0$ is abrupt in the sense that predictions based on the continuity of physical quantities are not effective.
Still, there is no ``spooky action at a distance'' because there are no comparable measurement values for $t\,{\leqq}\,0$. 
It is worth noting that the existence of deterministic relations between instantaneous values of position or momentum in individual trials is an experimental fact confirmed by Compton scattering experiments~\cite{Compton1925,Bothe1,Bothe2}.
\par
According to Ehrenfest's theorem, the classical equations of motion for the radiation field take the following form
\begin{align}
\frac{\partial}{\partial{t}}{\bm A}^{(j)}_{\mathrm{T}}({\bm x},t)
& = - {\bm E}^{(j)}_{\mathrm{T}}({\bm x},t),
\label{eeEqMotion11}\\
\frac{\partial}{\partial{t}}{\bm B}^{(j)}({\bm x},t)
& = - {\bm \nabla}\times{\bm E}^{(j)}_{\mathrm{T}}({\bm x},t),
\label{eeEqMotion12}\\
\frac{\partial}{\partial{t}}{\bm E}^{(j)}_{\mathrm{T}}({\bm x},t)
& = \frac{1}{c^{2}} {\bm \nabla} \times {\bm B}^{(j)}({\bm x},t)
- \frac{1}{c^{2}} {\bm J}^{(j)}_{\mathrm{T}}({\bm x},t),
\label{eeEqMotion13}
\end{align}
where we defined the instantaneous values of magnetic flux density ${\bm B}$ and the transverse component of current density ${\bm J}_{\mathrm{T}}$ by
\begin{align}
{\bm B}^{(j)}({\bm x},t)
& \equiv {\bm \nabla} \times {\bm A}^{(j)}_{\mathrm{T}}({\bm x},t),
\label{eeEqMotion14}\\
{\bm J}^{(j)}_{\mathrm{T}}({\bm x},t)
& \equiv \iiint {\bm \delta}^{{\mathrm{T}}}({\bm x}-{\bm x}')\cdot{\bm J}^{(j)}({\bm x}',t) d{\bm x}'.
\label{eeEqMotion15}
\end{align}
Also, the instantaneous values of current density ${\bm J}$ and charge density ${\sigma}$ are defined as
\begin{align}
{\bm J}^{(j)}({\bm x},t)
& \equiv {\mathrm{Tr}}\Bigg[ \left(\sum^{K}_{k=1} e_{k}\frac{\hat{\bm v}_{k} \delta({\bm x}-\hat{\bm r}_{k}) + \delta({\bm x}-\hat{\bm r}_{k}) \hat{\bm v}_{k}}{2} \right) \hat{\rho}\big(\,t\,\big|\,\mbox{`$\mathrm{history}^{(j)}_{t}$'}\,\big)\Bigg],
\label{eeEqMotion16}\\
\sigma^{(j)}({\bm x},t)
& \equiv {\mathrm{Tr}}\Bigg[ \left(\sum^{K}_{k=1} e_{k}\delta({\bm x}-\hat{\bm r}_{k}) \right) \hat{\rho}\big(\,t\,\big|\,\mbox{`$\mathrm{history}^{(j)}_{t}$'}\,\big)\Bigg].
\label{eeEqMotion17}
\end{align}
Combining these equations with the definition of the longitudinal component of the electric field ${\bm E}_{\mathrm{L}}$ and the electric field ${\bm E}$, namely
\begin{align}
{\bm \nabla}\cdot{\bm E}^{(j)}_{\mathrm{L}}({\bm x},t)
& = \frac{1}{\varepsilon_{0}}\sigma^{(j)}({\bm x},t)
,\quad
{\bm \nabla}\times{\bm E}^{(j)}_{\mathrm{L}}({\bm x},t)
= 0,
\label{eeEqMotion18}\\
{\bm E}^{(j)}({\bm x},t)
& \equiv {\bm E}^{(j)}_{\mathrm{T}}({\bm x},t) + {\bm E}^{(j)}_{\mathrm{L}}({\bm x},t),
\label{eeEqMotion19}
\end{align}
we obtain the Maxwell equations of classical electromagnetism.
It is worth noting that, according to the above formulation, the classical equation of motion holds for individual trials before statistical averaging.
Note also that when describing the double-slit experiment, the discreteness of individual events (i.e. the particle nature of light) is explained by the discrete response of the photodetector, and interference and diffraction (i.e. the wave nature of light) are explained by solving the field equations~\cite{Theory2}.

\subsubsection{Quantum fluctuations}
\label{sFluctuations}
When a classical approximation $F^{(j)}(t)$ exists for the instantaneous value of $F$, the instantaneous value of the quantum fluctuation of $F$ can be defined as
\begin{align}
\varDelta{F}^{(j)}(t) 
\equiv F^{(j)}(t) - \langle F(t) \rangle
,\quad \forall j.
\label{eeFluctuation1}
\end{align}
Here, $\langle F(t) \rangle$ is the expectation value of $F$ for the initial ensemble, while $F^{(j)}(t)$ is the instantaneous value as well as the expectation value of $F$ for the updated ensemble (Eq.\eqref{eeMeasClassical5}).
This highlights the importance of distinguishing between the two different statistical ensemble.
The power fluctuation can then be expressed as
\begin{align}
\big\langle{F}(t)^{2}\big\rangle
= \big| \langle{F}(t)\rangle \big|^{2}
+ \langle\varDelta{F}(t)^{2}\rangle,
\label{eeFluctuation5}
\end{align}
which has traditionally been interpreted as the wave and particle components of light according to light quantum hypothesis.
On the other hand, in our theory, the terms on the right-hand side merely represent the contributions from the expected value and the deviation from it.

\subsection{Bell's theorem and CHSH inequality}
\label{sBellCHSH}
\subsubsection{Deterministic relations between spins of subsystems}
\label{sBellCHSH1}
We now focus on two spin-1/2 systems in an entangled state to explain the relationship between our theory and well-known no-go theorems.
First, to summarize our theory's results, we denote the stationary state of the $k$-th subsystem ($k\,{\in}\,\{1,2\}$) as $|{\bm u}_{k},s_{k}\rangle^{\!k}$, with ${\bm u}_{k}$ as the unit vector of the measurement direction and $s_{k}\,{\in}\,\{\uparrow,\downarrow\}$ as the spin orientation. 
For $t\,{>}\,0$, when interactions can be ignored, we can define logical variables for the subsystems as
\begin{align}
\xjvar{`$\occupation{\state{{\bm u}_{k},s_{k}}{k}}{},t$'}{Big}
= \begin{cases}
1, & \mbox{$|{\bm u}_{k},s_{k}\rangle^{\!k}$ is occupied at time $t$},\\
0, & \mbox{$|{\bm u}_{k},s_{k}\rangle^{\!k}$ is not occupied at time $t$}
\end{cases}
,~~ k \in \{1,2\},
\label{eeExample4-1}
\end{align}
and the characteristic and partial characteristic operators take the following form:
\begin{align}
\xop{`$\occupation{\state{{\bm u}_{k},s_{k}}{k}}{}$'}{big} 
= \zop{`$\occupation{\state{{\bm u}_{k},s_{k}}{k}}{}$'}{big} 
= |{\bm u}_{k},s_{k}\rangle^{\!k}{}^{k}\!\langle {\bm u}_{k},s_{k}|
,\quad k \in \{1,2\}.
\label{eeExample4-2}
\end{align}
It follows that compatibility conditions hold between the following sets of propositions:
\begin{align}
\big\{
\mbox{`$\occupation{\state{{\bm u}_{1},\uparrow}{1}}{}$'},~
\mbox{`$\occupation{\state{{\bm u}_{1},\downarrow}{1}}{}$'},~
\mbox{`$\occupation{\state{{\bm u}_{2},\uparrow}{2}}{}$'},~
\mbox{`$\occupation{\state{{\bm u}_{2},\downarrow}{2}}{}$'}
\big\}
,\quad \forall {\bm u}_{1}
,\quad \forall {\bm u}_{2}.
\label{eeExample4-3}
\end{align}
Particularly, if the system is known to be in a singlet state, namely
\begin{align}
| 0 \rangle^{\!\mathrm{S}}
= \frac{1}{\sqrt{2}}\Big(\,
|{\bm u},\uparrow\rangle^{\!1} \otimes |{\bm u},\downarrow\rangle^{\!2}
- |{\bm u},\downarrow\rangle^{\!1} \otimes |{\bm u},\uparrow\rangle^{\!2}
\,\Big)
,\quad
{\bm u}\cdot{\bm u} = 1
,\quad
{\bm u} \in {\mathbb{R}}^{3},
\label{eeExample4-4}
\end{align}
then due to the general properties of entangled subsystems ({\S}\ref{sEntanglement}), the following deterministic relations exist when the measurement direction ${\bm u}$ is common:
\begin{align}
\xjvar{`$\occupation{\state{{\bm u},\uparrow}{2}}{},t$'}{big}
& = \xjvar{`$\occupation{\state{{\bm u},\downarrow}{1}}{},t$'}{big},
\label{eeExample4-5}\\
\xjvar{`$\occupation{\state{{\bm u},\downarrow}{2}}{},t$'}{big}
& = \xjvar{`$\occupation{\state{{\bm u},\uparrow}{1}}{},t$'}{big}.
\label{eeExample4-5a}
\end{align}
The formula for the joint probability ({\S}\ref{sEntanglement}) is the same as in quantum theory~\cite[Eq.(7.39)]{Peres}:
\begin{align}
{\mathrm{Pr}}\big(\mbox{`$\occupation{\state{{\bm u}_{1},s_{1}}{1}}{},t$'}\,{\wedge}\,\mbox{`$\occupation{\state{{\bm u}_{2},s_{2}}{2}}{},t$'}\big)
& \equiv \big\langle \xvar{`$\occupation{\state{{\bm u}_{1},s_{1}}{1}}{},t$'$\wedge$`$\occupation{\state{{\bm u}_{2},s_{2}}{2}}{},t$'}{big} \big\rangle
\label{eeExample4-6}\\
& = {}^{\mathrm{S}}\!\langle\,0\,|\,\xop{`$\occupation{\state{{\bm u}_{1},s_{1}}{1}}{}$'}{big}\,\xop{`$\occupation{\state{{\bm u}_{2},s_{2}}{2}}{}$'}{big}\,| 0 \rangle^{\!\mathrm{S}}
\label{eeExample4-8}\\
& = \frac{1 - {\bm u}_{1}\cdot{\bm u}_{2}}{4}.
\label{eeExample4-9}
\end{align}

\subsubsection{Relation to Bell's Inequality}
As is well known, Bell's paper on hidden variable theories addressed a model of local physical reality where the joint probability $P({\bm a}, {\bm b})$ is determined by~\cite[Eq.(2)]{Bell}
\begin{align} 
P({\bm a}, {\bm b})
\equiv \int A({\bm a},\lambda)\,B({\bm b},\lambda)\,\rho_{\lambda} d\lambda
,~~ \int\rho_{\lambda}d\lambda = 1
,~~ \begin{cases}
A({\bm a},\lambda) \in \{-1,+1\},\\
B({\bm b},\lambda) \in \{-1,+1\}.
\end{cases}
\label{eeExample4-41}
\end{align}
Here, $\lambda$ represents hidden variables, $\rho_{\lambda}$ is the probability density function of hidden variables, and $A({\bm a},\lambda)$ and $B({\bm b},\lambda)$ represent instantaneous spin values when measurement directions are ${\bm a}$ and ${\bm b}$, and the hidden variable is $\lambda$. 
Bell's inequality~\cite[Eq.(15)]{Bell}, namely
\begin{align} 
1 + P({\bm b}, {\bm c})
\geqq \left| P({\bm a}, {\bm b}) - P({\bm a}, {\bm c})\right|,
\label{eeExample4-42}
\end{align}
was derived from Eq.\eqref{eeExample4-41}, deliberately ignoring standard formulas in quantum theory, to elucidate the properties of hidden variables. 
In contrast, as our theory's joint probability equations are the same as quantum theory's, our theory is not a hidden variable theory of the type Bell assumed. 
Therefore, the violation of Bell's inequality is irrelevant to the validity of our theory.

\subsubsection{Relation to CHSH inequality}
The derivation of the CHSH inequality~\cite{CHSH} pointed out that assuming definite spin values in individual measurements leads to contradictions. 
If we define spin variables that take either $+1$ or $-1$ for measurement directions ${\bm a}$, ${\bm b}$, ${\bm c}$, ${\bm d}$ as
\begin{align}
& A^{(j)}({\bm a}) = 2 \xjvar{`$\occupation{\state{{\bm a},\uparrow}{{1}}}{}$'}{big} - 1
,\quad
C^{(j)}({\bm c}) = 2 \xjvar{`$\occupation{\state{{\bm c},\uparrow}{{1}}}{}$'}{big} - 1,
\label{eeExample4-43}\\
& B^{(j)}({\bm b}) = 2 \xjvar{`$\occupation{\state{{\bm b},\uparrow}{{2}}}{}$'}{big} - 1
,\quad
  D^{(j)}({\bm d}) = 2 \xjvar{`$\occupation{\state{{\bm d},\uparrow}{{2}}}{}$'}{big} - 1,
\label{eeExample4-44}
\end{align}
and forget about the constraints due to the compatibility condition, we can show that~\cite[{\S}2.6]{Nielsen2000}
\begin{align}
A^{(j)}({\bm a}) B^{(j)}({\bm b}) + B^{(j)}({\bm b}) C^{(j)}({\bm c}) + C^{(j)}({\bm c}) D^{(j)}({\bm d}) - D^{(j)}({\bm d}) A^{(j)}({\bm a})
= \pm 2
,~\forall{j}.
\label{eeExample4-45}
\end{align}
This led to the CHSH inequality~\cite{CHSH}:
\begin{align}
\Big| \big\langle A({\bm a}) B({\bm b}) \big\rangle
+ \big\langle B({\bm b}) C({\bm c}) \big\rangle
+ \big\langle C({\bm c}) D({\bm d}) \big\rangle
- \big\langle D({\bm d}) A({\bm a}) \big\rangle \Big|
\leqq 2
\label{eeExample4-46}
\end{align}
On the other hand, when
\begin{align}
{\bm a} = {\bm e}_{z}
,\quad
{\bm b} = -\frac{{\bm e}_{z} + {\bm e}_{x}}{\sqrt{2}}
,\quad
{\bm c} = {\bm e}_{x}
,\quad 
{\bm d} = \frac{{\bm e}_{z} - {\bm e}_{x}}{\sqrt{2}},
\label{eeExample4-48}
\end{align}
the joint probability formula of quantum theory (Eq.\eqref{eeExample4-9}) gives
\begin{align}
\big\langle A({\bm a}) B({\bm b}) \big\rangle
+ \big\langle B({\bm b}) C({\bm c}) \big\rangle
+ \big\langle C({\bm c}) D({\bm d}) \big\rangle
- \big\langle D({\bm d}) A({\bm a}) \big\rangle
= 2\sqrt{2},
\label{eeExample4-47}
\end{align}
contradicting Eq.\eqref{eeExample4-46}.
\par
In standard formulation of quantum theory, this contradiction was explained as arising from assuming integer spin values ($+1$ or $-1$). 
However, our theory suggests the true cause of contradiction is considering arithmetic operations that do not satisfy compatibility conditions. 
Indeed, using the expression of partial characteristic operator in Eq.\eqref{eeExample4-2}, we find compatibility conditions hold for combinations
\begin{align}
& \big\{ \mbox{`$\occupation{\state{{\bm a},\uparrow}{{1}}}{}$'},~
\mbox{`$\occupation{\state{{\bm b},\uparrow}{{2}}}{}$'} \big\}
,\quad
\big\{ \mbox{`$\occupation{\state{{\bm b},\uparrow}{{2}}}{}$'},~
\mbox{`$\occupation{\state{{\bm c},\uparrow}{{1}}}{}$'} \big\},
\label{eeExample4-51}\\
& \big\{ \mbox{`$\occupation{\state{{\bm c},\uparrow}{{1}}}{}$'},~
\mbox{`$\occupation{\state{{\bm d},\uparrow}{{2}}}{}$'} \big\}
,\quad
\big\{ \mbox{`$\occupation{\state{{\bm d},\uparrow}{{2}}}{}$'},~
\mbox{`$\occupation{\state{{\bm a},\uparrow}{{1}}}{}$'} \big\},
\label{eeExample4-52}
\end{align}
but not for
\begin{align}
& \big\{ \mbox{`$\occupation{\state{{\bm a},\uparrow}{{1}}}{}$'},~
\mbox{`$\occupation{\state{{\bm c},\uparrow}{{1}}}{}$'} \big\}
,\quad
\big\{ \mbox{`$\occupation{\state{{\bm b},\uparrow}{{2}}}{}$'},~
\mbox{`$\occupation{\state{{\bm d},\uparrow}{{2}}}{}$'} \big\}.
\label{eeExample4-53}
\end{align}
Therefore, while the multiplication of spin variables, namely
\begin{align}
A({\bm a})B({\bm b})
,\quad
B({\bm b})C({\bm c})
,\quad
C({\bm c})D({\bm d})
,\quad
D({\bm d})A({\bm a}),
\label{eeExample4-54}
\end{align}
can be defined without issue, the following addition violates compatibility conditions and cannot be defined:
\begin{align}
A({\bm a})B({\bm b})
+
B({\bm b})C({\bm c})
+
C({\bm c})D({\bm d})
+
D({\bm d})A({\bm a}).
\label{eeExample4-55}
\end{align}
In a nutshell, the CHSH inequality does not hold in our theory because the compatibility condition restricts arithmetic operations between instantaneous values.
Similarly, assuming compatibility conditions resolves other contradictions related to measurement contextuality, though we omit detailed analysis of individual cases for brevity.

\subsection{Issues with the Light Quantum Hypothesis}
\label{sLightQuantum}
Modern quantum optics theory explains the statistical properties of radiation fields solely through the quantization of atomic systems and radiation fields.
From a modern perspective, the most compelling evidence for the existence of photons is the observation of discontinuous individual events and their deterministic relations (See supplementary material).
Considering this, the essence of the light quantum hypothesis can be expressed in the following deterministic relation~\cite{Einstein1905}:
\begin{align}
\xjvar{`absorbed,\,$t$'}{big} = \xjvar{`ionized,\,$t$'}{big}
,\quad \forall j,
\label{eeLightQuantum1}
\end{align}
where `ionized,\,$t$' represents a proposition of the atomic system meaning `the atom is photoionized at time $t$', and `absorbed,\,$t$' represents a proposition of the radiation field meaning `a photon is absorbed by the atomic system at time $t$'. 
As demonstrated in Supplementary Material B, the proposition `ionized,\,$t$' can accurately reproduce the standard formula for light detection probability when defined as:
\begin{align}
\mbox{`ionized,\,$t$'}
= \mbox{`$\occupation{\stateset{+}}{},t$'$\wedge$`$\occupation{\stateset{-}}{},0$'}
,\quad t\,{>}\,0.
\label{eeLightQuantum2}
\end{align}
Here, $\stateset{-}$ and $\stateset{+}$ are the sets of all bound states and scattering states of the atomic system, respectively. 
However, for the proposition `absorbed,\,$t$', no representation of a partial characteristic operator consistent with quantum optical theory exists, highlighting a limitation of the light quantum hypothesis from our perspective.
\par
To illustrate this more concretely, let us first consider the case where the radiation field is initially in the $n\,{=}\,1$ Fock state $|1\rangle$. 
In this case, we define the proposition `absorbed,\,$t$' as
\begin{align}
\mbox{`absorbed,\,$t$'}
\equiv \mbox{`$\occupation{\state{\mbox{vac}}{}}{},t$'${\wedge}$`$\occupation{\state{1}{}}{},0$'}
,\quad t\,{>}\,0,
\label{eeLightQuantum3}
\end{align}
We can then show that
\begin{align}
{\mathrm{Tr}}\big[\,\xop{`absorbed,\,$t$'}{big}\,\hat{\rho}(\mbox{`${\mathrm{A}}_{0},t_{0}$'})\,\big]
= {\mathrm{Tr}}\big[\,\xop{`ionized,\,$t$'}{big}\,\hat{\rho}(\mbox{`${\mathrm{A}}_{0},t_{0}$'})\,\big],
\label{eeLightQuantum3a}
\end{align}
using the relations
\begin{align}
\xop{`absorbed,\,$t$'}{big}
& =  | 1 \rangle\langle 1 | \,\hat{U}(t)\, | \mbox{vac} \rangle\langle \mbox{vac} | \,\hat{U}(t)\, | 1 \rangle\langle 1 |,
\label{eeLightQuantum5}\\
\hat{\rho}(\mbox{`${\mathrm{A}}_{0},t_{0}$'})
& = |{\mathrm{1s}}\rangle\langle {\mathrm{1s}}| \otimes |1\rangle\langle 1|.
\label{eeLightQuantum4}
\end{align}
Here, $|\mbox{vac}\rangle$ denotes the vacuum state of the radiation field, $|{\mathrm{1s}}\rangle$ the bound state of the atomic system, and $\hat{U}(t)$ the time evolution operator (Supplementary Material B). 
This result (Eq.\eqref{eeLightQuantum3a}) aligns with the predictions of the light quantum hypothesis (Eq.\eqref{eeLightQuantum1}).
Then, by first-order perturbation calculation similar to that shown in Supplementary Material B, we can show that
\begin{align}
& {\mathrm{Tr}}\big[\,\xop{`absorbed,\,$t$'}{big}\,\hat{\rho}(\mbox{`${\mathrm{A}}_{0},t_{0}$'})\,\big]
= {\mathrm{Tr}}_{\bm r}\big[\,
\langle\psi(t)|\mbox{vac}\rangle\langle\mbox{vac}| \psi(t)\rangle\,\big],
\label{eeLightQuantum6}\\
& |\psi(t)\rangle
\equiv \hat{U}(t) \big|{\mathrm{1s}}\big\rangle|1\rangle
\cong c_{\mathrm{1s}}(t) \big|{\mathrm{1s}}\big\rangle\big|1\big\rangle
+ \iiint c_{{\bm p}}(t) \big|{\bm p}\big\rangle|\mbox{vac}\rangle\,d{\bm p},
\label{eeLightQuantum7}
\end{align}
where ${\mathrm{Tr}}_{\bm r}$ is the trace operation for the atomic system, $|\psi(t)\rangle$ is the state vector of the atomic system, and $|{\bm p}\rangle$ is the scattering state of the atomic system.
From the right-hand side of Eq.\eqref{eeLightQuantum7}, we can see that the atomic system and radiation field are entangled at $t\,{>}\,0$ when the radiation field is initially in the $n\,{=}\,1$ Fock state.
Additionally, due to the general properties of entangled subsystems ({\S}\ref{sEntanglement}), there exist a deterministic relation between the stationary states of the atomic system and the radiation field, which is the the light quantum hypothesis of Eq.\eqref{eeLightQuantum1}.
\par
On the other hand, when the radiation field is initially in a coherent state $|\tilde{\alpha}\rangle$, the state vector in Eq.\eqref{eeLightQuantum7} will have the following form:
\begin{align}
|\psi(t)\rangle
\equiv \hat{U}(t) \big|{\mathrm{1s}}\big\rangle |\tilde{\alpha}\rangle
\cong \left( c_{\mathrm{1s}}(t) \big|{\mathrm{1s}}\big\rangle + \iiint c_{{\bm p}}(t) \big|{\bm p}\big\rangle\,d{\bm p} \right) |\tilde{\alpha}\rangle.
\label{eeLightQuantum11}
\end{align}
In this case, the atomic system and radiation field are not entangled, and substituting Eq.\eqref{eeLightQuantum11} into equation Eq.\eqref{eeLightQuantum6} yields
\begin{align}
{\mathrm{Tr}}\big[\,\xop{`absorbed,\,$t$'}{big}\,\hat{\rho}(\mbox{`${\mathrm{A}}_{0},t_{0}$'})\,\big]
& = \big| \langle\mbox{vac}|\tilde{\alpha}\rangle \big|^{2}\,
\iiint \big| c_{{\bm p}}(t) \big|^{2}\,d{\bm p}
\label{eeLightQuantum13}\\
& = \big| \langle\mbox{vac}|\tilde{\alpha}\rangle \big|^{2}\,
{\mathrm{Tr}}\big[\,\xop{`ionized,\,$t$'}{big}\,\hat{\rho}(\mbox{`${\mathrm{A}}_{0},t_{0}$'})\,\big],
\label{eeLightQuantum14}
\end{align}
which contradicts Eq.\eqref{eeLightQuantum3a} as required by the light quantum hypothesis of Eq.\eqref{eeLightQuantum1}. 
Therefore, the definition of proposition 'absorbed,\,$t$' by Eq.\eqref{eeLightQuantum3} does not hold generally, and in this sense, the light quantum hypothesis is limited as a quantitative model of reality.

\section{Discussion}
\label{cDiscussion}
Given the pivotal role of stationary states in our theory, it is imperative to elucidate the unresolved issues pertaining to their definition. 
We begin by expounding on the rationale behind the definition of stationary states as delineated in Assumptions B1--B14.

\paragraph*{Rationale for the definition of stationary states}
In this paper, we have defined stationary states using the non-interacting Hamiltonian:
\begin{align}
\hat{H}^{{\mathsf{U}}}_{0} 
\equiv {\hat{\mathsf{U}}}^{\dagger}\,\left(\hat{H}_{\mathrm{A}} + \hat{H}_{\mathrm{F}} \right) {\hat{\mathsf{U}}},
\label{eeDiscussion1-1}
\end{align}
where $\hat{H}_{\mathrm{A}}$ and $\hat{H}_{\mathrm{F}}$ represent the Hamiltonians of the atomic system and radiation field, respectively, and $\hat{\mathsf{U}}$ denotes a time-independent unitary transformation.
Our decision to eschew the total Hamiltonian:
\begin{align}
\hat{H}^{{\mathsf{W}}} 
\equiv {\hat{\mathsf{W}}}^{\dagger}\,\left(\hat{H}_{\mathrm{A}} + \hat{H}_{\mathrm{F}} + \hat{H}_{\mathrm{I}} \right) {\hat{\mathsf{W}}}
\label{eeDiscussion1-2}
\end{align}
in defining stationary states is predicated on the experimental findings of Dehmelt et al.~\cite{QuantumJump1}, which demonstrated that the frequency of quantum jumps approximates the inverse of the radiative lifetime. 
This experimental observation aligns with the definition of stationary states as eigenstates of $\hat{H}^{{\mathsf{U}}}_{0}$, but is incongruent with a definition based on the eigenstates of $\hat{H}^{{\mathsf{W}}}$.
The inclusion of the radiation field Hamiltonian $\hat{H}_{\mathrm{F}}$ in Eq.\eqref{eeDiscussion1-1} is motivated by the experiments of Haroche et al.~\cite{Haroche1}, which observed quantum jumps between Fock states of radiation fields. 
Furthermore, it has become apparent that representing light with well-defined phase (e.g., weak laser light) through the occupation of a finite set of Fock states is problematic. 
Consequently, the definition of stationary states must encompass quantum states with small phase uncertainty, such as coherent states. 
This necessitates the incorporation of the time-independent unitary transformation ${\hat{\mathsf{U}}}$ in Eq.\eqref{eeDiscussion1-1}.

\paragraph*{Ambiguity in the definition of unitary transformation}
The introduction of the time-independent unitary transformation ${\hat{\mathsf{U}}}$ raises new questions regarding the permissible range of unitary transformations. 
In fact, for the purpose of including coherent states, it suffices to consider the vacuum displacement operator:
\begin{align}
{\hat{\mathsf{U}}}
= \prod_{{\bm k}\lambda} e^{\tilde{\alpha}^{*}_{{\bm k}\lambda} \hat{a}_{{\bm k}\lambda} - \tilde{\alpha}_{{\bm k}\lambda} \hat{a}^{\dagger}_{{\bm k}\lambda}}
,\quad \tilde{\alpha}_{{\bm k}\lambda}\,{\in}\,{\mathbb{C}},
\label{eeDiscussion1-3}
\end{align}
where $\hat{a}_{{\bm k}\lambda}$ and $\hat{a}^{\dagger}_{{\bm k}\lambda}$ are the ladder operators for the ${\bm k}\lambda$-mode of radiation fields.
However, the choice of ${\hat{\mathsf{U}}}$ depends on the choice of ${\hat{\mathsf{W}}}$ in Eq.\eqref{eeDiscussion1-2}. 
The selection of ${\hat{\mathsf{W}}}$ is intricately linked to subtle issues, including the choice between minimal coupling and multipolar Hamiltonians~\cite[{\S}14.1.3]{Mandel}, methods for incorporating radiative corrections~\cite{Cohen,Milonni1993,Hopfield1958}, and considerations related to gauge degrees of freedom.
Consequently, the extent to which the definition of ${\hat{\mathsf{U}}}$ should be expanded remains ambiguous.
Conversely, limiting the scope of ${\hat{\mathsf{U}}}$ effectively constrains the permissible models of reality, rendering the choice of ${\hat{\mathsf{U}}}$ physically significant. 
Therefore, in this paper, we have refrained from specifying the allowable ranges of ${\hat{\mathsf{U}}}$ and ${\hat{\mathsf{W}}}$, leaving their forms undetermined.

\paragraph*{Challenges in Relativistic Extension}
Our theory is further constrained by its non-relativistic nature. 
The concept of stationary states lacks relativistic invariance, and relativistic theories face the problem of gauge ghosts. 
Consequently, any attempt to extend our approach to a relativistic framework would likely necessitate fundamental modifications.
More specifically, our theory involves two types of time evolution: one based on the non-interacting Hamiltonian $\hat{H}^{{\mathsf{U}}}_{0}$ and another on the total Hamiltonian $\hat{H}^{{\mathsf{W}}}$.
However, at present, we lack a clear path to reformulate the former time evolution in a relativistic manner.
In this sense, our theory can be viewed as a phenomenological description focused solely on representing individual events observed in quantum optics experiments. 
It should be noted, however, that non-relativistic theories are not without merit, as evidenced by the fact that quantum optical theories can effectively account for the effects of propagation delay~\cite{Milonni2004}.
Furthermore, even in the non-relativistic case, the standard formulation fails to represent the world of reality (i.e., integer functions obtained in individual trials). 
Therefore, at this juncture, a phenomenological approach is undeniably necessary to represent empirical laws of reality.

\paragraph*{Insufficient information on the laws of reality}
A more fundamental issue is the current paucity of knowledge regarding the laws of reality.
As explained in Section \ref{cIntro}, the laws of reality envisioned by our theory are not differential equations for continuous functions, but rather constraints on propositions at different times or for different subsystems.
Examples of laws of reality include quantum jumps from bound states to scattering states in photoexcited atoms, the validity of classical equations of motion when representing charged particle motion at low precision, and the existence of deterministic relations in EPR-inspired experiments.
The light quantum hypothesis can also be considered a law of reality, albeit one that is valid only for specific initial conditions.
Whether there are other empirical laws of reality remains a matter for debate, and indeed, the utility of knowledge about these laws is yet to be fully understood.
Moreover, to represent the laws of reality, it is necessary to specify the stationary state of the entire system at each instant using a subspace $\stateset{{\mathrm{C}}_{\Omega}}$, with the condition ${\mathrm{C}}_{\Omega}$ not being uniquely determined.
Nevertheless, our theory suggests that it is possible to assess the validity of representations using compatibility conditions.
Therefore, we argue that a phenomenological hidden variable theory remains effective in representing experimental data of integer functions, even if it deviates from the standard formulation.

\section{Conclusions}
\label{cConclusion}
This paper focuses on propositions with probabilities predictable by quantum theory, formulating a hidden variable theory that describes the occupation of stationary states using Boolean logical variables.
A key feature of this theory is a novel selection rule for propositions, derived from the need to harmonize Boolean logic with quantum theory.
This selection rule extends the definition of joint probability, allowing deterministic behavior to be derived from the stochastic laws of quantum theory. 
Simultaneously, it avoids the constraints of major no-go theorems by restricting arithmetic operations between physical quantities.
In this theory, the most fundamental hidden variables are logical variables specifying the occupied stationary state $|\Omega^{(j)}(t)\rangle$ at each moment. 
However, such a sharp representation offers no special advantage due to the stochastic time variation of $|\Omega^{(j)}(t)\rangle$
On the other hand, in a non-sharp representation based on a set of stationary states $\stateset{{\mathrm{C}}_{\Omega}}$, logical variables may exhibit deterministic behavior, allowing deterministic relations to be established between logical variables of subsystems.
Significantly, these deterministic relations may represent laws of reality that the standard formulation of quantum theory cannot capture.
In describing the laws of reality, the crucial aspect is not the time variation of logical variables, but rather the interrelationships of conditions ${\mathrm C}_{\Omega}$ representing the status of subsystems.
Using propositions whose probabilities are predictable by quantum theory, it becomes possible to describe integer functions and physical quantities whose expectation values can be predicted by quantum theory.
In our theory, the existence of simultaneous values of physical quantities is determined not by the commutativity of operators, but by the compatibility condition.
We also defined proxy representations for the instantaneous values of physical quantities that cannot be measured without additional measuring instruments, demonstrating how subtle experimental facts about measurement contextuality are accounted for.
Based on these results, we argue that a hidden variable theory for non-relativistic QED is not a theory of 
point particles such as photons, but rather a theory of stationary states and individual events.
However, the definition of stationary states requires further investigation, limiting our theory to a phenomenological approach that aims solely to represent experimental data of integer functions.

\clearpage
\appendix
\addtocontents{toc}{\protect\setcounter{tocdepth}{1}}
\section{Issues with representing individual trials using density operators}
\label{sDensity}
In this section, we show that assuming binary logical variables and their expectation values can be represented by density operators leads to inconsistencies, even without assuming the differentiability of density operators.
To this end, consider the following set of assumptions:
\begin{enumerate}
\renewcommand{\labelenumi}{(\roman{enumi})~}
\item The occupation of a stationary state $|\phi\rangle$ can be represented by a binary logical variable:
\begin{align}
\xjvar{`$\occupation{\state{\phi}{}}{},t$'}{big}
& \equiv \begin{cases}
1, & \mbox{$\phi = \Omega^{(j)}(t)$}\\
0, & \mbox{$\phi \neq \Omega^{(j)}(t)$}
\end{cases}
,\quad 1 \leqq j \leqq N.
\label{eeParadox11}
\end{align}
\item The expectation value of the logical variable is defined in a statistical sense, so that
\begin{align}
\big\langle \xvar{`$\occupation{\state{\phi}{}}{},t$'}{big} \big\rangle 
= \lim_{N\to\infty}\frac{1}{N}\sum_{j=1}^{N} \xjvar{`$\occupation{\state{\phi}{}}{},t$'}{big}.
\label{eeParadox12}
\end{align}
\item The expectation value of the logical variable can be expressed using the density operator $\hat{\rho}(t)$ as follows:
\begin{align}
\big\langle \xvar{`$\occupation{\state{\phi}{}}{},t$'}{big} \big\rangle 
= {\rm{Tr}}\Big[\,\xop{`$\occupation{\state{\phi}{}}{}$'}{big}\,\hat{\rho}(t)\,\Big].
\label{eeParadox13}
\end{align}
\item The operator $\xop{`$\occupation{\state{\phi}{}}{}$'}{}$ is a projection operator onto the stationary state $|\phi\rangle$, namely
\begin{align}
\xop{`$\occupation{\state{\phi}{}}{}$'}{big} \equiv |\phi\rangle\langle\phi|.
\label{eeParadox14}
\end{align}
\item The logical variable can be represented using the density operator $\check{\rho}^{(j)}(t)$ for individual trials as
\begin{align}
\xjvar{`$\occupation{\state{\phi}{}}{},t$'}{big} 
= {\rm{Tr}}\Big[\,\xop{`$\occupation{\state{\phi}{}}{}$'}{big}\,\check{\rho}^{(j)}(t)\,\Big].
\label{eeParadox15}
\end{align}
\end{enumerate}
Here, $j$ is the trial number, $N$ is the total number of trials, and $\phi$ represents a complete set of quantum numbers of the stationary state. 
$|\Omega^{(j)}(t)\rangle$ is the stationary state occupied at time $t$ in the $j$-th trial and satisfies the normalization condition $\big|\langle\Omega^{(j)}(t)|\Omega^{(j)}(t)\rangle\big|\,{=}\,1$.
From assumptions (ii), (iii), (iv), and (v), for any $|\phi\rangle$, we have
\begin{align}
\left\langle\phi\,\Big|\,\hat{\rho}(t)\,\Big|\phi\right\rangle
= \lim_{N\to\infty} \left\langle\phi\,\Bigg|\frac{1}{N}\sum_{j=1}^{N} \,\check{\rho}^{(j)}(t)\,\Bigg|\phi\right\rangle.
\label{eeParadox16}
\end{align}
If the order of the inner product and the limit operations is interchangeable, we obtain
\begin{align}
\hat{\rho}(t)
= \lim_{N\to\infty}\frac{1}{N}\sum_{j=1}^{N} \check{\rho}^{(j)}(t).
\label{eeParadox17}
\end{align}
On the other hand, if we assume that the density operator $\hat{\rho}(t)$ is pure, i.e.,
\begin{align}
\hat{\rho}(t) = |\psi(t)\rangle\langle\psi(t)|
,\quad
\langle\psi(t)|\psi(t)\rangle = 1,
\label{eeParadox18}
\end{align}
then by substituting
\begin{align}
\check{\rho}^{(j)}(t) = |\Omega^{(j)}(t)\rangle\langle \Omega^{(j)}(t)|
\label{eeParadox19}
\end{align}
into Eq.\eqref{eeParadox17} and taking $\langle\psi(t)|\ldots|\psi(t)\rangle$ on both sides, we get
\begin{align}
1 = \lim_{N\to\infty}\frac{1}{N}\sum_{j=1}^{N} \left|\langle \psi(t) | \Omega^{(j)}(t)\rangle \right|^{2}.
\label{eeParadox21}
\end{align}
Since the term in the sum on the right-hand side has an absolute value of 1 or less by definition, Eq.\eqref{eeParadox19} implies 
\begin{align}
\left| \langle \psi(t) | \Omega^{(j)}(t)\rangle \right|^{2} = 1
,\quad \forall j.
\label{eeParadox22}
\end{align}
Therefore, if we simultaneously require assumptions (i)--(v), for a pure system, we obtain
\begin{align}
|\psi(t)\rangle = \exp({\mathrm{i}}\theta^{(j)})\,|\Omega^{(j)}(t)\rangle
,\quad \theta^{(j)} \in {\mathbb{R}}
,\quad\forall j.
\label{eeParadox23}
\end{align}
This conclusion contradicts the well-known fact that the state vector can be a superposition of stationary states, indicating that there must be an unreasonable assumption among (i)-(v).
\par
Note that there are two completely different interpretations of this fact.
According to the first interpretation that denies assumption (i), the occupation of stationary states cannot be represented by logical variables. 
This interpretation corresponds to the conventional explanation that regards Gleason's theorem as a no-go theorem for hidden variable theories~\cite{Peres}. 
On the other hand, the second interpretation that denies assumption (v), the occupation of stationary states can be described using logical variables if individual trials are not represented by density operators.
Interestingly, in the classical approximation ({\S}\ref{sClassical}), individual trials can be described by density operators ($\check{\rho}^{(j)}(t)\,{=}\,\hat{\rho}(t)$) because assumption (ii) is removed by updating the ensemble according to measurement results.

\section{Details of compatibility conditions}
\label{sAppModel}
\subsection{Propositions that are always true for specific ensembles}
\label{sAppendix100}
In this section, we show that compatibility conditions always hold between propositions that are true with 100\% probability.
For this purpose, in the definition of sets of stationary states,
\begin{align}
\stateset{{\mathrm{C}}_{i}}
\equiv \left\{\,|\phi_{i}\rangle\,\Big|
~\hat{H}_{0}|\phi_{i}\rangle\,{=}\,E(\phi_{i})|\phi_{i}\rangle
,~\langle \phi_{i} | \phi'_{i} \rangle\,{=}\,\delta_{\phi_{i},\phi'_{i}}
,~\mbox{$|\phi_{i}\rangle$ satisfy $\mathrm{C}_{i}$}
\right\}
,\quad 1\,{\leqq}\,i\,{\leqq}\,I,
\label{eeGroup1}
\end{align}
we consider the case where (i) different sets of stationary states are not necessarily orthogonalized (i.e., not necessarily $\langle \phi_{i} | \phi_{i'} \rangle\,{=}\,\delta_{\phi_{i},\phi_{i'}}$ when $|\phi_{i}\rangle\,{\in}\,\stateset{{\mathrm{C}}_{i}}$ and $|\phi_{i'}\rangle\,{\in}\,\{\!\!\{{\mathrm{C}}_{i'}\}\!\!\}$), and (ii) all $\{\!\!\{{\mathrm{C}}_{i}\}\!\!\}$ ($1\,{\leqq}\,i\,{\leqq}\,I$) are occupied with 100\% probability at a given time $t_{i}$.
\par
Then, due to assumption (ii), we have
\begin{align}
\xjvar{\mbox{`$\occupation{\stateset{{\mathrm{C}}_{i}}}{},t_{i}$'}}{big}
= 1
,\quad \forall{j}
,\quad 1\,{\leqq}\,i\,{\leqq}\,I.
\label{eeGroup2}
\end{align}
Therefore, by the definition of expectation values (Assumption B9) and the laws of quantum theory (Assumptions B10, B11, B13), we obtain
\begin{align}
\big\langle \xvar{\mbox{`$\occupation{\stateset{{\mathrm{C}}_{i}}}{},t_{i}$'}}{big} \big\rangle
& = {\mathrm{Tr}}\Big[\,\zop{\mbox{`$\occupation{\stateset{{\mathrm{C}}_{i}}}{}$'}}{big}\,\hat{\rho}(\,t_{i}\,|\mbox{`$\mathrm{A}_{0},t_{0}$'})\,\zopd{\mbox{`$\occupation{\stateset{{\mathrm{C}}_{i}}}{}$'}}{big}\,\Big]
\label{eeGroup3}\\
& = \sum_{|\phi_{\mathcal{I}}\rangle}^{\stateset{{\mathrm{C}}_{\mathcal{I}}}}
\langle\,\phi_{\mathcal{I}}\,|\,\hat{\rho}(\,t_{i}\,|\mbox{`$\mathrm{A}_{0},t_{0}$'})\,|\,\phi_{\mathcal{I}}\,\rangle
\label{eeGroup4}\\
& = 1
,\quad 1\,{\leqq}\,i\,{\leqq}\,I.
\label{eeGroup5}
\end{align}
Also, from the definition of logical variables (Assumption B5) and the definition of expectation values (Assumption B9), we see
\begin{align}
& 0 \leqq \big\langle \xvar{\mbox{`$\occupation{\stateset{{\mathrm{C}}_{i}}}{},t_{i}$'}}{big} \big\rangle
\leqq 1
,\quad 1\,{\leqq}\,i\,{\leqq}\,I
\label{eeGroup6}
\end{align}
and from the Hermiticity of the density operator (Assumption B13), we see that
\begin{align}
& \langle\,\phi_{\mathcal{I}}\,|\,\hat{\rho}(\,t_{i}\,|\mbox{`$\mathrm{A}_{0},t_{0}$'})\,|\,\phi_{\mathcal{I}}\,\rangle \geqq 0
\label{eeGroup7}
\end{align}
Therefore, when assumption (ii) holds, matrix elements of the density operator where $|\phi_{\mathcal{I}}\rangle\,{\notin}\,\stateset{{\mathrm{C}}_{\mathcal{I}}}$ or $|\phi'_{\mathcal{I}}\rangle\,{\notin}\,\stateset{{\mathrm{C}}_{\mathcal{I}}}$ are always zero:
\begin{align}
0 
= \big\langle\,\phi_{\mathcal{I}}\,|\,\hat{\rho}(\,t_{i}\,|\mbox{`$\mathrm{A}_{0},t_{0}$'})\,|\,\phi'_{\mathcal{I}}\,\big\rangle
= \big\langle\,\phi'_{\mathcal{I}}\,|\,\hat{\rho}(\,t_{i}\,|\mbox{`$\mathrm{A}_{0},t_{0}$'})\,|\,\phi_{\mathcal{I}}\,
\big\rangle.
\label{eeGroup8}
\end{align}
\par
Thus, when transforming the left-hand side of the compatibility condition (Eq.\eqref{eeGen5}) as
\begin{align}
& {\rm{Tr}}\Big[\,\hat{K}_{i_{1},\ldots,i_{\mathcal{I}}}\,\hat{\rho}(\,t\,|\mbox{`$\mathrm{A}_{0},t_{0}$'})\,\hat{K}^{\dagger}_{i_{1},\ldots,i_{\mathcal{I}}}\,\Big]
\nonumber\\
& = {\rm{Tr}}\Big[\,\hat{K}^{\dagger}_{i_{1},\ldots,i_{\mathcal{I}-1}}\,\hat{K}_{i_{1},\ldots,i_{\mathcal{I}-1}}\,\zop{`$\occupation{\stateset{{\mathrm{C}}_{\mathcal{I}}}}{}$'}{big}\,\hat{\rho}(\,t\,|\mbox{`$\mathrm{A}_{0},t_{0}$'})\,\zopd{`$\occupation{\stateset{{\mathrm{C}}_{\mathcal{I}}}}{}$'}{big}\,\Big]
\label{eeGroup11}\\
& = \sum_{|\phi_{\mathcal{I}}\rangle}^{\stateset{{\mathrm{C}}_{\mathcal{I}}}} \sum_{|\phi'_{\mathcal{I}}\rangle}^{\stateset{{\mathrm{C}}_{\mathcal{I}}}} 
\langle \phi_{\mathcal{I}}\,|\,\hat{K}^{\dagger}_{i_{1},\ldots,i_{\mathcal{I}-1}}\,\hat{K}_{i_{1},\ldots,i_{\mathcal{I}-1}}\,|\,\phi'_{\mathcal{I}} \rangle 
\langle\,\phi'_{\mathcal{I}}\,|\,\hat{\rho}(\,t\,|\mbox{`$\mathrm{A}_{0},t_{0}$'})\,|\,\phi_{\mathcal{I}}\,\rangle,
\label{eeGroup12}
\end{align}
based on Eq.\eqref{eeGroup8}, the sum on the right-hand side of Eq.\eqref{eeGroup12} can be replaced with a sum over the entire Hilbert space:
\begin{align}
& = \sum_{|\phi_{\mathcal{I}}\rangle} \sum_{|\phi'_{\mathcal{I}}\rangle}
\langle \phi_{\mathcal{I}}\,|\,\hat{K}^{\dagger}_{i_{1},\ldots,i_{\mathcal{I}-1}}\,\hat{K}_{i_{1},\ldots,i_{\mathcal{I}-1}}\,|\,\phi'_{\mathcal{I}} \rangle 
\langle\,\phi'_{\mathcal{I}}\,|\,\hat{\rho}(\,t\,|\mbox{`$\mathrm{A}_{0},t_{0}$'})\,|\,\phi_{\mathcal{I}}\,\rangle
\label{eeGroup15}\\
& = {\rm{Tr}}\Big[\,\hat{K}_{i_{1},\ldots,i_{\mathcal{I}-1}}\,\hat{\rho}(\,t\,|\mbox{`$\mathrm{A}_{0},t_{0}$'})\,\hat{K}^{\dagger}_{i_{1},\ldots,i_{\mathcal{I}-1}}\,\Big].
\label{eeGroup16}
\end{align}
By repeatedly using Eq.\eqref{eeGroup16}, we obtain
\begin{align}
{\rm{Tr}}\Big[\,\hat{K}^{\dagger}_{i_{1},\ldots,i_{\mathcal{I}}}\,\hat{K}_{i_{1},\ldots,i_{\mathcal{I}}}\,\hat{\rho}(\,t\,|\mbox{`$\mathrm{A}_{0},t_{0}$'})\,\Big]
= {\rm{Tr}}\Big[\,\hat{\rho}(\,t\,|\mbox{`$\mathrm{A}_{0},t_{0}$'})\,\Big]
\label{eeGroup17}
\end{align}
and find that the left-hand side of Eq.\eqref{eeGroup7} is independent of the order of partial characteristic operators. 
Therefore, when the sets $\stateset{{\mathrm{C}}_{i}}$ ($1\,{\leqq}\,i\,{\leqq}\,I$) include all stationary states with finite occupation probability, compatibility conditions always hold between elementary propositions `$\occupation{\stateset{{\mathrm{C}}_{i}}}{},t_{i}$' ($1\,{\leqq}\,i\,{\leqq}\,I$). 

\subsection{Compatibility conditions for arbitrary number of propositions}
\label{sGeneralizations}
In the main text, we derived compatibility conditions for two propositions `$\mathrm{A},\,t_{\mathrm{A}}$' and `$\mathrm{B},\,t_{\mathrm{B}}$'. 
Below, we derive compatibility conditions for $I$ propositions more generally. 
For this purpose, we make the following assumption:
\begin{myAssumptionA}
When propositions `${\mathrm{A}}_{i},t_{i}$' ($1\,{\leqq}\,i\,{\leqq}\,I$) are elementary propositions (defined at a single time), the probability of the compound proposition defined by
\begin{align}
\mbox{`$\mathrm{A},\{t\}$'}
\equiv \bigwedge^{I}_{i=1} \mbox{`${\mathrm{A}}_{i},t_{i}$'}
\equiv \mbox{`${\mathrm{A}}_{1},t_{1}$'$\wedge$`${\mathrm{A}}_{2},t_{2}$'$\wedge\ldots\wedge$`${\mathrm{A}}_{I},t_{\mathrm{I}}$'}
\label{eeGen1}
\end{align}
can be calculated by the following equation:
\begin{align}
\big\langle \xvar{`A,$\{t\}$'}{} \big\rangle
= {\rm{Tr}}\Big[\,\xop{`A,$\{t\}$'}{}\,\hat{\rho}(\mbox{`$\mathrm{A}_{0},t_{0}$'})\,\Big]
= {\rm{Tr}}\Big[\,\zop{`A,$\{t\}$'}{}\,\hat{\rho}(\mbox{`$\mathrm{A}_{0},t_{0}$'})\,\zopd{`A,$\{t\}$'}{}\,\Big].
\label{eeGen2}
\end{align}
Here, we abbreviate the set of times $t_{1},\ldots,t_{I}$ as $\{t\}$.
\end{myAssumptionA}
\noindent
By substituting Eq.\eqref{eeGen1} into Eq.\eqref{eeGen2} and repeatedly applying the laws of Boolean operations (Assumption A5), we find that when all $t_{1},\ldots,t_{I}$ are different from each other, $\zop{`A,$\{t\}$'}{}$ in Eq.\eqref{eeGen2} takes the following form:
\begin{align}
\zop{`A,$\{t\}$'}{}
= \hat{K}({\mathrm{A}}_{1},t_{1};\ldots;{\mathrm{A}}_{I},t_{I})
\equiv \mathcal{T} \prod^{I}_{i=1} \zop{`${\mathrm{A}}_{i},t_{i}$'}{leftright}.
\label{eeGen3}
\end{align}
To clarify the correspondence with the Consistent Histories approach, we introduced the Chain operator $\hat{K}({\mathrm{A}}_{1},t_{1};\ldots;{\mathrm{A}}_{I},t_{I})$.
Substituting Eq.\eqref{eeGen3} into Eq.\eqref{eeGen2}, we find the following relationship:
\begin{align}
\big\langle \xvar{`A,$\{t\}$'}{} \big\rangle
= {\rm{Tr}}\left[\,\hat{K}({\mathrm{A}}_{1},t_{1};\ldots;{\mathrm{A}}_{I},t_{I})\,\hat{\rho}(\mbox{`$\mathrm{A}_{0},t_{0}$'}) \hat{K}({\mathrm{A}}_{1},t_{1};\ldots;{\mathrm{A}}_{I},t_{I})^{\dagger}\,\right].
\label{eeGen4}
\end{align}
For this joint probability to be uniquely defined at $t_{1}\,{=}\,\ldots\,{=}\,t_{I}\,{=}\,t$, the right-hand side of Eq.{eeGen4} must be equal regardless of the time order of partial characteristic operators at $t_{1}\,{=}\,\ldots\,{=}\,t_{I}\,{=}\,t$.
Therefore, when assuming A9, we need to also assume the following:
\begin{myAssumptionA}
Let $\{i_{1},\,{\ldots}\,,i_{\mathcal{I}}\}$ ($i_{1}\,{<}\,{\ldots}\,{<}\,i_{\mathcal{I}}$) be any combination of natural numbers extracted without duplication from the set of natural numbers $\{1,\,{\ldots}\,,I\}$. 
Then, elementary propositions `${\mathrm{A}}_{i},t_{i}$' ($1\,{\leqq}\,i\,{\leqq}\,I$) need to be chosen so that the product of partial characteristic operators
\begin{align}
\hat{K}_{i_{1},\ldots,i_{\mathcal{I}}}
\equiv \zop{`${\mathrm{A}}_{i_{1}}$'}{} \cdots \zop{`${\mathrm{A}}_{i_{\mathcal{I}}}$'}{}
,\quad 1\,{\leqq}\,\mathcal{I}\,{\leqq}\,I
\label{eeGen6}
\end{align}
satisfies the following relationship:
\begin{align}
& {\rm{Tr}}\Big[\,\hat{K}_{i_{1},\ldots,i_{\mathcal{I}}}\,\hat{\rho}(\,t\,|\mbox{`$\mathrm{A}_{0},t_{0}$'})\,\hat{K}_{i_{1},\ldots,i_{\mathcal{I}}}{}^{\dagger}\,\Big]
\nonumber\\
& = {\rm{Tr}}\Big[\,\left({\mathcal{P}}\hat{K}_{i_{1},\ldots,i_{\mathcal{I}}}\right)\,\hat{\rho}(\,t\,|\mbox{`$\mathrm{A}_{0},t_{0}$'})\,\left({\mathcal{P}}\hat{K}_{i_{1},\ldots,i_{\mathcal{I}}}\right)^{\dagger}\,\Big]
,\quad 2\,{\leqq}\,\mathcal{I}\,{\leqq}\,I.
\label{eeGen5}
\end{align}
Here, $\hat{\rho}(\,t\,|\mbox{`$\mathrm{A}_{0},t_{0}$'})$ is a density operator, and ${\mathcal{P}}\hat{K}_{i_{1},\ldots,i_{\mathcal{I}}}$ represents the operator obtained by rearranging the partial characteristic operators of $\hat{K}_{i_{1},\ldots,i_{\mathcal{I}}}$ in any order.
\end{myAssumptionA}
\noindent
Writing out Eq.\eqref{eeGen5} explicitly, we find that for $\mathcal{I}\,{=}\,2$, it becomes the same as Eq.\eqref{eeComp15} in the main text, and for $\mathcal{I}\,{=}\,3$, it takes the following form:
\begin{align}
& {\rm{Tr}}\Big[\,\zopd{`${\mathrm{A}}_{3}$'}{} \zopd{`${\mathrm{A}}_{2}$'}{} \zopd{`${\mathrm{A}}_{1}$'}{} \zop{`${\mathrm{A}}_{1}$'}{} \zop{`${\mathrm{A}}_{2}$'}{} \zop{`${\mathrm{A}}_{3}$'}{}\,\hat{\rho}(\,t\,|\mbox{`$\mathrm{A}_{0},t_{0}$'})\,\Big]
\nonumber\\
& ={\rm{Tr}}\Big[\,\zopd{`${\mathrm{A}}_{2}$'}{} \zopd{`${\mathrm{A}}_{3}$'}{} \zopd{`${\mathrm{A}}_{1}$'}{} \zop{`${\mathrm{A}}_{1}$'}{} \zop{`${\mathrm{A}}_{3}$'}{} \zop{`${\mathrm{A}}_{2}$'}{}\,\hat{\rho}(\,t\,|\mbox{`$\mathrm{A}_{0},t_{0}$'})\,\Big]
\label{eeGen11}\\
& ={\rm{Tr}}\Big[\,\zopd{`${\mathrm{A}}_{3}$'}{} \zopd{`${\mathrm{A}}_{1}$'}{} \zopd{`${\mathrm{A}}_{2}$'}{} \zop{`${\mathrm{A}}_{2}$'}{} \zop{`${\mathrm{A}}_{1}$'}{} \zop{`${\mathrm{A}}_{3}$'}{}\,\hat{\rho}(\,t\,|\mbox{`$\mathrm{A}_{0},t_{0}$'})\,\Big]
\label{eeGen12}\\
& ={\rm{Tr}}\Big[\,\zopd{`${\mathrm{A}}_{1}$'}{} \zopd{`${\mathrm{A}}_{3}$'}{} \zopd{`${\mathrm{A}}_{2}$'}{} \zop{`${\mathrm{A}}_{2}$'}{} \zop{`${\mathrm{A}}_{3}$'}{} \zop{`${\mathrm{A}}_{1}$'}{}\,\hat{\rho}(\,t\,|\mbox{`$\mathrm{A}_{0},t_{0}$'})\,\Big]
\label{eeGen13}\\
& ={\rm{Tr}}\Big[\,\zopd{`${\mathrm{A}}_{2}$'}{} \zopd{`${\mathrm{A}}_{1}$'}{} \zopd{`${\mathrm{A}}_{3}$'}{} \zop{`${\mathrm{A}}_{3}$'}{} \zop{`${\mathrm{A}}_{1}$'}{} \zop{`${\mathrm{A}}_{2}$'}{}\,\hat{\rho}(\,t\,|\mbox{`$\mathrm{A}_{0},t_{0}$'})\,\Big]
\label{eeGen14}\\
& ={\rm{Tr}}\Big[\,\zopd{`${\mathrm{A}}_{1}$'}{} \zopd{`${\mathrm{A}}_{2}$'}{} \zopd{`${\mathrm{A}}_{3}$'}{} \zop{`${\mathrm{A}}_{3}$'}{} \zop{`${\mathrm{A}}_{2}$'}{} \zop{`${\mathrm{A}}_{1}$'}{}\,\hat{\rho}(\,t\,|\mbox{`$\mathrm{A}_{0},t_{0}$'})\,\Big].
\label{eeGen15}
\end{align}
Using the equation for conditional change of the density operator (Eq.\eqref{eeJoint3a}), these equations can be written in a form similar to the case of $\mathcal{I}\,{=}\,2$:
\begin{align}
& {\rm{Tr}}\Big[\,\zopd{`${\mathrm{A}}_{2}$'}{} \zopd{`${\mathrm{A}}_{3}$'}{} \zop{`${\mathrm{A}}_{3}$'}{} \zop{`${\mathrm{A}}_{2}$'}{}\,\hat{\rho}(\mbox{`${\mathrm{A}}_{1},t$'})\,\Big]
\nonumber\\
& ={\rm{Tr}}\Big[\,\zopd{`${\mathrm{A}}_{3}$'}{} \zopd{`${\mathrm{A}}_{2}$'}{} \zop{`${\mathrm{A}}_{2}$'}{} \zop{`${\mathrm{A}}_{3}$'}{}\,\hat{\rho}(\mbox{`${\mathrm{A}}_{1},t$'})\,\Big]
\label{eeGen16}\\
& ={\rm{Tr}}\Big[\,\zopd{`${\mathrm{A}}_{3}$'}{} \zopd{`${\mathrm{A}}_{1}$'}{} \zop{`${\mathrm{A}}_{1}$'}{} \zop{`${\mathrm{A}}_{3}$'}{}\,\hat{\rho}(\mbox{`${\mathrm{A}}_{2},t$'})\,\Big]
\label{eeGen17}\\
& ={\rm{Tr}}\Big[\,\zopd{`${\mathrm{A}}_{1}$'}{} \zopd{`${\mathrm{A}}_{3}$'}{} \zop{`${\mathrm{A}}_{3}$'}{} \zop{`${\mathrm{A}}_{1}$'}{}\,\hat{\rho}(\mbox{`${\mathrm{A}}_{2},t$'})\,\Big]
\label{eeGen18}\\
& ={\rm{Tr}}\Big[\,\zopd{`${\mathrm{A}}_{2}$'}{} \zopd{`${\mathrm{A}}_{1}$'}{} \zop{`${\mathrm{A}}_{1}$'}{} \zop{`${\mathrm{A}}_{2}$'}{}\,\hat{\rho}(\mbox{`${\mathrm{A}}_{3},t$'})\,\Big]
\label{eeGen19}\\
& ={\rm{Tr}}\Big[\,\zopd{`${\mathrm{A}}_{1}$'}{} \zopd{`${\mathrm{A}}_{2}$'}{} \zop{`${\mathrm{A}}_{2}$'}{} \zop{`${\mathrm{A}}_{1}$'}{}\,\hat{\rho}(\mbox{`${\mathrm{A}}_{3},t$'})\,\Big].
\label{eeGen20}
\end{align}
Therefore, when considering the meaning of compatibility conditions, second-order compatibility conditions are particularly important. 
Note that while the right-hand side of Eq.\eqref{eeGen4} is similar to the definition of correlation functions~\cite[{\S}11.12]{Mandel} or many-body Green's functions~\cite[{\S}3.7]{Fetter}, it reduces to those equations only when partial characteristic operators (projection operators) can be replaced by creation and annihilation operators.

\clearpage
\section{Summary of notations} 
\label{sAppNotation}
\begin{table}[H]
\label{ttNotation1}
\centering
\begin{tabular}{cl}
\toprule
\raisebox{0em}{\bettershortstack[c]{0.75}{2.5}{Symbol}} & 
  \raisebox{0em}{\bettershortstack[c]{0.75}{2.5}{~~~~~~~~~~~~~~~~~~~~~~~~~~~~~~~~~~~~~~~~~~Meaning}} \\
\midrule
\raisebox{0em}{\bettershortstack[l]{0.75}{2.5}{`A'}} & 
  \raisebox{0em}{\bettershortstack[l]{0.75}{2.5}{`A is true'}} \\
\raisebox{0em}{\bettershortstack[l]{0.75}{2.5}{$\mbox{`A,\,$t$'}$}} & 
  \raisebox{0em}{\bettershortstack[l]{0.75}{2.5}{`A is true at time $t$'}} \\
\raisebox{0em}{\bettershortstack[l]{0.75}{2.5}{$\mbox{`o($\phi$)'}$}} & 
  \raisebox{0em}{\bettershortstack[l]{0.75}{2.5}{`A stationary state $\phi$ is occupied'}} \\
\raisebox{0em}{\bettershortstack[l]{0.75}{2.5}{$\mbox{`o($\{\!\!\{{\mathrm{C}}\}\!\!\}$)'}$}} & 
  \raisebox{0em}{\bettershortstack[l]{0.75}{2.5}{`One of the stationary states in set $\{\!\!\{{\mathrm{C}}\}\!\!\}$ is occupied'}} \\
\raisebox{0em}{\bettershortstack[l]{0.75}{2.5}{$\{\!\!\{{\mathrm{C}}\}\!\!\}$}} & 
  \raisebox{0em}{\bettershortstack[l]{0.75}{2.5}{A set of stationary state satisfying condition C (when $\hat{\mathsf{U}}\,{=}\,\hat{1}$)}} \\
\raisebox{0em}{\bettershortstack[l]{0.75}{2.5}{$\stateset{\mathrm{C}}_{\hat{\mathsf{U}}}$}} & 
  \raisebox{0em}{\bettershortstack[l]{0.75}{2.5}{A set of stationary state satisfying condition C (when $\hat{\mathsf{U}}\,{\neq}\,\hat{1}$)}} \\
\raisebox{0em}{\bettershortstack[l]{0.75}{2.5}{$\xjvar{`A,\,$t$'}{}$}} & 
  \raisebox{0em}{\bettershortstack[l]{0.75}{2.5}{The truth value of `A,\,$t$' in the $j$-th trial (${\in}\,\{0,1\}$)}} \\
\raisebox{0em}{\bettershortstack[l]{0.75}{2.5}{$\langle\xvar{`A,\,$t$'}{}\rangle$}} & 
  \raisebox{0em}{\bettershortstack[l]{0.75}{2.5}{The expectation value of $\xjvar{`A,\,$t$'}{}$ (the probability of `A,\,$t$')}} \\
\raisebox{0em}{\bettershortstack[l]{0.75}{2.5}{$\xop{`A,\,$t$'}{}$}} & 
  \raisebox{0em}{\bettershortstack[l]{0.75}{2.5}{The characteristic operator of `A,\,$t$' (Heisenberg picture)}} \\
\raisebox{0em}{\bettershortstack[l]{0.75}{2.5}{$\zop{`A,\,$t$'}{}$}} & 
  \raisebox{0em}{\bettershortstack[l]{0.75}{2.5}{The partial characteristic operator of `A,\,$t$' (Heisenberg picture)}} \\
\raisebox{0em}{\bettershortstack[l]{0.5}{2.5}{$\hat{x}(\mbox{`A'})$}} & 
  \raisebox{0em}{\bettershortstack[l]{0.5}{2.5}{The characteristic operator of `A' (Schr\"{o}dinger picture)}} \\
\raisebox{0em}{\bettershortstack[l]{0.5}{2.5}{$\zop{`A'}{}$}} & 
  \raisebox{0em}{\bettershortstack[l]{0.5}{2.5}{The partial characteristic operator of `A' (Schr\"{o}dinger picture)}}~~~~~~~~~ \\
\raisebox{0em}{\bettershortstack[l]{0.5}{2.5}{$\{t\}$}} & 
  \raisebox{0em}{\bettershortstack[l]{0.5}{2.5}{A set of times $\{t_{1},\ldots,t_{I}\}$ ($I\,{\in}\,{\mathbb{N}}$)}}\\
\bottomrule
\end{tabular}
\end{table}
\begin{table}[H]
\label{ttNotation2}
\centering
\begin{tabular}{clc}
\toprule
~~~Symbol~~~ & ~~~~~~~~~~~~~~~~~~~~~~~~~~~~~~~~~Meaning~~~~~~~~~~~~~~~~~~~~~~~~~~~~~~~~~ & ~~~Range~~~ \\ 
\midrule
${\mathrm{i}}$ & Imaginary unit & -\\
$h$ & Planck's constant & -\\
$\hbar$ & Planck's constant divided by $2\pi$ & -\\
$t$ & Time & $\mathbb{R}$\\
$\nu$ & Frequency & $\mathbb{R}$\\
$\omega$ & Angular frequency (${=}\,2\pi{\nu}$) & $\mathbb{R}$\\
${\bm k}$ & Wavevector (${=}\,(k_{x},k_{y},k_{z})$) & $\mathbb{R}^{3}$\\
$\lambda$ & Polarization & $\{1,2\}$\\
${\bm x}$ & Cartesian coordinate (${=}\,(x,y,z)$) & $\mathbb{R}^{3}$ \\
${\bm r}$ & Relative position (for an idealized system) & $\mathbb{R}^{3}$ \\
${\bm p}$ & Relative momentum (for an idealized system)  & $\mathbb{R}^{3}$ \\
${\bm R}$ & Center-of-mass position (for an idealized system) & $\mathbb{R}^{3}$ \\
${\bm P}$ & Center-of-mass momentum (for an idealized system)  & $\mathbb{R}^{3}$ \\
${\bm r}_{k}$ & Position of the $k$-th particle (for an idealized system) & $\mathbb{R}^{3}$ \\
${\bm p}_{k}$ & Momentum of the $k$-th particle (for an idealized system)  & $\mathbb{R}^{3}$ \\
${\mathcal{R}}_{k}$ & Position of the $k$-th particle (for a realistic system) & $\mathbb{R}^{3}$ \\
${\mathcal{P}}_{k}$ & Momentum of the $k$-th particle (for a realistic system) & $\mathbb{R}^{3}$ \\
$q_{l}$ & Canonical variable of the $l$-th degree of freedom & $\mathbb{R}$ \\
$p_{l}$ & Canonical momentum of the $l$-th degree of freedom & $\mathbb{R}$ \\
$m_{k}$ & Mass of the $k$-th particle & $\mathbb{R}$\\
$e_{k}$ & Charge of the $k$-th particle & $\mathbb{R}$\\
$g_{k}$ & Lande $g$-factor of the $k$-th particle & $\mathbb{R}$\\
$c$ & Light speed in vacuum & -\\
$\varepsilon_{0}$ & Vacuum permittivity & -\\
$\mu_{0}$ & Vacuum permiability & -\\
$\delta_{\phi \phi'}$ & Kronecker delta or delta function & -\\
$\delta^{{\mathrm{T}}}_{\xi \xi'}({\bm x}-{\bm x}')$ & Transverse delta function~\cite{Mandel} & -\\
\bottomrule
\end{tabular}
\end{table}

\begin{table}[H]
\label{ttNotation3}
\centering
\begin{tabular}{clc}
\toprule
~~~Symbol~~~ & ~~~~~~~~~~~~~~~~~~~~~~~~~~~~~~~~~Meaning~~~~~~~~~~~~~~~~~~~~~~~~~~~~~~~~~ & ~~~Range~~~ \\ 
\midrule
$\hat{H}$ & Total Hamiltonian & (operator)\\
$\hat{H}_{0}$ & Non-interacting Hamiltonian & (operator)\\
$\hat{H}_{\mathrm{A}}$ & Hamiltonian of the atomic system & (operator)\\
$\hat{H}_{\mathrm{F}}$ & Hamiltonian of the radiation field & (operator)\\
$\hat{H}_{\mathrm{I}}$ & Hamiltonian of the atom-field interaction & (operator)\\
$E$ & Stationary state energy (the eigenvalue of $\hat{H}_{0}$) & ${\mathbb{R}}$\\
$\mathcal{E}$ & Total energy & ${\mathbb{R}}$\\
$\hat{\mathcal{H}}$ & Hamiltonian of the extended system with meas. instruments\!\!\!\!\! & (operator)\\
$\hat{\mathcal{H}}_{0}$ & Non-interacting Hamiltonian of the extended system & (operator)\\
$\hat{H}^{(l)}$ & Hamiltonian of the $l$-th subsystem & (operator)\\
$\hat{\bm A}_{\mathrm{T}}({\bm x})$ & Vector potential operator (transverse part) & (operator)\\
$\hat{\bm E}_{\mathrm{T}}({\bm x})$ & Electric field operator (transverse part) & (operator)\\
$\hat{\bm B}({\bm x})$ & Magnetic flux density operator & (operator)\\
$\hat{\bm S}$ & Spin angular momentum operator & (operator)\\
$\hat{\bm D}$ & Electric dipole moment operator & (operator)\\
$\hat{\bm \kappa}$ & Operator coefficient & (operator)\\
$\hat{a}_{{\bm k}\lambda}$ & Annihilation operator of the ${{\bm k}\lambda}$-mode & (operator)\\
$\hat{a}^{\dagger}_{{\bm k}\lambda}$ & Creation operator of the ${{\bm k}\lambda}$-mode & (operator)\\
$\hat{\boldsymbol\sigma}_{k}$ & Pauli operators of the $k$-th particle & (operator)\\
$\hat{U}$ & Time evolution operator & (operator)\\
$\hat{\mathsf{W}}$ & Time-independent unitary operator (Eq.\eqref{eeModel11}) & (operator)\\
$\hat{\mathsf{U}}$ & Time-independent unitary operator (Eq.\eqref{eeModel22}) & (operator)\\
$\hat{K}_{1,\ldots,\mathcal{I}}$ & Chain operator (Eq.\eqref{eeModel132}) & (operator)\\
$\hat{\rho}(\mbox{`${\mathrm{A}}_{0}$'})$ & Density operator for an ensemble specified by `$\mathrm{A}_{0}$' & (operator)\\
$\hat{\rho}(\mbox{`$\mathrm{A}_{0},t_{0}$'})$ & Density operator (Schr\"{o}dinger picture) & (operator)\\
$|~~\rangle$ & Quantum state (bra vector) & -\\
$\langle~~|$ & Quantum state (ket vector) & -\\
$\langle~~\rangle$ & Expectation value for the initial ensemble & -\\
$\langle~~|\!|\mbox{`${\mathrm{A}},t$'}\rangle$ & Conditional expectation value conditioned by `${\mathrm{A}},t$' & -\\
$\mathrm{Tr}$ & Trace & -\\
$\dagger$ & Hermitian conjugate & -\\
$::\,::$  & Normal-ordering symbol for partial characteristic operators  & -\\
$\mathcal{T}$ & Time-ordering symbol for partial characteristic operators & -\\
$\mathcal{P}$ & Permutation symbol for partial characteristic operators & -\\
$\mathcal{L}$ & Liouville operator & -\\
$\otimes$ & Tensor product & -\\
$\lnot$ & Negation & -\\
$\wedge$ & Conjunction & -\\
$\vee$ & Disjunction & -\\
$\overline{\phantom{w}}$ & Complement & -\\
$\cup$ & Union & -\\
$\cap$ & Intersection & -\\
$\mathrm{Pr}(\mbox{`A'})$ & The probability of `A' & $[0,1]$\\
$\mathrm{Pr}(\mbox{`A'}{\wedge}\mbox{`B'})$ & The joint probability of `A' and `B' & $[0,1]$\\
$\mathrm{Pr}(\mbox{`A'}|\mbox{`B'})$ & The conditional probability of `A' under `B' & $[0,1]$\\
\bottomrule
\end{tabular}
\end{table}

\begin{table}[H]
\label{ttNotation4}
\centering
\begin{tabular}{clc}
\toprule
~~~Symbol~~~ & ~~~~~~~~~~~~~~~~~~~~~~~~~~~~~~~~~Meaning~~~~~~~~~~~~~~~~~~~~~~~~~~~~~~~~~ & ~~~Range~~~ \\ 
\midrule
$i$ & Index of times & $\{1,\ldots,I\}$\\
$j$ & Index of trials & $\{1,\ldots,N\}$\\
$k$ & Index of charged particles & $\{1,\ldots,K\}$\\
$l$ & Index of degrees of freedom & $\{1,\ldots,L\}$\\
$\mu$ & Index of conditions & ${\mathbb{Z}}$\\
$h$ & Index of the cells containing water droplets (Eq.\eqref{eeExample2-3}) & ${\mathbb{Z}}$\\
$n$ & General integers (number of counts, etc.) & $\mathbb{N}$\\
${\mathcal{I}}$ & Order of compatibility condition & $1\,{\leqq}\,{\mathcal{I}}\,{\leqq}\,I$\\
$\xi$ & Cartesian index & $\{x,y,z\}$\\
$\phi$ & Index of stationary state & -\\
$\chi$ & Index of stationary state (degeneracy index) & -\\
$|\phi\rangle$ & Stationary state (general) & -\\
$|\Phi\rangle$ & Stationary state (realistic system) & -\\
$|\varphi\rangle$ & Eigenstate of an observable (idealized system) & -\\
$\{\!\!\{+\}\!\!\}$ & Entire set of scattering states (relative motion)\!\!\!\!\!\!\!\!\!\! & -\\
$\{\!\!\{-\}\!\!\}$ & Entire set of bound states (relative motion)\!\!\!\!\!\!\!\!\!\! & -\\
${\bm u}_{k}$ & Spin measurement direction for the $k$-th particle & ${\mathbb{R}}^{3}$\\
${\bm u}_{\xi}$ & Unit vector in the $\xi$ direction & ${\mathbb{R}}^{3}$\\
$\gamma$ & Photoionization rate & -\\
$\tau$ & Time delay from photoionization to detected signal & -\\
$\tilde{\alpha}$ & Mode amplitude (complex) & ${\mathbb{C}}$\\
`True' & Logical constant $\top$ & -\\
`False' & Logical constant $\bot$ & -\\
`TN' & True-Negative confusion & -\\
`FP' & False-Positive confusion & -\\
`ionized,\,$t$' & Proposition `The atom was photoionized by time $t$' & -\\
`output,\,$t$' & Proposition `The photodetector responded by time $t$' & -\\
`dead,\,$t$' & Proposition `The cat is dead at time $t$' & -\\
`alive,\,$t$' & Proposition `The cat is alive at time $t$' & -\\
`absorbed,\,$t$' & Proposition `A photon is absorbed by time $t$' & -\\
`o(${\uparrow}_{x}$)' & Proposition `Spin is up in $x$-direction' & -\\
`o(${\downarrow}_{x}$)' & Proposition `Spin is down in $x$-direction' & -\\
`o(${\uparrow}_{y}$)' & `Spin is up in $y$-direction' & -\\
`o(${\downarrow}_{y}$)' & Proposition `Spin is down in $y$-direction' & -\\
$\delta{F}, \delta{G}$ & Intervals of discretization grids for $F$ and $G$ & ${\mathbb{R}}$\\
$\hat{F}, \hat{G}$ & Physical quantities (system w/o measuring instruments) & (operator)\\
$\hat{\mathcal{F}}, \hat{\mathcal{G}}$ & Physical quantities (system with measuring instruments) & (operator)\\
${F}^{(j)}, {G}^{(j)}$ & Instantaneous values of $F$ and $G$ & ${\mathbb{R}}$ or ${\mathbb{C}}$\\
${\mathcal{F}}^{(j)}, {\mathcal{G}}^{(j)}$ & Instantaneous values of $\mathcal{F}$ and $\mathcal{G}$ & ${\mathbb{R}}$ or ${\mathbb{C}}$\\
$\Delta{F}^{(j)}, \Delta{G}^{(j)}$ & Quantum fluctuation of $F$ and $G$ & ${\mathbb{R}}$ or ${\mathbb{C}}$\\
${\mathrm{Var}}_{F}, {\mathrm{Var}}_{G}$ & Variance of $F$ and $G$ & ${\mathbb{R}}$ or ${\mathbb{C}}$\\
\bottomrule
\end{tabular}
\end{table}

\bibliographystyle{unsrt}
\bibliography{main}

\clearpage
\addtocontents{toc}{\protect\setcounter{tocdepth}{-1}}
\section*{Supplementary material A: Model Details}
\setcounter{section}{1}
\setcounter{subsection}{0}
\subsection{The importance of exclusivity of occupation}
\label{sRelevant}
This section supplements the role of exclusivity of occupation in the model of the world of reality shown in Section \ref{cReality}. 
First, we show that the assumption of exclusivity of occupation (Assumption B7) is necessary for Assumptions B1--B14 to be consistent. 
It is sufficient to look at a simple example, so we consider the case where there are only two stationary states ($\stateset{\mathrm{C}}\,{=}\,\{\phi_{1},\phi_{2}\}$ and $\langle\phi_{1}|\phi_{2}\rangle\,{=}\,0$).
Then, by the definition of elementary propositions (Assumption B4), the definition of logical variables (Assumption B5), and the laws of Boolean logic (Assumption B6), we have
\begin{align}
\xjvar{`$\occupation{\stateset{{\mathrm{C}}}}{},t$'}{big}
& = \xjvar{`$\occupation{\state{\phi_{1}}{}}{},t$'$\vee$`$\occupation{\state{\phi_{2}}{}}{},t$'}{big}
,\quad {\forall}j
\label{eeExclusive3}\\
& = \xjvar{`$\occupation{\state{\phi_{1}}{}}{},t$'}{big} + \xjvar{`$\occupation{\state{\phi_{2}}{}}{}$'}{big}
- \xjvar{`$\occupation{\state{\phi_{1}}{}}{},t$'$\wedge$`$\occupation{\state{\phi_{2}}{}}{},t$'}{big}
,\quad {\forall}j.
\label{eeExclusive4}
\end{align}
On the other hand, from the trace formula (Assumption B10), the Born rule (Assumption B11), and the laws of Boolean logic (Assumption B12), we find that
\begin{align}
\langle \xvar{`$\occupation{\stateset{{\mathrm{C}}}}{},t$'}{big} \rangle
= \langle \xvar{`$\occupation{\state{\phi_{1}}{}}{},t$'}{big} \rangle + \langle \xvar{`$\occupation{\state{\phi_{2}}{}}{},t$'}{big} \rangle.
\label{eeExclusive1}
\end{align}
This relationship needs to hold for any initial ensemble `${\mathrm{A}}_{0},\,t_{0}$', so by the definition of expectation values (Assumption B9), the following relationship needs to hold:
\begin{align}
\xjvar{`$\occupation{\stateset{{\mathrm{C}}}}{},t$'}{big}
= \xjvar{`$\occupation{\state{\phi_{1}}{}}{},t$'}{big} + \xjvar{`$\occupation{\state{\phi_{2}}{}}{},t$'}{big}
,\quad {\forall}j.
\label{eeExclusive2}
\end{align}
Comparing Eq.\eqref{eeExclusive4} and Eq.\eqref{eeExclusive2}, we can see that for Assumptions B4, B5, B6, B10, B12, and B13 to be consistent, the exclusivity of occupation (Assumption B7) must hold as
\begin{align}
& \xjvar{`$\occupation{\state{\phi_{1}}{}}{},t$'$\wedge$`$\occupation{\state{\phi_{2}}{}}{},t$'}{big} = 0
\quad\mbox{for}\quad
\phi_{1} \neq \phi_{2},
\label{eeExclusive5}\\
& \xjvar{`$\occupation{\state{\phi_{1}}{}}{},t$'$\wedge$`$\occupation{\state{\phi_{2}}{}}{},t$'}{big}
= \xjvar{`$\occupation{\state{\phi_{1}}{}}{},t$'}{big}
= \xjvar{`$\occupation{\state{\phi_{2}}{}}{},t$'}{big}
\quad\mbox{for}\quad
\phi_{1} = \phi_{2}.
\label{eeExclusive6}
\end{align}
\par
Next, we see that the assumption of exclusivity of occupation (Assumption B7) is also necessary to rationally define elementary propositions.
To do this, we consider the case where the sets of stationary states $\stateset{\mathrm{C}^{\mu}}$ representing groups $\{\stateset{\mathrm{C}^{\mu}}\,|\,\mu\,{\in}\,{\mathbb{Z}}\}$ of elementary propositions are mutually orthogonalized. 
In this case, for the complement $\overline{\phantom{w}}$, intersection $\cap$, and union $\cup$ of the sets of stationary states to have the properties:
\begin{align}
\stateset{\overline{\mathrm{C}}^{\mu}}
& = \left\{\,|\phi\rangle\,\Big|\,|\phi\rangle\,{\in}\,\stateset{\mathrm{E}}
,\,\mbox{$|\phi\rangle$ satisfies $\lnot\mathrm{C}^{\,\mu}$} \right\}
,\quad \mu\,{\in}\,{\mathbb{Z}}
\label{eeExclusive11}\\
\stateset{{\mathrm{C}}^{\mu}}\,{\cap}\,\stateset{{\mathrm{C}}^{\mu'}}
& = \left\{\,|\phi\rangle\,\Big|\,|\phi\rangle\,{\in}\,\stateset{\mathrm{E}}
,\,\mbox{$|\phi\rangle$ satisfies $\mathrm{C}^{\mu}\,{\wedge}\,\mathrm{C}^{\mu'}$} \right\}
,\quad \mu, \mu' \in {\mathbb{Z}}
\label{eeExclusive12}\\
\stateset{{\mathrm{C}}^{\mu}}\,{\cup}\,\stateset{{\mathrm{C}}^{\mu'}}
& = \left\{\,|\phi\rangle\,\Big|\,|\phi\rangle\,{\in}\,\stateset{\mathrm{E}}
,\,\mbox{$|\phi\rangle$ satisfies $\mathrm{C}^{\mu}\,{\vee}\,\mathrm{C}^{\mu'}$} \right\}
,\quad \mu, \mu' \in {\mathbb{Z}}
\label{eeExclusive13}
\end{align}
(where $\stateset{\mathrm{E}}$ is the complete system of stationary states), the following relations need to hold between propositions defined at the same time:
\begin{align}
\xjvar{$\lnot$`$\occupation{\stateset{{\mathrm{C}}^{\mu}}}{},t$'}{big}
& = \xjvar{`$\occupation{\stateset{\overline{\mathrm{C}}^{\mu}}}{},t$'}{big}
\label{eeExclusive14}\\
\xjvar{`$\occupation{\stateset{{\mathrm{C}}^{\mu}}}{},t$'$\wedge$`$\occupation{\stateset{{\mathrm{C}}^{\mu'}}}{},t$'}{big}
& = \xjvar{`$\occupation{\stateset{{\mathrm{C}}^{\mu}\,{\wedge}\,{\mathrm{C}}^{\mu'}}}{},t$'}{big}
= \xjvar{`$\occupation{\stateset{{\mathrm{C}}^{\mu}}\,{\cap}\,\stateset{{\mathrm{C}}^{\mu'}}}{},t$'}{big}
\label{eeExclusive16}\\
\xjvar{`$\occupation{\stateset{{\mathrm{C}}^{\mu}}}{},t$'$\vee$`$\occupation{\stateset{{\mathrm{C}}^{\mu'}}}{},t$'}{big}
& = \xjvar{`$\occupation{\stateset{{\mathrm{C}}^{\mu}\,{\vee}\,{\mathrm{C}}^{\mu'}}}{},t$'}{big}
= \xjvar{`$\occupation{\stateset{{\mathrm{C}}^{\mu}}\,{\cup}\,\stateset{{\mathrm{C}}^{\mu'}}}{},t$'}{big}.
\label{eeExclusive18}
\end{align}
However, Eq.\eqref{eeExclusive16} reduces to the assumption of exclusivity of occupation (Assumption B7) when there are only two stationary states, so if we require Eq.\eqref{eeExclusive11}--\eqref{eeExclusive13} and \eqref{eeExclusive14}--\eqref{eeExclusive18}, we need to assume the exclusivity of occupation (Assumption B7). 
From the above, we have found that the exclusivity of occupation (Assumption B7) is necessary for the result of applying any Boolean operation to a group $\{\mbox{`$\occupation{\stateset{{\mathrm{C}}^{\mu}_{i}}}{},t_{i}$'}\,|\,\mu\,{\in}\,{\mathbb{Z}}\}$ of elementary propositions to be an elementary proposition (`$\occupation{\stateset{{\mathrm{C}}_{i}}}{},t_{i}$').

\subsection{Form of partial characteristic operators and deterministic relations}
\label{sPTL}
We show that when the partial characteristic operators of given propositions `A' and `B' are projection operators and are known to be equal for any density operator, there exists a deterministic relation between propositions `A' and `B'.
First, we consider the case where the following relationship holds by assumption:
\begin{align}
\zop{`A'}{} = \zop{`B'}{}
,\quad
\zop{`A'}{}^{2} = \zop{`A'}{}
,\quad
\zop{`B'}{}^{2} = \zop{`B'}{}.
\label{eeInverse1}
\end{align}
In this case, since the compatibility condition holds between propositions `A' and `B', we can consider the conjunction `A,\,$t$'$\wedge$`B,\,$t$'. 
Also, by the trace formula (Assumption B10), operator transformation rule (Assumption B12), and equation of motion (Assumption B13), we have\begin{align}
\langle \xvar{`A,\,$t$'$\wedge$`B,\,$t$'}{} \rangle
= \langle \xvar{`A,\,$t$'}{} \rangle
= \langle \xvar{`B,\,$t$'}{} \rangle.
\label{eeInverse3}
\end{align}
Therefore, according to the discussions in {\S}\ref{sDeterministic}, the following deterministic relation holds between propositions `A,\,$t$' and `B,\,$t$':
\begin{align}
\xjvar{`A,\,$t$'}{} = \xjvar{`B,\,$t$'}{}
,\quad \forall j.
\label{eeInverse4}
\end{align}

\subsection{Instantaneous value of total energy}
In Dirac's 1927 paper~\cite[{\S}7]{Dirac1927}, it was pointed out that the energy conservation law for elementary processes does not hold in quantum theory due to the non-commutativity of the non-interacting Hamiltonian $\hat{H}_{0}\,{=}\,\hat{H}_{\mathrm{A}}\,{+}\,\hat{H}_{\mathrm{F}}$ and the interaction Hamiltonian $\hat{H}_{\mathrm{I}}$.
This assertion is valid within the framework of the standard formulation and serves as an appropriate reason for rejecting the light quantum hypothesis. 
In our theory, however, there is a possibility of simultaneous values existing for physical quantities with non-commuting operators, necessitating a reconsideration of this explanation.
In particular, the instantaneous value of the total energy stands as the sole exception to the definition of instantaneous values of physical quantities as described in the main text. 
Therefore, we provide additional explanation below.
\par
First, we define the instantaneous values of the total energy ${\mathcal{E}}^{(j)}(t)$ and the energy of stationary states $E^{(j)}(t)$ by the following equations:
\begin{align}
& {\mathcal{E}}^{(j)}(t)
= \sum_{{\mathcal{E}}} {\mathcal{E}}\,\xjvar{`$\occupation{\stateset{\mathcal{E}}}{},t$'}{leftright}
,\quad \forall j,
\label{eeAppendixEnergy-1}\\
& E^{(j)}(t)
= \sum_{E} E\,\xjvar{`$\occupation{\stateset{E}}{},t$'}{leftright}
,\quad \forall j,
\label{eeAppendixEnergy-2}
\end{align}
where $\stateset{\mathcal{E}}$ is a set of orthogonalized quantum states with total energy in the range $\mathcal{E}\,{\pm}\,{\delta{\mathcal{E}}}/{2}$\footnote{Examples of quantum states with definite total energy include polariton states in infinite crystals~\cite{Hopfield1958} and dressed states of two-level atoms~\cite{Cohen}.}, and $\stateset{E}$ is a set of stationary states with energy in the range $E\,{\pm}\,\delta{E}/{2}$, namely
\begin{align}
& \stateset{\mathcal{E}}
\equiv \left\{\,|\Phi\rangle\,\Bigg|\,\hat{H}|\Phi\rangle = H |\Phi\rangle
,\,\langle\Phi|\Phi'\rangle = \delta_{\Phi,\Phi'},\,
{\mathcal{E}}-\frac{\delta{\mathcal{E}}}{2} \leqq H < {\mathcal{E}}+\frac{\delta{\mathcal{E}}}{2}\,\right\}
\label{eeAppendixEnergy-3}\\
& \stateset{E}
\equiv \left\{\,|\phi\rangle\,\Bigg|\,\hat{H}_{0}|\phi\rangle = H_{0} |\phi\rangle
,\,\langle\phi|\phi'\rangle = \delta_{\phi,\phi'},\,
E-\frac{\delta{E}}{2} \leqq H_{0} < E+\frac{\delta{E}}{2}\,\right\}.
\label{eeAppendixEnergy-4}
\end{align}
Then, the following relation holds for any ensemble `${\mathrm{A}}_{0},t_{0}$':
\begin{align}
\frac{d}{dt}\big\langle\,{\mathcal{E}}(t)\,\big\rangle
& = \frac{d}{dt}{\mathrm{Tr}}\left[ \hat{H}(t) \hat{\rho}(\mbox{`${\mathrm{A}}_{0},t_{0}$'})\right]
= \frac{1}{{\mathrm{i}}\hbar} {\mathrm{Tr}}\left[[\hat{H},\hat{H}]_{-}\hat{\rho}(\mbox{`${\mathrm{A}}_{0},t_{0}$'})\right]
= 0.
\label{eeAppendixEnergy-5}
\end{align}
\par
On the other hand, in the classical approximation, if we set the precision of representation ($\delta{\mathcal{E}}$ and $\delta{E}$) to satisfy the constraint of the complementarity principle (Eq.\eqref{eeComplementarity8}),
\begin{align}
\delta{\mathcal{E}}\,\delta{E}
\gg \frac{1}{2}\,\Bigg| {\mathrm{Tr}}\left[\,\left[\hat{H},\hat{H}_{0}\right]_{-}\hat{\rho}(\mbox{`${\mathrm{A}}_{0}$'},t)\right] \Bigg|
=\frac{\hbar}{2}\,\Bigg| {\mathrm{Tr}}\left[\,\frac{d\hat{H}_{\mathrm{I}}(t)}{dt}\,\hat{\rho}(\mbox{`${\mathrm{A}}_{0},t_{0}$'})\,\right] \Bigg|,
\label{eeAppendixEnergy-7}
\end{align}
then the compatibility condition holds between propositions `$\occupation{\stateset{\mathcal{E}}}{},t$' and `$\occupation{\stateset{E}}{},t$'.
Therefore, if we take time $t$ before the start or after the end of the interaction, the right-hand side of Eq.\eqref{eeAppendixEnergy-7} becomes zero, allowing us to make $\delta\mathcal{E}$ and $\delta{E}$ arbitrarily small.
Consequently, if we define the instantaneous value of the total energy ${\mathcal{E}}^{(j)}(t)$ as a constant that coincides with the energy of the stationary state ${E}^{(j)}(t)$ at times when the interaction energy can be ignored, then ${\mathcal{E}}^{(j)}(t)$ takes a constant value for an initial ensemble `${\mathrm{A}}_{0},t_{0}$' where ${\mathcal{E}}^{(j)}(t)$ has a single definite value.
Therefore, in our theory, while the light quantum hypothesis does not hold, it is possible to specify the instantaneous value of the total energy ${\mathcal{E}}^{(j)}(t)$ with arbitrary precision.
Note that the instantaneous value of the total energy has this exceptional property because the total Hamiltonian $\hat{H}$ plays a peculiar role in Assumptions B1--B14.

\subsection{Comparison with the light quantum hypothesis}
\begin{table}[t]
\caption{Explanation of various phenomena by the light quantum hypothesis, quantum optics, and the theory proposed in this paper. $\checkmark$: Quantitative explanation, ($\checkmark$): Semi-quantitative explanation, (---): Qualitative explanation, ---: Difficult to explain.}
\label{tLightQuantum1}
\centering
\begin{tabular}{p{56mm}p{22mm}p{24mm}p{31mm}}
\toprule
\raisebox{0em}{\bettershortstack[c]{0.7}{2.5}{\small ~~~~~~~~~~~~~~~~~Item}} & 
  \raisebox{0em}{\bettershortstack[c]{0.7}{2.5}{\small Light-quantum}} & 
  \raisebox{0em}{\bettershortstack[c]{0.7}{2.5}{\small Quantum optics}} &
  \raisebox{0em}{\bettershortstack[c]{0.7}{2.5}{\small ~~~~~~This work}} \\
\midrule
\raisebox{0em}{\bettershortstack[l]{0.7}{2.5}{\small \!\!\!\!\!Photoelectric effect (spectrum)}} & 
  \raisebox{0em}{\bettershortstack[c]{0.7}{2.5}{\small ($\checkmark$) \cite{Einstein1905}}} &
  \raisebox{0em}{\bettershortstack[c]{0.7}{2.5}{\small $\checkmark$ \cite[{\S}14.2]{Mandel}}} &
  \raisebox{0em}{\bettershortstack[c]{0.7}{2.5}{\small $\checkmark$ Eqs.\eqref{eeExample1-35}--\eqref{eeExample1-36}}} \\ 
\raisebox{0em}{\bettershortstack[l]{0.7}{2.5}{\small \!\!\!\!\!Photoelectric effect (angular spectrum)}} & 
  \raisebox{0em}{\bettershortstack[c]{0.7}{2.5}{\small ---}} &
  \raisebox{0em}{\bettershortstack[c]{0.7}{2.5}{\small $\checkmark$ \cite[{\S}5.6]{Loudon}}} &
  \raisebox{0em}{\bettershortstack[c]{0.7}{2.5}{\small $\checkmark$ Eqs.\eqref{eeExample1-35}--\eqref{eeExample1-36}}} \\
\raisebox{0em}{\bettershortstack[l]{0.7}{2.5}{\small \!\!\!\!\!Blackbody radiation (spectrum)}} & 
  \raisebox{0em}{\bettershortstack[c]{0.7}{2.5}{\small $\checkmark$ \cite{Einstein1906,Einstein1917}}} &
  \raisebox{0em}{\bettershortstack[c]{0.7}{2.5}{\small $\checkmark$ \cite[{\S}13.1]{Mandel}}} &
  \raisebox{0em}{\bettershortstack[c]{0.7}{2.5}{\small $\checkmark$ ($\leftarrow$)}} \\
\raisebox{0em}{\bettershortstack[l]{0.7}{2.5}{\small \!\!\!\!\!Blackbody radiation (fluctuations)}} & 
  \raisebox{0em}{\bettershortstack[c]{0.7}{2.5}{\small ($\checkmark$) \cite{Einstein1909a,Einstein1909b}}} &
  \raisebox{0em}{\bettershortstack[c]{0.7}{2.5}{\small $\checkmark$ \cite[{\S}13.3]{Mandel}}} &
  \raisebox{0em}{\bettershortstack[c]{0.7}{2.5}{\small $\checkmark$ Eq.\eqref{eeFluctuation5}}} \\ 
\raisebox{0em}{\bettershortstack[l]{0.7}{2.5}{\small \!\!\!\!\!Bose-Einstein Condensation (laser)}} & 
  \raisebox{0em}{\bettershortstack[c]{0.7}{2.5}{\small ($\checkmark$) \cite{Bose1924,Einstein1924}}} &
  \raisebox{0em}{\bettershortstack[c]{0.7}{2.5}{\small $\checkmark$ \cite[Chap.7]{Loudon}}} &
  \raisebox{0em}{\bettershortstack[c]{0.7}{2.5}{\small $\checkmark$ ($\leftarrow$)}} \\
\raisebox{0em}{\bettershortstack[l]{0.7}{2.5}{\small \!\!\!\!\!Compton effect (angular spectrum)}} & 
  \raisebox{0em}{\bettershortstack[c]{0.7}{2.5}{\small ($\checkmark$) \cite{Einstein1917}}} &
  \raisebox{0em}{\bettershortstack[c]{0.7}{2.5}{\small ($\checkmark$) \cite{Yariv}\!\!\!\!\!\!}} &
  \raisebox{0em}{\bettershortstack[c]{0.7}{2.5}{\small ($\checkmark$) ($\leftarrow$)\!\!}} \\
\midrule
\raisebox{0em}{\bettershortstack[l]{0.7}{2.5}{\small \!\!\!\!\!Photoelectric effect (discontinuity)}} & 
  \raisebox{0em}{\bettershortstack[c]{0.7}{2.5}{\small ($\checkmark$) \cite{Einstein1905}}} &
  \raisebox{0em}{\bettershortstack[c]{0.7}{2.5}{\small (---) (``photon'')}} &
  \raisebox{0em}{\bettershortstack[c]{0.7}{2.5}{\small $\checkmark$ Eq.\eqref{eeExample1-2}}} \\
\raisebox{0em}{\bettershortstack[l]{0.7}{2.5}{\small \!\!\!\!\!Compton effect (coincidence)}} & 
  \raisebox{0em}{\bettershortstack[c]{0.7}{2.5}{\small ($\checkmark$) \cite{Einstein1917}}} &
  \raisebox{0em}{\bettershortstack[c]{0.7}{2.5}{\small (---) (``photon'')}} &
  \raisebox{0em}{\bettershortstack[c]{0.7}{2.5}{\small ($\checkmark$) Eqs.\eqref{eeExample2-39}--\eqref{eeExample2-40}}}\\
\raisebox{0em}{\bettershortstack[l]{0.7}{2.5}{\small \!\!\!\!\!EPR correlations (coincidence)}} & 
  \raisebox{0em}{\bettershortstack[c]{0.7}{2.5}{\small ---~\cite{EPR}}} &
  \raisebox{0em}{\bettershortstack[c]{0.7}{2.5}{\small ($\checkmark$) }} &
  \raisebox{0em}{\bettershortstack[c]{0.7}{2.5}{\small $\checkmark$ Eqs.\eqref{eeExample4-5}--\eqref{eeExample4-5a}}} \\
\raisebox{0em}{\bettershortstack[l]{0.7}{2.5}{\small \!\!\!\!\!Quantum jump (discontinuity)}} & 
  \raisebox{0em}{\bettershortstack[c]{0.7}{2.5}{\small (---) \cite{Einstein1905}}} &
  \raisebox{0em}{\bettershortstack[c]{0.7}{2.5}{\small (---) (``photon'')}} &
  \raisebox{0em}{\bettershortstack[c]{0.7}{2.5}{\small $\checkmark$ Eq.\eqref{eeExample1-2}}} \\
\bottomrule
\end{tabular}
\end{table}
This section supplements the experimental basis for the light quantum hypothesis. 
Einstein's 1905 paper~\cite{Einstein1905} was the first to point out that individual events in the photoelectric effect cannot be adequately represented by classical electromagnetism. 
Einstein used this discovery as a starting point to construct the light quantum hypothesis based on energy conservation in individual trials and wave-particle duality~\cite{Einstein1905,Einstein1906,Einstein1917,Bose1924,Einstein1924}. 
On the other hand, Bohr's 1913 paper~\cite{Bohr1913} was the first to suggest that individual events in the photoelectric effect could be explained without the light quantum hypothesis. 
The traditional explanation of the light quantum hypothesis is that it was verified when the Bohr-Kramers-Slater (BKS) theory~\cite{Bohr1924} proposed by Bohr was experimentally refuted~\cite{Compton1925,Bothe1,Bothe2}. 
However, with the subsequent quantization of atomic systems and radiation fields by non-relativistic QED~\cite{Dirac1927,Fermi1932}, and the development of wave representations of radiation fields through the quantum theory of coherence~\cite{Glauber1963} and semi-classical theory of radiation~\cite{Jaynes1963,Lamb1964,Lamb1968}, it is now possible to explain the statistical properties of radiation fields without using the light quantum hypothesis~\cite{Antiphoton,Knight1983}. 
To confirm this, the upper part of Table \ref{tLightQuantum1} shows literature on explaining the statistical properties of radiation fields. 
As the purpose is to confirm known facts, citations are limited to Einstein's paper and standard textbooks. While relativistic calculations~\cite{Bethe,Klein-Nishina} are necessary for computing the spectral shape of Compton scattering, the frequency-angle spectrum of inelastic scattering~\cite{Yariv} can be semi-quantitatively explained based on phase matching conditions even within the scope of non-relativistic QED.
\par
Therefore, in modern times, the only experimental facts that could serve as evidence for the existence of photons are the discontinuity of individual events and deterministic relations between individual events. 
The lower part of Table \ref{tLightQuantum1} summarizes the status of explanations for the discontinuity of individual events and deterministic relations. 
The problem with modern quantum optics theory is that, due to constraints of standard formulation (Section \ref{cIntro}), it cannot directly represent the discontinuous changes in the numbers of the emitted photoelectron or the deterministic relations in scattering directions in the Compton effect. 
This is likely why the concept of photons could not be eliminated from quantum optics theory despite repeated criticisms. 
This situation is aptly expressed by Glauber's statement, ``A photon is what a detector detects.'' 
However, as the light quantum hypothesis belongs to old quantum theory, its explanations of deterministic relations between spins and quantum jumps between stationary states are insufficient. 
In contrast, our theory can explicitly represent discontinuous changes in the occurrence of individual events (Supplementary Material B) and various deterministic relations ({\S}\ref{sBellCHSH1}, Supplementary Material B) by using logical variables to represent the occupation of stationary states. 
Therefore, while our theory has constraints due to the non-relativistic approximation, it is considered to be a potential replacement for the light quantum hypothesis as a substructure of non-relativistic QED.

\section*{Supplementary material B: Examples}
\setcounter{section}{2}
\setcounter{subsection}{0}
\subsection{Photoelectric effect in hydrogen atom}
\begin{figure}[t]
\centering
\begin{tikzpicture}
  \draw (-4,0)--(-0.3,0);
  \node (Incident) at (-2.3,0.3) {\small Incident light};
  \fill[gray] (0,0) circle (0.1);
  \draw[->] (30:0.25) -- (30:2.5);
  \draw[->] (210:0.25) -- (210:0.8);
  \node (E) at (2.5,1.4) {\small $e^{-}$};
  \node (P) at (-1,-0.6) {\small $p^{+}$};
  \node (R) at (0.4,-0.4) {\small ${\bm R}={\bm 0}$};
  \node (R) at (2.9,1.0) {\small ${\bm r}_{\mathrm{e}}$};
  \node (R) at (-0.6,-1.0) {\small ${\bm r}_{\mathrm{p}}$};
  \node[draw,rectangle] (Sca) at (-7.7,1) {\begin{tabular}{p{57mm}}\small ~\\\small $\{\!\!\{+\}\!\!\}$:~Scattering states ($E(\phi^{\bm r})\,{>}\,0$)\\\small ~\end{tabular}};
  \node[draw,rectangle] (Bou) at (-7.7,-0.7) {\begin{tabular}{p{57mm}}\small ~\\\small ~~$\{\!\!\{-\}\!\!\}$:~Bound states ($E(\phi^{\bm r})\,{\leqq}\,0$)\\\small ~\end{tabular}};
  \draw[->,very thick] (-7.7,-0.2)--(-7.7,0.5);
\end{tikzpicture}
\caption{Schematic of the photoionization of a hydrogen atom. A hydrogen atom, initially in one of the bound states, undergoes a discontinuous transition to one of the scattering states upon exposure to monochromatic light. The left figure shows a simplified energy diagram of the relative motion. $\stateset{-}$ and $\stateset{+}$ denote the entire sets of the bound and scattering states, where $-$ and $+$ stand for the sign of the eigenenergy $E(\phi^{\bm r})$. The right figure schematically shows the trajectories of the emitted electron and proton detected, for example, by a cloud chamber. The coordinates of the electron and proton are denoted by ${\bm r}_{\mathrm{e}}$ and ${\bm r}_{\mathrm{p}}$, respectively. The emission directions are correlated even in a single trial, as shown in Compton scattering experiments~\cite{Compton1925,Bothe1,Bothe2}.}
\label{figHydrogen1}
\end{figure}
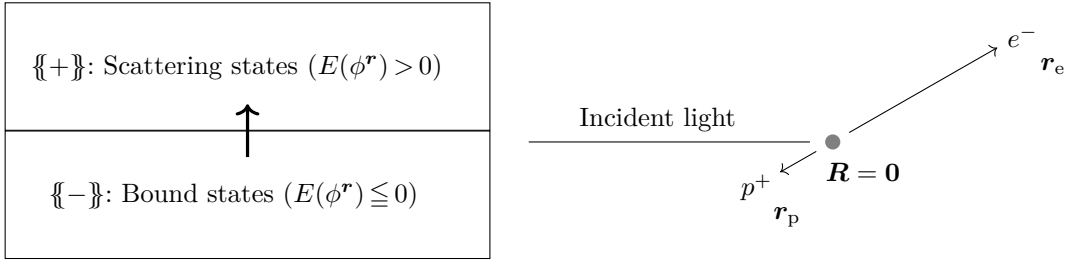
%
\subsubsection{General formula}
In this section, we demonstrate the calculation of the photoelectric effect in a hydrogen atom~\cite{Loudon,Bethe} as an example where analytical calculation is possible.
First, based on our model of reality, we model the individual events of the photoelectric effect in a hydrogen atom as a transition from a bound state to a scattering state of the hydrogen atom (Assumptions B1-B8):
\begin{align}
\xjvar{`ionized,\,$t$'}{big}
& \equiv \begin{cases}
  1, & \mbox{The atom has been photoionized by $t$}\\
  0, & \mbox{The atom has not been photoionized by $t$}
\end{cases}
\label{eeExample1-1}\\
& = \xjvar{`$\occupation{\stateset{+}}{},t$'$\wedge$`$\occupation{\stateset{-}}{},0$'}{Big}
,\quad t\,{>}\,0
,\quad j\,{\in}\,{\mathbb{N}}.
\label{eeExample1-2}
\end{align}
Here, $j$ is the trial number, and $\stateset{-}$ and $\stateset{+}$ represent all bound states and scattering states of the atom, respectively (Fig.\ref{figHydrogen1}). 
The expectation value of this logical variable is defined for the initial ensemble specified by the proposition `${\mathrm{A}}_{0},0$' and density operator $\hat{\rho}(\mbox{`${\mathrm{A}}_{0},0$'})$ as (Assumption B9):
\begin{align}
\big\langle \xvar{`$\mathrm{ionized},\,t$'}{big} \big\rangle
\equiv \frac{\sum_{j}^{\mbox{\scriptsize `${\mathrm{A}}_{0},0$'}} \xjvar{`$\mathrm{ionized},\,t$'}{big}}{\sum_{j}^{\mbox{\scriptsize `${\mathrm{A}}_{0},0$'}} 1}.
\label{eeExample1-3}
\end{align}
This has the meaning of the probability of the proposition `ionized,\,$t$'. 
Therefore, to find the rate of occurrence of the photoelectric effect, it is sufficient to calculate the expectation value of the logical variable using the laws of quantum theory (Assumptions B10-B13) and Eq.\eqref{eeExample1-2}, and then find its time derivative. 
The specific calculation process is as follows:
\begin{align}
& \big\langle \xvar{`$\mathrm{ionized},\,t$'}{big} \big\rangle
\nonumber\\
& = {\mathrm{Tr}}\Big[\,\xop{`$\mathrm{ionized},\,t$'}{big}\,\hat{\rho}(\mbox{`${\mathrm{A}}_{0},0$'})\,\Big]
\label{eeExample1-4}\\
& = {\mathrm{Tr}}\Big[\,\xop{`$\occupation{\stateset{+}}{},t$'${\wedge}$`$\occupation{\stateset{-}}{},0$'}{Big}\,\hat{\rho}(\mbox{`${\mathrm{A}}_{0},0$'})\,\Big]
\label{eeExample1-5}\\
& = {\mathrm{Tr}}\Big[\,\zopd{`$\occupation{\stateset{-}}{},0$'}{Big}\,\zopd{`$\occupation{\stateset{+}}{},t$'}{Big}\,\zop{`$\occupation{\stateset{+}}{},t$'}{Big}\,\zop{`$\occupation{\stateset{-}}{},0$'}{Big}\,\hat{\rho}(\mbox{`${\mathrm{A}}_{0},0$'})\,\Big]
\label{eeExample1-6}\\
& = {\mathrm{Tr}}\Big[\,\hat{U}^{\dagger}(t)\,\zop{`$\occupation{\stateset{+}}{}$'}{Big}\,\hat{U}(t)\,\hat{\rho}(\mbox{`${\mathrm{A}}_{0},0$'})\,\Big]
\label{eeExample1-7}
\end{align}
Here, $\hat{U}(t)\,{\equiv}\,\exp({t}\hat{H}/{{\rm{i}}\hbar})$ is the time evolution operator, and we know that the partial characteristic operator takes the following form according to the Born rule:
\begin{align}
\zop{`$\occupation{\stateset{+}}{}$'}{Big} 
= \sum_{|\phi\rangle}^{\stateset{+}} |\phi\rangle\langle\phi|
,\quad
\zop{`$\occupation{\stateset{-}}{}$'}{Big} 
= \sum_{|\phi\rangle}^{\stateset{-}} |\phi\rangle\langle\phi|
\label{eeExample1-8}
\end{align}
To obtain Eq.\eqref{eeExample1-7}, we assumed that the hydrogen atom is in a bound state at the initial time.
Also, due to the orthogonality of partial characteristic operators, the compatibility condition (Assumption B14) holds between propositions `$\occupation{\stateset{+}}{}$' and `$\occupation{\stateset{-}}{}$', so the use of logical product ($\wedge$) in Eq.\eqref{eeExample1-2} and Eq.\eqref{eeExample1-5} is justified. Eq.\eqref{eeExample1-7} can be evaluated under the first-order perturbation approximation
\begin{align}
& \hat{U}(t)
\cong \hat{U}_{0}(t) + \hat{U}_{1}(t)
\label{eeExample1-11}\\
& \hat{U}_{0}(t) \equiv \exp\left(\frac{t}{{\mathrm{i}}\hbar}\hat{H}_{0}\right)
,\quad
\hat{U}_{1}(t) \equiv \hat{U}_{0}(t)\,\frac{1}{{\rm{i}}\hbar}\int_{0}^{t} \!\!\tilde{H}_{\mathrm{I}}(t')dt'
\label{eeExample1-12}\\
& \tilde{H}_{\mathrm{I}}(t) 
\equiv \hat{U}^{\dagger}_{0}(t)\hat{H}_{\mathrm{I}}\hat{U}_{0}(t)
\label{eeExample1-13}
\end{align}
and the long-wavelength approximation, resulting in the following form~\cite{Theory2}:
\begin{align}
& \frac{d}{dt}\big\langle \xvar{`$\mathrm{ionized},\,t$'}{big} \big\rangle
\cong {\rm{Tr}}\left[\,\Big(\,{\tilde{\bm A}^{(-)}_{{\mathrm{T}}}}({\bm R},t)\,{\cdot}\,\hat{\bm \kappa}\,{\cdot}\,{\tilde{\bm A}^{(+)}_{{\mathrm{T}}}}({\bm R},t)\,\Big)\,\hat{\rho}(\mbox{`${\mathrm{A}}_{0},0$'})\,\right]
\phantom{\Big|}
\label{eeExample1-14}\\
& \hat{\bm \kappa}
\cong 2\pi\sum_{|\phi\rangle}^{\stateset{+}}\sum_{|\phi'\rangle}^{\stateset{-}} 
(\hbar\omega_{\phi}-\hbar\omega_{\phi'})^{2} 
\langle{\phi'}|\hat{\bm D}|\phi\rangle \langle\phi|\hat{\bm D}|\phi'\rangle
\, |\phi'\rangle\langle{\phi'}|
\label{eeExample1-15}
\end{align}
Here, ${\bm R}$ (${\in}\,{\mathbb{R}^{3}}$) is the position of the center of mass of the hydrogen atom, $\hat{\bm D}\equiv\sum_{k}^{{\mathrm{e}},{\mathrm{p}}} e_{k}\hat{\bm r}_{k}$ is the electric dipole moment operator of the hydrogen atom, and ${\tilde{\bm A}^{(+)}_{{\mathrm{T}}}}$ and ${\tilde{\bm A}^{(-)}_{{\mathrm{T}}}}$ are the positive and negative oscillation parts of the transverse component of the vector potential in the interaction picture. 
Eq.\eqref{eeExample1-15} matches the formula for light detection probability known in the field of quantum optics~\cite[{\S}14.2]{Mandel}, confirming the validity of the modeling in Eq.\eqref{eeExample1-2}.

\subsubsection{Frequency-angle spectrum of the photoelectric effect in hydrogen atom}
For more specific calculations, we need to separate the Hamiltonian of the hydrogen atom $\hat{H}_{\mathrm{A}}$ and the Hamiltonian of the radiation field $\hat{H}_{\mathrm{F}}$, determine the specific stationary states, and further specify the form of the density operator.
Here, we consider the case where, at the initial time, the relative motion of the hydrogen atom is in the 1s state $|\mathrm{1s}\rangle^{\!\bm r}$, the spin is in the singlet state $|0\rangle^{\!\mathrm{S}}$, and the center of mass is in a quantum state $|{\bm 0}\rangle^{\!\bf R}$ near the origin at rest. 
The quantum state $|{\bm 0}\rangle^{\!\bf R}$ is not an eigenstate of the hydrogen atom Hamiltonian $\hat{H}_{\mathrm{A}}$, but a proxy representation ({\S}\ref{sProxy}) of the stationary state of the entire system localized by interaction with the external environment.
Also, we assume that only the ${\bm k}\lambda$ mode of the radiation field is in the stationary state $|\mbox{rad}\rangle^{\!{\bm k}\lambda}$, and other modes are in the vacuum state $|\mbox{vac}\rangle^{\!\overline{{\bm k}\lambda}}$.
Here, ${\bm k}\,{\in}\,{\mathbb{R}^{3}}$ is the wave vector, and $\lambda\,{\in}\,\{1,2\}$ represents the polarization component. 
\begin{align}
\hat{\rho}(\mbox{`${\mathrm{A}}_{0},0$'})
& = |{\mathrm{1s}}\rangle^{\!\bm r}{}^{\bm r}\!\langle{\mathrm{1s}}| 
\otimes |0\rangle^{\!\mathrm{S}}{}^{\mathrm{S}}\!\langle 0|
\otimes |{\bm 0}\rangle^{\!\bf R}{}^{\bf R}\!\langle{\bm 0}| 
\otimes |\mbox{rad}\rangle^{\!{\bm k}\lambda}{}^{{\bm k}\lambda}\!\langle\mbox{rad}|
\otimes |\mbox{vac}\rangle^{\!\overline{{\bm k}\lambda}}\,{}^{\!\overline{{\bm k}\lambda}}\!\langle\mbox{vac}|
\label{eeExample1-21}
\end{align}
In the following, to separate the motion of the atomic system and the radiation field, we consider cases where the radiation field $|\mbox{rad}\rangle^{\!{\bm k}\lambda}$ at the initial time is either the $n\,{=}\,1$ Fock state $|\mbox{1}\rangle^{\!{\bm k}\lambda}$ or the coherent state $|\tilde{\alpha}\rangle^{\!{\bm k}\lambda}$ ($\tilde{\alpha}\,{\in}\,{\mathbb{C}}$).
\par
Then, by substituting Eq.\eqref{eeExample1-8} and Eq.\eqref{eeExample1-21} into equation Eq.\eqref{eeExample1-7}, under the first-order perturbation approximation and long-wavelength approximation, we obtain
\begin{align}
\big\langle \xvar{`ionized,$t$'}{big} \big\rangle
= \iiint {}^{\bm r}\!\langle{\bm p}|\,\hat{\rho}^{\bm r}(\,t\,|\mbox{`${\mathrm{A}}_{0},0$'})\,|{\bm p}\rangle^{\!\bm r}\,d{\bm p}.
\label{eeExample1-22}
\end{align}
Here, we denoted the scattering state with momentum ${\bm p}$ as $|{\bm p}\rangle^{\!\bm r}$. 
The scattering state $|{\bm p}\rangle^{\!\bm r}$ includes a plane wave with wave number ${\bm p}/\hbar$ and a near-field term decaying as ${\approx}\,1/\|{\bm r}\|$~\cite{Bethe}. 
Also, the reduced density operator for relative motion $\hat{\rho}^{\bm r}(\,t\,|\mbox{`${\mathrm{A}}_{0},0$'})$ can be expressed under the above conditions as~\cite{Theory2}
\begin{align}
\hat{\rho}^{\bm r}(\,t\,|\mbox{`${\mathrm{A}}_{0},t_{0}$'}) 
\cong \hat{U}^{\bm r}(t)|\mathrm{1s}\rangle\langle\mathrm{1s}|\hat{U}^{\bm r}(t)
= |\psi(t)\rangle^{\!\bm r} {}^{\bm r}\!\langle\psi(t)\big|.
\label{eeExample1-23}
\end{align}
Here, we defined the time evolution operator for relative motion $\hat{U}^{\bm r}(t)$ as
\begin{align}
& \hat{U}^{\bm r}(t) 
\equiv \begin{cases}
  \displaystyle  
    {}^{{\bm k}\lambda}\!\langle 1|\,\hat{U}^{{\bm r}{{\bm k}\lambda}}_{0}(t)\,|1\rangle^{\!{\bm k}\lambda} 
    + {}^{{\bm k}\lambda}\!\langle{0}|\,\hat{U}^{{\bm r}{{\bm k}\lambda}}_{1}(t)\,|1\rangle^{\!{\bm k}\lambda}
  \phantom{\Big|}
  & ~~ \mbox{the case $|\mbox{rad}\rangle^{\!{\bm k}\lambda} = |1\rangle^{\!{\bm k}\lambda}$}\\
  \displaystyle 
    {}^{{\bm k}\lambda}\!\langle\tilde{\alpha}|\,\hat{U}^{{\bm r}{{\bm k}\lambda}}_{0}(t)\,|\tilde{\alpha}\rangle^{\!{\bm k}\lambda} 
    + {}^{{\bm k}\lambda}\!\langle\tilde{\alpha}|\,\hat{U}^{{\bm r}{{\bm k}\lambda}}_{1}(t)\,|\tilde{\alpha}\rangle^{\!{\bm k}\lambda}
  \phantom{\Big|}
  & ~~ \mbox{the case $|\mbox{rad}\rangle^{\!{\bm k}\lambda} = |\tilde{\alpha}\rangle^{\!{\bm k}\lambda}$}.
\end{cases}
\label{eeExample1-26}
\end{align}
with

\begin{align}
& \hat{U}^{{\bm r}{{\bm k}\lambda}}_{0}(t)
\equiv 
{}^{\mathrm{S}}\!\langle 0| 
{}^{\bf R}\!\langle{\bm 0}| 
\,{}^{\!\overline{{\bm k}\lambda}}\!\langle\mbox{vac}|
\,\hat{U}_{0}(t)\,
|\mbox{vac}\rangle^{\!\overline{{\bm k}\lambda}}
|{\bm 0}\rangle^{\!\bf R}
|0\rangle^{\!\mathrm{S}}
\label{eeExample1-24}\\
& \hat{U}^{{\bm r}{{\bm k}\lambda}}_{1}(t)
\equiv 
{}^{\mathrm{S}}\!\langle 0| 
{}^{\bf R}\!\langle{\bm 0}| 
\,{}^{\!\overline{{\bm k}\lambda}}\!\langle\mbox{vac}|
\,\hat{U}_{1}(t)\,
|\mbox{vac}\rangle^{\!\overline{{\bm k}\lambda}}
|{\bm 0}\rangle^{\!\bf R}
|0\rangle^{\!\mathrm{S}}.
\label{eeExample1-25}
\end{align}
This time evolution operator for relative motion corresponds to the time evolution operator in the semiclassical theory of radiation, as the operators of the radiation field are replaced by matrix elements. 
Therefore, the state vector of the relative motion of the hydrogen atom in this representation
\begin{align}
& |\psi(t)\rangle^{\!\bm r}
\equiv \hat{U}^{\bm r}(t)|\mathrm{1s}\rangle
,\quad t>0
\label{eeExample1-27}
\end{align}
becomes a pure state vector of the atomic system.
\par
Particularly, when the frequency of the radiation field is sufficiently high such that $\nu\,{\gg}\,|E({\mathrm{1s}})|{/}h$, we can express the state vector of the hydrogen atom $|\psi(t)\rangle^{\!\bm r}$ as a superposition of the 1s state and scattering states:
\begin{align}
|\psi(t)\rangle^{\!\bm r} 
= c_{\mathrm{1s}}(t)\,|\mathrm{1s}\rangle^{\!\bm r} + \iiint c_{\bm p}(t)\,|{\bm p}\rangle^{\!\bm r}\,d{\bm p}
,\quad t>0
\label{eeExample1-28}
\end{align}
In this case, due to the unitarity of $\hat{U}^{\bm r}(t)$, the following relation holds:
\begin{align}
\big|c_{\mathrm{1s}}(t)\big|^{2} + \iiint \big|c_{\bm p}(t)\big|^{2}\,d{\bm p} 
= 1.
\label{eeExample1-29}
\end{align}
Also, by substituting Eq.\eqref{eeExample1-23} and Eq.\eqref{eeExample1-28} into Eq.\eqref{eeExample1-22} and differentiating with respect to time, we obtain
\begin{align}
\frac{d}{dt}\big\langle \xvar{`$\mathrm{ionized},\,t$'}{big} \big\rangle
& = \frac{d}{dt}\iiint |c_{\bm p}(t)\big|^{2}\,d{\bm p}
\label{eeExample1-31}\\
& = \iiint \lim_{\Delta{t}{\to}0} \frac{1}{\Delta{t}} \left(|c_{\bm p}(t + \Delta{t})\big|^{2}- |c_{\bm p}(t)\big|^{2}\right)\,d{\bm p}.
\label{eeExample1-32}
\end{align}
Furthermore, by rewriting Eq.\eqref{eeExample1-31} using Eq.\eqref{eeExample1-29}, and evaluating $c_{\bm p}(t\,{+}\,\Delta{t})$ in Eq.\eqref{eeExample1-32} using first-order perturbation approximation, we obtain the following equation after going through the equation transformations described in references \cite{Loudon} and \cite{Bethe}:
\begin{align}
\frac{d}{dt}\big\langle \xvar{`$\mathrm{ionized},\,t$'}{big} \big\rangle
& = - \frac{d}{dt}\big|c_{\mathrm{1s}}(t)\big|^{2}
\label{eeExample1-33}\\
& \approx \mbox{const.} \iiint \frac{\cos^{2}({\bm p},{\bm e}_{\lambda})}{\nu^{7/2}}\,\delta\left( \frac{\|{\bm p}\|^{2}}{2m} + \Big|E_{\mathrm{1s}}\Big| - h\nu \right) d{\bm p}
\,\mbox{(light power)} \big|c_{\mathrm{1s}}(t)\big|^{2}.
\label{eeExample1-34}
\end{align}
Here, ${\bm e}_{\lambda}$ on the right-hand side is the polarization vector (direction of the electric field), and $({\bm p},{\bm e}_{\lambda})$ represents the angle between ${\bm p}$ and ${\bm e}_{\lambda}$. 
By integrating Eq.\eqref{eeExample1-34} with respect to time and applying the laws of Boolean logic (Assumption B6), we find that the probability of the hydrogen atom not being ionized decays according to the following equation:
\begin{align}
& \big\langle \xvar{$\lnot$`$\mathrm{ionized},\,t$'}{big} \big\rangle
= \exp\big(\,{-}\,\gamma_{\mathrm{1s}}{t}\,\big)
\label{eeExample1-35}\\
& \gamma_{\mathrm{1s}} \approx \mbox{const.} \iiint \frac{\cos^{2}({\bm p},{\bm e}_{\lambda})}{\nu^{7/2}}\,\delta\left( \frac{\|{\bm p}\|^{2}}{2m} + \Big|E_{\mathrm{1s}}\Big| - h\nu \right) d{\bm p}
\,{\times}\,\mbox{(light power)}.
\label{eeExample1-36}
\end{align}
This result correctly reproduces the main features of the photoelectric effect: the exponential decay of the probability of the proposition $\lnot\mbox{`$\mathrm{ionized},\,t$'}$ and the proportionality of the ionization rate to $\nu^{-7/2}\cos^{2}({\bm p},{\bm e}_{\lambda})$~\cite{Bethe}. 
Thus, we have rationally explained both the continuous (Eq.\eqref{eeExample1-35}) and discontinuous (Eq.\eqref{eeExample1-2}) aspects of the photoelectric effect in hydrogen atoms within the framework of non-relativistic QED, based on Assumptions B1-B14.

\subsection{Classical trajectriess and EPR paradox}
\subsubsection{Phenomenological representation}
\begin{figure}[t]
\centering
\includegraphics{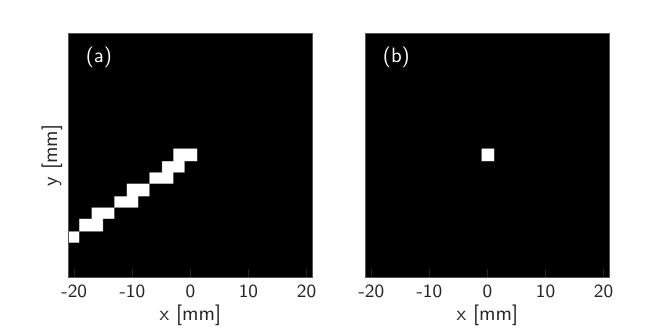}
\caption{Schematic illustration of the measurement of the electron position. (a) Presence or absence of water droplets as represented by $x^{(j)}(\mbox{`$\mathrm{droplet}_{{\mathcal{R}}},t$'})$ (black:0, white:1). (b) The position of the latest water droplet as represented by $x^{(j)}(\mbox{`o$(\{\!\!\{{\mathcal{R}}\}\!\!\}),t$'})$ (black:0, white:1).}
\label{figElectron}
\end{figure}
Next, we consider the representation of motion trajectories of electrons and protons produced in the photoelectric effect of hydrogen atoms. 
Traditionally, the representation of motion trajectories of electrons and photons produced in the Compton effect has been problematic. 
However, as this paper limits its consideration to non-relativistic theory, we will consider electrons and protons produced in the photoelectric effect instead of the Compton effect. 
This setting is not realistic experimentally due to the effect of electrons and protons being deflected by interaction with polar molecules in the cloud chamber. 
However, theoretically, there is a clear correspondence with the EPR paradox, as shown below.
\par
First, by dividing real space into discrete cells and denoting the center coordinates of each cell as ${\mathcal{R}}$, we can represent whether there is a water droplet in the cell at position ${\mathcal{R}}$ using the following logical variable (Fig.\ref{figElectron}(a)):
\begin{align}
\xjvar{`$\mathrm{droplet}, {{\mathcal{R}}},t$'}{big}
=\begin{cases}
1, & \mbox{When a water droplet exists in cell ${\mathcal{R}}$ at time $t$},\\
0, & \mbox{Otherwise}.
\end{cases}
\label{eeExample2-1}
\end{align}
Using this logical variable, we can express the latest water droplet position ${\mathcal{R}}^{(j)}(t)$ in the following form:
\begin{align}
& {\mathcal{R}}^{(j)}(t_{h})
= \sum_{{\mathcal{R}}} {\mathcal{R}}\,\xjvar{`$\occupation{\stateset{\mathcal{R}}}{},t_{h}$'}{big}
,\quad
h \in {\mathbb{N}}
\label{eeExample2-2}\\
& \xjvar{`$\occupation{\stateset{\mathcal{R}}}{},t_{h}$'}{big}
= \xjvar{`$\mathrm{droplet}, {{\mathcal{R}}}, t_{h}$'}{big}
- \xjvar{`$\mathrm{droplet}, {{\mathcal{R}}}, t_{h-1}$'}{big}
\label{eeExample2-3}
\end{align}
Here, $h$ (${\in}\,{\mathbb{N}}$) is the sequential number of the cell where the water droplet appeared, and $t_{h}$ represents the time when a water droplet appeared in the $h$-th cell. 
$\stateset{\mathcal{R}}$ is the set of stationary states of the entire system including charged particles and water droplets where a water droplet exists at ${\mathcal{R}}$. 
$\sum_{{\mathcal{R}}}$ represents the sum over all cells.

\subsubsection{Physical quantities in systems with and without measurement instruments}
Furthermore, we consider cases where different charged particle trajectories can be distinguished based on the continuity of water droplet trajectories.
In this case, we can phenomenologically write the electron position ${\mathcal{R}}^{(j)}_{\mathrm{e}}(t)$ and proton position ${\mathcal{R}}^{(j)}_{\mathrm{p}}(t)$ in the following form:
\begin{align}
{\mathcal{R}}^{(j)}_{k}(t)
& = \sum_{{\mathcal{R}}_{k}} {\mathcal{R}}_{k} \, \xjvar{`$\occupation{\stateset{{\mathcal{R}}_{k}}}{},t$'}{big}
,\quad
k \in \{\mathrm{e},\mathrm{p}\}
\label{eeExample2-4}
\end{align}
Here, $\stateset{\mathcal{R}_{k}}$ is the set of stationary states of the entire system including water droplets where the water droplet created by the $k$-th charged particle is at ${\mathcal{R}_{k}}$.
Particularly, in measurements with an applied magnetic field, by substituting the curvature radius of the trajectory and the magnetic field into the classical mechanics formula, we can estimate the electron momentum ${\mathcal{P}}^{(j)}_{\mathrm{e}}(t)$ and proton momentum ${\mathcal{P}}^{(j)}_{\mathrm{p}}(t)$:
\begin{align}
{\mathcal{P}}^{(j)}_{k}(t)
& = \sum_{{\mathcal{P}}_{k}} {\mathcal{P}}_{k} \, \xjvar{`$\occupation{\stateset{{\mathcal{P}}_{k}}}{},t$'}{big}
,\quad
k \in \{\mathrm{e},\mathrm{p}\}.
\label{eeExample2-5}
\end{align}
Here, $\stateset{\mathcal{P}_{k}}$ is the set of stationary states of the entire system corresponding to the water droplet configuration that formed the basis for the estimated momentum value ${\mathcal{P}_{k}}$.
While it is difficult to explicitly write down the functional form of the stationary states constituting $\stateset{\mathcal{R}_{k}}$ and $\stateset{\mathcal{P}_{k}}$, the stationary states constituting $\stateset{\mathcal{R}_{k}}$ and $\stateset{\mathcal{P}_{k}}$ are orthogonalized to each other by definition. Therefore, the compatibility condition holds between the propositions `$\occupation{\stateset{{\mathcal{R}}_{k}}}{},t$' ($k\,{\in}\,\{\mathrm{e},\mathrm{p}\}$) and `$\occupation{\stateset{{\mathcal{P}}_{k'}}}{},t$' ($k'\,{\in}\,\{\mathrm{e},\mathrm{p}\}$):
\begin{align}
& {\rm{Tr}}\Big[\,\zopd{`$\occupation{\stateset{{\mathcal{P}}_{k'}}^{k'}}{big}$'}{big}\,\zopd{`$\occupation{\stateset{\mathcal{R}}^{k}}{big}$'}{big}\,\zop{`$\occupation{\stateset{\mathcal{R}}^{k}}{big}$'}{big}\,\zop{`$\occupation{\stateset{{\mathcal{P}}_{k'}}^{k'}}{big}$'}{big}\,\hat{\wp}(\,t\,|\mbox{${\mathcal{A}}_{0},0$})\,\Big]
\nonumber\\
& = {\rm{Tr}}\Big[\,\zopd{`$\occupation{\stateset{\mathcal{R}}^{k}}{big}$'}{big}\,\zopd{`$\occupation{\stateset{{\mathcal{P}}_{k'}}^{k'}}{big}$'}{big}\,\zop{`$\occupation{\stateset{{\mathcal{P}}_{k'}}^{k'}}{big}$'}{big}\,\zop{`$\occupation{\stateset{\mathcal{R}}^{k}}{big}$'}{big}\,\hat{\wp}(\,t\,|\mbox{${\mathcal{A}}_{0},0$}) \,\Big],
\nonumber\\
& \quad k \in \{\mathrm{e,p}\},~k' \in \{\mathrm{e,p}\}.
\label{eeExample2-6}
\end{align}
Here, $\hat{\wp}(\,t\,|\mbox{${\mathcal{A}}_{0},0$})$ represents the density operator of the entire system including water droplets.
While Eq.\eqref{eeExample2-6} is a formal equation, it is consistent with the fact that position and momentum can be estimated with finite precision in experiments.
\par
On the other hand, the objects of theoretical calculations are the positions and momenta of charged particles in an idealized system without water droplets.
The instantaneous values of these quantities can be defined by the following equations:
\begin{align}
{\bm r}^{(j)}_{k}(t) 
& = \sum_{{\bm r}_{k}} {\bm r}_{k}\,\xjvar{`$\occupation{\stateset{{\bm r}_{k}}^{k}}{},t$'}{}
,\quad
k \in \{{\mathrm{e}},\,{\mathrm{p}}\}
\label{eeExample2-11}\\
{\bm p}^{(j)}_{k}(t) 
& = \sum_{{\bm p}_{k}} {\bm p}_{k}\,\xjvar{`$\occupation{\stateset{{\bm p}_{k}}^{k}}{},t$'}{}
,\quad
k \in \{{\mathrm{e}},\,{\mathrm{p}}\}
\label{eeExample2-12}
\end{align}
Here, $\stateset{{\bm r}_{k}}^{k}$ is the set of eigenstates of the position operator $\hat{\bm r}_{k}$, $\stateset{{\bm p}_{k}}^{k}$ is the set of eigenstates of the momentum operator $\hat{\bm p}_{k}$, defined by
\begin{align}
\stateset{{\bm r}_{k}}^{k}
& \equiv \left\{\,|\varphi\rangle\,\Big|\, 
\hat{{\bm r}}_{k}\,|\varphi\rangle\,{=}\,{\bm r}(\varphi)|\varphi\rangle
,~\langle\varphi|\varphi'\rangle\,{=}\,\delta_{\varphi,\varphi'}
,~{\bm r}(\varphi)\,{\in}\,\mbox{cell containing ${\bm r}$}\right\}
\label{eeExample2-13}\\
\stateset{{\bm p}_{k}}^{k}
& \equiv \left\{\,|\varsigma\rangle\,\Big|\, 
\hat{{\bm p}}_{k}\,|\varsigma\rangle\,{=}\,{\bm p}(\varsigma)|\varsigma\rangle
,~\langle\varsigma|\varsigma'\rangle\,{=}\,\delta_{\varsigma,\varsigma'}
,~{\bm p}(\varsigma)\,{\in}\,\mbox{cell containing ${\bm p}$}\right\}.
\label{eeExample2-14}
\end{align}
Also, the characteristic operators and partial characteristic operators of the propositions `$\occupation{\stateset{{\bm r}_{k}}^{k}}{},t$' and `$\occupation{\stateset{{\bm p}_{k}}^{k}}{},t$' have the following form:
\begin{align}
& \xop{`$\occupation{\stateset{{\bm r}_{k}}^{\!k}}{leftright}$'}{big}
= \zop{`$\occupation{\stateset{{\bm r}_{k}}^{\!k}}{leftright}$'}{big}
= \iiint_{\mbox{\scriptsize{cell ${\bm r}$}}} \!\!\! |{\bm r}'\rangle^{\!k}{}^{k}\!\langle {\bm r}'|\,d{\bm r}'
\label{eeExample2-15}\\
& \xop{`$\occupation{\stateset{{\bm p}_{k'}}^{\!k'}}{leftright}$'}{big}
= \zop{`$\occupation{\stateset{{\bm p}_{k'}}^{\!k'}}{leftright}$'}{big}
= \iiint_{\mbox{\scriptsize{cell ${\bm p}$}}} \!\!\! |{\bm p}'\rangle^{\!k'}{}^{k'}\!\langle {\bm p}'|\,d{\bm p}'
\label{eeExample2-16}
\end{align}

\subsubsection{Deterministic Relations}
In experiments using cloud chambers, it's common to identify the charged particle trajectories ${\bm r}^{(j)}_{k}$ and ${\bm p}^{(j)}_{k}$ (Eqs.\eqref{eeExample2-11}--\eqref{eeExample2-12}) with ${\mathcal{R}}^{(j)}_{k}$ and ${\mathcal{P}}^{(j)}_{k}$ (Eqs.\eqref{eeExample2-4}--\eqref{eeExample2-5}) determined by the water droplet configuration. 
This relation can be expressed by the following deterministic relation:
\begin{align}
& \xjvar{`$\occupation{\stateset{{\bm r}_{k}}^{k}}{},t$'}{}
= \xjvar{`$\occupation{\stateset{{\mathcal{R}}_{k}}}{},t$'}{}
~~\mbox{for}~~
{\bm r}_{k} = {\mathcal{R}}_{k}
,~~ k \in \{\mathrm{e},\mathrm{p}\}
,~~ {\forall}j
\label{eeExample2-23}\\
& \xjvar{`$\occupation{\stateset{{\bm p}_{k}}^{k}}{},t$'}{}
= \xjvar{`$\occupation{\stateset{{\mathcal{P}}_{k}}}{},t$'}{}
~~\mbox{for}~~
{\bm p}_{k} = {\mathcal{P}}_{k}
,~~ k \in \{\mathrm{e},\mathrm{p}\}
,~~ {\forall}j.
\label{eeExample2-24}
\end{align}
Also, by substituting Eqs.\eqref{eeExample2-23}--\eqref{eeExample2-24} into Eqs.\eqref{eeExample2-4}--\eqref{eeExample2-5} and Eqs.\eqref{eeExample2-11}--\eqref{eeExample2-12}, we find the following relationships:
\begin{align}
{\bm r}^{(j)}_{k}(t)
& = {\mathcal{R}}^{(j)}_{k}(t)
,~~ k \in \{\mathrm{e},\mathrm{p}\}
,~~ {\forall}j
\label{eeExample2-25}\\
{\bm p}^{(j)}_{k}(t)
& = {\mathcal{P}}^{(j)}_{k}(t)
,~~ k \in \{\mathrm{e},\mathrm{p}\}
,~~ {\forall}j.
\label{eeExample2-26}
\end{align}
For the joint probabilities of physical quantities ${\bm r}_{k}$ and ${\bm p}_{k}$ in the idealized system without water droplets to match the joint probabilities of physical quantities ${\mathcal{R}}_{k}$ and ${\mathcal{P}}_{k}$ in the realistic system with water droplets, the following compatibility condition must hold even in the subsystem without water droplets:
\begin{align}
& {\rm{Tr}}\Big[\,\zopd{`$\occupation{\stateset{{\bm p}_{k'}}^{k'}}{big}$'}{big}\,\zopd{`$\occupation{\stateset{{\bm r}_{k}}^{k}}{big}$'}{big}\,\zop{`$\occupation{\stateset{{\bm r}_{k}}^{k}}{big}$'}{big}\,\zop{`$\occupation{\stateset{{\bm p}_{k'}}^{k'}}{big}$'}{big}\,\hat{\rho}(\,t\,|\mbox{`${\mathrm{A}}_{0},0$'})\,\Big]
\nonumber\\
& = {\rm{Tr}}\Big[\,\zopd{`$\occupation{\stateset{{\bm r}_{k}}^{k}}{big}$'}{big}\,\zopd{`$\occupation{\stateset{{\bm p}_{k'}}^{k'}}{big}$'}{big}\,\zop{`$\occupation{\stateset{{\bm p}_{k'}}^{k'}}{big}$'}{big}\,\zop{`$\occupation{\stateset{{\bm r}_{k}}^{k}}{big}$'}{big}\,\hat{\rho}(\,t\,|\mbox{`${\mathrm{A}}_{0},0$'}) \,\Big],
\nonumber\\
& \quad k \in \{\mathrm{e,p}\},~k' \in \{\mathrm{e,p}\},
\label{eeExample2-28}
\end{align}
This equation holds unconditionally due to the commutativity of partial characteristic operators when the charged particles are different ($k'\,{\neq}\,k$). 
However, when the charged particles are the same ($k'\,{=}\,k$), because the position eigenstates and momentum eigenstates are not orthogonal, it's necessary to lower the precision of the approximate representation of the subsystem's physical quantities ${\bm r}_{k}$ and ${\bm p}_{k}$ to satisfy the compatibility condition.

\subsubsection{Probabilistic interpretation of wave functions and deterministic relations}
There are cases where the compatibility condition generally holds. 
First, when the position uncertainty $\delta{\bm r}\,{\equiv}\,\delta{\bm r}_{\mathrm{e}}\,{=}\,\delta{\bm r}_{\mathrm{p}}$ or momentum uncertainty $\delta{\bm p}\,{\equiv}\,\delta{\bm p}_{\mathrm{e}}\,{=}\,\delta{\bm p}_{\mathrm{p}}$ is infinite. 
When the momentum uncertainty $\delta{\bm p}$ is infinite, $\hat{z}(\mbox{`$\occupation{\stateset{{\bm p}_{k}}^{k}}{}$'})\,{\cong}\,\hat{1}$, so the compatibility condition holds even when the position uncertainty $\delta{\bm r}$ is arbitrarily small. 
Particularly, in the problem of photoionization of hydrogen atoms, because the density operator (Eq.\eqref{eeExample1-23}) has the form
\begin{align}
& \hat{\rho}(\,t\,|\mbox{`${\mathrm{A}}_{0},0$'})
\cong |\psi(t)\rangle^{\!\bm r}{}^{\bm r}\!\langle\psi(t)|
,\quad
|\psi(t)\rangle^{\!\bm r} 
= c_{\mathrm{1s}}(t)\,|\mathrm{1s}\rangle^{\!\bm r} + \iiint c_{\bm p}(t)\,|{\bm p}\rangle^{\!\bm r}\,d{\bm p},
\label{eeExample2-31}
\end{align}
we can calculate the joint probability of electron and proton positions using the following equation:
\begin{align}
\big\langle \xvar{`$\occupation{\stateset{{\bm r}_{\mathrm{e}}}^{\!{\mathrm{e}}}}{},t$'$\wedge$`$\occupation{\stateset{{\bm r}_{\mathrm{p}}}^{\!{\mathrm{p}}}}{},t$'}{big}\,\big\rangle
& = {\rm{Tr}}\Big[\,\xop{`$\occupation{\stateset{{\bm r}_{\mathrm{e}}}^{\!{\mathrm{e}}}}{}$'}{big}\,\xop{`$\occupation{\stateset{{\bm r}_{\mathrm{p}}}^{\!{\mathrm{p}}}}{}$'}{big}\,\hat{\rho}(\,t\,|\mbox{`${\mathrm{A}}_{0},0$'})\,\Big]
\label{eeExample2-32}\\
& = \iiint_{\mbox{\scriptsize{cell ${\bm r}_{\mathrm{e}}$}}} \!\!\!\!\!\!\!\!\! d{\bm r}_{\mathrm{e}}
\iiint_{\mbox{\scriptsize{cell ${\bm r}_{\mathrm{p}}$}}} \!\!\!\!\!\!\!\!\! d{\bm r}_{\mathrm{p}}
\left| \psi({\bm r}_{\mathrm{e}}, {\bm r}_{\mathrm{p}},t) \right|^{2},
\label{eeExample2-33}
\end{align}
where, we defined the two-particle wave function as
\begin{align}
\psi({\bm r}_{\mathrm{e}}, {\bm r}_{\mathrm{p}},t)
\equiv {}^{\mathrm{p}}\!\langle {\bm r}_{\mathrm{p}} |{}^{\mathrm{e}}\!\langle {\bm r}_{\mathrm{e}} | \psi(t) \rangle^{\!\bm r}|{\bm 0}\rangle^{\bf R}
= {}^{\bm R}\!\langle {\bm R} | {}^{\bm r}\!\langle {\bm r} |\psi(t) \rangle^{\!\bm r}|{\bm 0}\rangle^{\bf R}
= \psi^{\bm r}({\bm r},t)\,\delta_{{\bm R},{\bm 0}}.
\label{eeExample2-34}
\end{align}
This shows that the compatibility condition holds in the position representation ($\delta{\bm p}\,{\to}\,\infty$, $\delta{\bm r}\,{\to}\,0$), and the probabilistic interpretation of the two-particle wave function is valid. 
However, in this case, because the momentum uncertainty is infinite, the joint probability of electron and proton momenta is meaningless.
Similarly, when the position uncertainty $\delta{\bm r}$ is infinite, the compatibility condition holds even when the momentum uncertainty $\delta{\bm p}$ is small, and the probabilistic interpretation of the wave function in the momentum representation is valid.
\par
On the other hand, in the classical approximation ({\S}\ref{sClassical}), if we take the representation precisions $\delta{\bm r}\,{=}\,(\delta{x},\delta{y},\delta{z})$ and $\delta{\bm p}\,{=}\,(\delta{p}_{x},\delta{p}_{y},\delta{p}_{z})$ to satisfy the complementarity principle:
\begin{align}
\delta{x}\,\delta{p}_{x} \gg \frac{\hbar}{2}
,\quad
\delta{y}\,\delta{p}_{y} \gg \frac{\hbar}{2}
,\quad
\delta{z}\,\delta{p}_{z} \gg \frac{\hbar}{2},
\label{eeExample2-35}
\end{align}
the compatibility condition holds between any physical quantities defined as functions of position and momentum. 
Particularly, in the problem of photoionization of hydrogen atoms, because the two-particle wave function is separable into relative motion and center-of-mass motion, we can write the state vector as
\begin{align}
| \psi(t) \rangle^{\!{\mathrm{e}},{\mathrm{p}}}
& = | \psi(t) \rangle^{\!\bm r}|{\bm 0}\rangle^{\bf R}
= \sum_{{\bm r}_{\mathrm{e}}} \sum_{{\bm r}_{\mathrm{p}}} 
\psi({\bm r}_{\mathrm{e}}, {\bm r}_{\mathrm{p}},t)
\big| {\bm r}_{\mathrm{e}} \big\rangle^{\!{\mathrm{e}}}
\big| {\bm r}_{\mathrm{p}} \big\rangle^{\!{\mathrm{p}}}
\label{eeExample2-36}\\
& = \sum_{\bm r} \psi^{\bm r}({\bm r},t)
\Big| \frac{m_{\mathrm{p}}}{m_{\mathrm{e}}+m_{\mathrm{p}}}{\bm r} \Big\rangle^{\!\!{\mathrm{e}}}
\Big| -\frac{m_{\mathrm{e}}}{m_{\mathrm{e}}+m_{\mathrm{p}}}{\bm r} \Big\rangle^{\!\!{\mathrm{p}}},
\label{eeExample2-37}
\end{align}
showing that a entanglement occurs between the electron and proton positions.
\par
Therefore, for the positions and momenta of electrons and protons produced by photoionization, due to the general properties of subsystems in quantum entanglement ({\S}\ref{sEntanglement}), the following deterministic relations hold:
\begin{align}
\xjvar{`$\occupation{\stateset{{\bm r}_{\mathrm{e}}}^{\mathrm{e}}}{leftright},t$'}{big}
= \xjvar{`$\occupation{\stateset{{\bm r}_{\mathrm{p}}}^{\mathrm{p}}}{leftright},t$'}{big}
~~\mbox{for}~~
{\bm r}_{\mathrm{e}} = -\frac{m_{\mathrm{p}}}{m_{\mathrm{e}}}{\bm r}_{\mathrm{p}}
,\quad {\forall}j
\label{eeExample2-38}
\end{align}
Consequently, the following deterministic relation exists:
\begin{align}
{\bm r}^{(j)}_{\mathrm{e}}(t)
= -\frac{m_{\mathrm{p}}}{m_{\mathrm{e}}}\,{\bm r}^{(j)}_{\mathrm{p}}(t)
,\quad {\forall}j.
\label{eeExample2-39}
\end{align}
Similarly, for momentum, we can say that the following deterministic relation holds:
\begin{align}
{\bm p}^{(j)}_{\mathrm{e}}(t)
= - {\bm p}^{(j)}_{\mathrm{p}}(t)
,\quad {\forall}j
\label{eeExample2-40}
\end{align}
Thus, in our theory, the deterministic relations problematized in the EPR paradox hold. 
However, because the instantaneous values of position and physical quantities are defined by Eqs.\eqref{eeExample2-11}--\eqref{eeExample2-12}, and the precision of the propositions used in the definition is limited by the compatibility condition, positions and momenta do not have definite values in the sense of classical mechanics. 
Additionally, in the Copenhagen interpretation, considering that position and momentum take definite values contradicts the premises of the interpretation itself. 
In contrast, our theory is not based on the Copenhagen interpretation, so even if position and momentum take definite values, no logical contradiction arises.

\subsection{Photomultiplier tube and Schr\"{o}dinger's cat}
One of the main advantages of our theory is the ability to explicitly represent deterministic relations between successive individual events.
To see this, let us consider the case of amplifying an electron produced by the photoelectric effect using a photomultiplier tube. 
In this case, we can phenomenologically represent the presence or absence of the photomultiplier tube's output signal using the following logical variable:
\begin{align}
\xjvar{`output,$t+\tau^{(j)}$'}{big}
& = \xjvar{$\big(\lnot$`TN'$\wedge$`ionized,$t$'$\big)\,{\vee}\,$`FP'}{Big}
\label{eeExample3-1}\\
& = \xjvar{$\lnot$`TN'}{big}\,\xjvar{`ionized',$t$'}{big} + \xjvar{`FP'}{big}
\label{eeExample3-2}
\end{align}
Here, $\tau^{(j)}$ is the time delay between photoionization and the detection signal, and `TN' and `FP' are propositions of misdetection. 
Physically, `TN' represents a type of misdetection where no output signal is generated by the photomultiplier tube despite photoionization occurring (signal loss), and `FP' represents a type of misdetection where an output signal is generated despite no photoionization occurring (amplification noise). The compatibility condition between propositions `ionized', `TN', and `FP' holds if the characteristic operators of `TN' and `FP' have the form $\xop{`TN'}{}\,{=}\,p_{\mathrm{TN}}\hat{1}$, $\xop{`FP'}{}\,{=}\,p_{\mathrm{FP}}\hat{1}$ ({\S}\ref{sDeterministic}). 
Assuming that atomic photoionization and electron number amplification are independent phenomena ({\S}\ref{sDeterministic}), and ignoring the variation in time delay $\tau^{(j)}$, we can evaluate the expectation value of Eq.\eqref{eeExample3-1} as follows:
\begin{align}
\big\langle \xvar{`output,$t+\tau$'}{big} \big\rangle
& \cong p_{\mathrm{TN}}\,\big\langle \xvar{`ionized,$t$'}{big} \big\rangle
+ p_{\mathrm{FP}}.
\label{eeExample3-3}
\end{align}
Eqs.\eqref{eeExample3-1}--\eqref{eeExample3-3} appropriately represent both the discontinuous and continuous aspects of the photodetector response.
\par
Next, let us consider Schrödinger's cat thought experiment triggered by the output pulse of a photomultiplier tube. 
In our theory, we can phenomenologically represent the life or death of the cat using the following logical variable:
\begin{align}
\xjvar{`dead,$t$'}{}
\equiv \begin{cases}
  1, & \mbox{The cat has been dead by time $t$},\\
  0, & \mbox{Otherwise}.
\end{cases}
\label{eeExample3-4}
\end{align}
Also, we can represent the assumption of the thought experiment (the cat dies if the atom is photoionized) by the following deterministic relation:
\begin{align}
\xjvar{`dead,$t$'}{big}
= \xjvar{`output,$t$'}{big}
= \xjvar{`ionized,$t$'}{big}
,\quad \forall{j}
\label{eeExample3-5}
\end{align}
For simplicity, we have omitted the propositions of erroneous determinations.
Here, if we define the proposition `alive' as the negation of the proposition `dead', we see that `dead' and `alive' are mutually exclusive ({\S}\ref{sDeterministic}) by the laws of Boolean logic (Assumption B6):
\begin{align}
\xjvar{`dead,$t$'$\wedge$`alive,$t$'}{big}
= \xjvar{`dead,$t$'$\wedge\lnot$`dead,$t$'}{big}
= 0
,\quad \forall{j}
\label{eeExample3-6}
\end{align} 
By taking the expectation value of Eq.\eqref{eeExample3-5} and using the calculation result of the photoionization rate (), we obtain
\begin{align}
\big\langle \xvar{`dead,$t$'}{big} \big\rangle
= \big\langle \xvar{`output,$t$'}{big} \big\rangle
= \big\langle \xvar{`ionized,$t$'}{big} \big\rangle
= 1 - \exp\big(\,{-}\,\gamma_{\mathrm{1s}}{t}\,\big).
\label{eeExample3-8}
\end{align}
This allows us to explain the exponential decay of the cat's survival probability ($\langle \xvar{`alive,$t$'}{} \rangle$ $\,{=}\,1\,{-}\,\langle \xvar{`dead,$t$'}{} \rangle$) based only on the thought experiment's assumption and Assumptions B1-B14.
\par
Furthermore, it is natural to write the stationary states of the entire system including the cat and hydrogen atom as
\begin{align}
|\mathrm{alive},\mathrm{1s}\rangle
\equiv |\mathrm{alive}\rangle^{\!\mathrm{cat}} |\mathrm{1s}\rangle^{\!\bm r}
,\quad
|\mathrm{dead},{\bm p}\rangle
\equiv |\mathrm{dead}_{\bm p}\rangle^{\!\mathrm{cat}} |{\bm p}\rangle^{\!\bm r}
\label{eeExample3-12}
\end{align}
and express the state vector of the entire system $|\psi(t)\rangle$ as a superposition of stationary states:
\begin{align}
| \psi(t) \rangle
& = c_{\mathrm{alive},\mathrm{1s}}(t) |\mathrm{alive},\mathrm{1s}\rangle 
+ \iiint c_{\mathrm{dead},{\bm p}}(t) |\mathrm{dead},{\bm p}\rangle d{\bm p}
\label{eeExample3-13}\\
& = c_{\mathrm{alive},\mathrm{1s}}(t) |\mathrm{alive}\rangle^{\!\mathrm{cat}} |\mathrm{1s}\rangle^{\!\bm r}
+ \iiint c_{\mathrm{dead},{\bm p}}(t) |\mathrm{dead}_{\bm p}\rangle^{\!\mathrm{cat}} |{\bm p}\rangle^{\!\bm r} d{\bm p}.
\label{eeExample3-14}
\end{align}
In this case, by defining the density operator of the entire system as
\begin{align}
\hat{\rho}(\,t\,|\mbox{`${\mathrm{A}}_{0},0$'})
= \big| \psi(t) \big\rangle \big\langle \psi(t) \big|
\label{eeExample3-15}
\end{align}
and using the general properties of entangled subsystems, we see that the deterministic relation in equation (399) must exist.
\par
In addition, in our theory, the density operator, evolving continuously according to the quantum equation of motion, changes discontinuously only when the ensemble is updated ({\S}\ref{sJoint}). 
When we condition the density operator on the life or death of the cat, we obtain equations equivalent to the projection postulate, namely
\begin{align}
\hat{\rho}\,\big(\,t_{\mathrm{m}}\,|\,\mbox{`alive,\,$t_{\mathrm{m}}$'}\,\big)
& \equiv \frac{| {\mathrm{alive}} \rangle^{\!\mathrm{cat}} {}^{\mathrm{cat}}\!\langle {\mathrm{alive}} |
\,\hat{\rho}(\,t\,|\mbox{`${\mathrm{A}}_{0},0$'})\,
| {\mathrm{alive}} \rangle^{\!\mathrm{cat}} {}^{\mathrm{cat}}\!\langle {\mathrm{alive}} |}
{{\mathrm{Tr}}\big[\,| {\mathrm{alive}} \rangle^{\!\mathrm{cat}} {}^{\mathrm{cat}}\!\langle {\mathrm{alive}} |
\,\hat{\rho}(\,t\,|\mbox{`${\mathrm{A}}_{0},0$'})\,
| {\mathrm{alive}} \rangle^{\!\mathrm{cat}} {}^{\mathrm{cat}}\!\langle {\mathrm{alive}} |\,\big]}
\label{eeExample3-21}\\
& = | {\mathrm{alive}} \rangle^{\!\mathrm{cat}} {}^{\mathrm{cat}}\!\langle {\mathrm{alive}} |
\otimes |\mathrm{1s}\rangle^{\!\bm r}{}^{\!\bm r}\!\langle\mathrm{1s}|
\label{eeExample3-22}\\ 
\hat{\rho}\,\big(\,t_{\mathrm{m}}\,\big|\,\mbox{`dead,\,$t_{\mathrm{m}}$'}\,\big)
& \equiv \frac{| {\mathrm{dead}} \rangle^{\!\mathrm{cat}} {}^{\mathrm{cat}}\!\langle {\mathrm{dead}} |
\,\hat{\rho}(\,t\,|\mbox{`${\mathrm{A}}_{0},0$'})\,
| {\mathrm{dead}} \rangle^{\!\mathrm{cat}} {}^{\mathrm{cat}}\!\langle {\mathrm{dead}} |}
{{\mathrm{Tr}}\big[\,| {\mathrm{dead}} \rangle \langle {\mathrm{dead}} |
\,\hat{\rho}(\,t\,|\mbox{`${\mathrm{A}}_{0},0$'})\,
| {\mathrm{dead}} \rangle^{\!\mathrm{cat}} {}^{\mathrm{cat}}\!\langle {\mathrm{dead}} |\,\big]}
\label{eeExample3-23}\\
& = | {\mathrm{dead}} \rangle^{\!\mathrm{cat}} {}^{\mathrm{cat}}\!\langle {\mathrm{dead}} |
\otimes |\mathrm{sca}\rangle^{\!\bm r}{}^{\!\bm r}\!\langle\mathrm{sca}|.
\label{eeExample3-24}
\end{align}
Here, we defined the state vector of the atom's scattering state by
\begin{align}
|\mathrm{sca}\rangle^{\!\bm r}
\equiv \iiint c_{\mathrm{dead},{\bm p}}(t) |{\bm p}\rangle^{\!\bm r} d{\bm p}.
\label{eeExample3-25}
\end{align}
\par
Thus, we have explained the following properties of Schr\"{o}dinger's cat thought experiment:
\begin{itemize}
\item Deterministic relationship between continuous individual events (Eq.\eqref{eeExample3-5})
\item Exclusive relationship between the cat's life and death (`dead,\,$t$' and `alive,\,$t$') in individual measurements (Eq.\eqref{eeExample3-6})
\item Exponential decay of the probability that the cat is alive (${\mathrm{Pr}(\mbox{`alive,\,$t$'})}$) (Eq.\eqref{eeExample3-8})
\item Coherent superposition of state vectors (Eq.\eqref{eeExample3-14})
\item Discontinuous change of the density operator accompanying information update (Eqs.\eqref{eeExample3-21}--\eqref{eeExample3-24})
\end{itemize}
Therefore, in our theory, assumptions of non-causal collapse of the state vector, world splitting, or decoherence by the external environment are unnecessary to explain Schr\"{o}dinger's cat thought experiment.

\end{document}